\documentclass[final,3p,onecolumn,times]{elsarticle}

\usepackage{amssymb,amsmath,amsfonts,amsthm}

\usepackage{hyperref}

\usepackage{graphicx}

 \usepackage{threeparttable}
 \usepackage{multirow}
\usepackage{CJK} 

\usepackage[usenames]{color}
\usepackage{cancel}
\usepackage[normalem]{ulem}

\begin{document}
\begin{CJK}{GBK}{song}  

\begin{frontmatter}

\title{Cycling in stochastic general equilibrium 
}
\author{Zhijian Wang$^{1}$ and Bin Xu$^{1}$}
\cortext[cauthor]{Corresponding author. email: wangzj@zju.edu.cn. }
\address{
$^{1}$   Experimental Social Science Laboratory, Zhejiang University, Hangzhou 310058, China \\
 }

\date{\today}
\begin{abstract}
By generalizing the measurements on the game experiments of mixed strategy Nash equilibrium, we study the dynamical pattern in a representative dynamic stochastic general equilibrium (DSGE). The DSGE model describes the entanglements of the three variables (output gap [$y$], inflation [$\pi$] and nominal interest rate [$r$]) which can be presented in 3D phase space. We find that, even though the trajectory of $\pi\!-\!y\!-\!r$ in phase space appears  highly stochastic, it can be visualized and quantified. It exhibits as clockwise cycles,  counterclockwise cycles and weak cycles, respectively, when projected onto $\pi\!-\!y$, $y\!-\!r$ and $r\!-\!\pi$ phase planes. We find also that empirical data of United State (1960-2013) significantly exhibit same cycles. The resemblance between the cycles in general equilibrium and the cycles in mixed strategy Nash equilibrium suggest that, there generally exists dynamical fine structures accompanying with equilibrium. The fine structure, describing the entanglement of the non-equilibrium (the constantly deviating from the equilibrium), displays as endless cycles.
\end{abstract}



\end{frontmatter}
~~\\
~~\\
~~\\
~~\\
\begin{flushright}
   Cycles are not, like tonsils, separable
things that \\ might be treated by themselves, but are, \\like
the beat of the heart, of the essence \\ of the organism that
displays them. \\~\\| J. A. Schumpeter (1939) \cite{schumpeter1939business}
\end{flushright}

\clearpage

\tableofcontents
\clearpage

\section{Introduction}

\subsection{The general equilibrium in academical study and policy engineering}
Equilibrium is the central concept in economics. Like Nash equilibrium in game theory, general equilibrium theory is essential in macroeconomics. 
As with all models, general equilibrium theory is an abstraction from a real economy; it is proposed as being a useful model, both by considering equilibrium as long-term expectations and by considering actual fluctuations 
as deviations 
from equilibrium \cite{Mankiw2010,gali2009monetary,de2012lectures}. 

In past several decades, basing on the models of general equilibrium
(e.g., real business cycles, New Keynesian model),  many developments have occurred in
the ivory towers of academia.  In recent years many models
were implemented empirically as tools for policy
analysis. For example, the Dynamic Stochastic General Equilibrium (DSGE)
models  are increasingly applied in central banks \cite{gali2009monetary,de2012lectures,tovar2009dsge}.
However, these models, together with the empirical data, have not been well identified  \cite{fernandez2010econometrics,canova2007methods,consolo2009statistical,koop2013identification}.
From the standpoint of macroeconomic engineering, the foundation of the applications is not solid enough \cite{gali2009monetary,de2012lectures,mankiw2006macroeconomist}.

DSGE model, describing the entanglements of the three variables (output gap [$y$], inflation [$\pi$] and nominal interest rate [$r$]), is a model of general equilibrium. Its solution can be approximately regarded as a \emph{mixed equilibrium steady state} 
in long run. 
In short run, however,
the three variables could constantly deviate from equilibrium due to instinct noise.
Once one variables deviates, the others would response to this deviation,
and then the response itself appears also as a deviation.
Such deviations and responses will entangle and the dynamical pattern could emerge. We will visualize and quantify the pattern in this note.

\subsection{Aims, contents and structure}

In this note, by generalizing the measurements introduced in the experimental investigations of
 Matching Pennies games
\cite{XuWang2011ICCS}\cite{xu2012periodic}\cite{wang2012evolutionary} and Rock Paper Scissors
games \cite{XuZhouWang2013}\cite{WangXuZhou2014socialcycling} on mixed strategy Nash equilibrium,
we promote a set of visualized and quantified measurements
to illustrate the cycles | the entanglement of the deviations from the general equilibrium among the three variables ($\pi, y, r$).
Specifically,
 (1) we illustrate the long-run cycling and distribution with the time series generated by the DSGE model (for more details, see section \ref{SI:modelandsimulation} in SI), with which (2) we test whether or not the same cycling exists in empirical data. Instead of evaluating the precision of the DSGE model, 
we focus on identifying the patterns in DSGE and in empirical data.

We illustrate our results with theoretical and empirical time series.
The theoretical time series are generated with a representative DSGE model \cite{de2012lectures}\cite{Linnemann2014}\cite{gatti2013lectures} rooting in macroeconomics textbooks ~\cite{gali2009monetary,Mankiw2010}. The empirical data comes from World Bank database (United States, 1960-2013).
 In Section \ref{sec:results}, we visualize and quantify the cyclic patterns in the DSGE first,
and then visualize and quantify the cycles in empirical data. 
 In Section \ref{sec:discussion}, we discuss the nature of the commensal of cycle and equilibrium, the related  literatures and the further questions.
 The method and material, including the details on the model, the simulation, the empirical data, the measurements and additional results, are at last.

\section{Results\label{sec:results}}
 Result mainly includes two points. First, the cycling can be visualized and quantified in DSGE model. Second,  same cycling can be observed in empirical data. 
 We hope this result is helpful to establish the picture | In general equilibrium, there exists cyclic pattern which can be visualized and quantified by definitive measurements.

\begin{figure}
    \begin{center}
     \includegraphics[width=.40\linewidth]{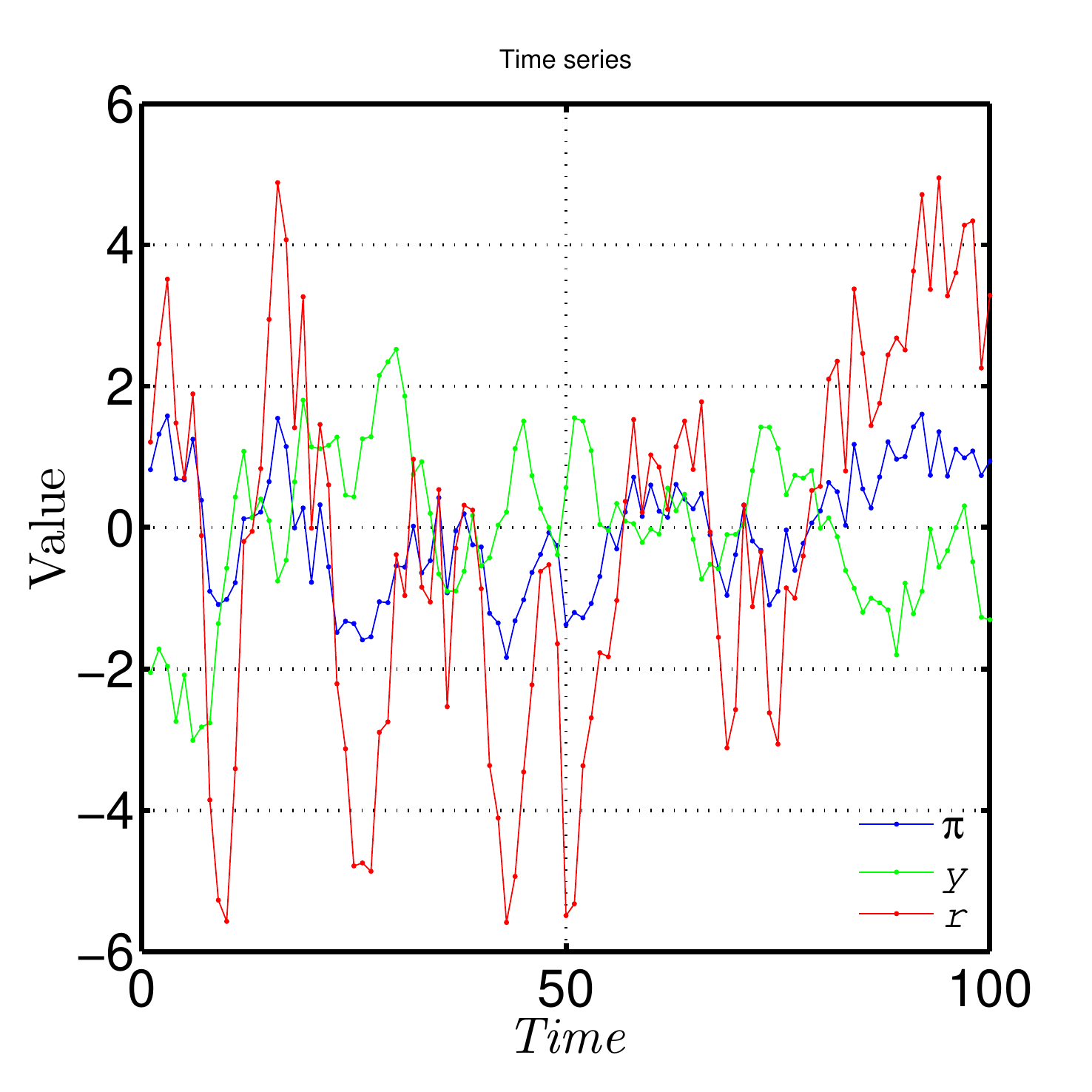}
     \includegraphics[width=.40\linewidth]{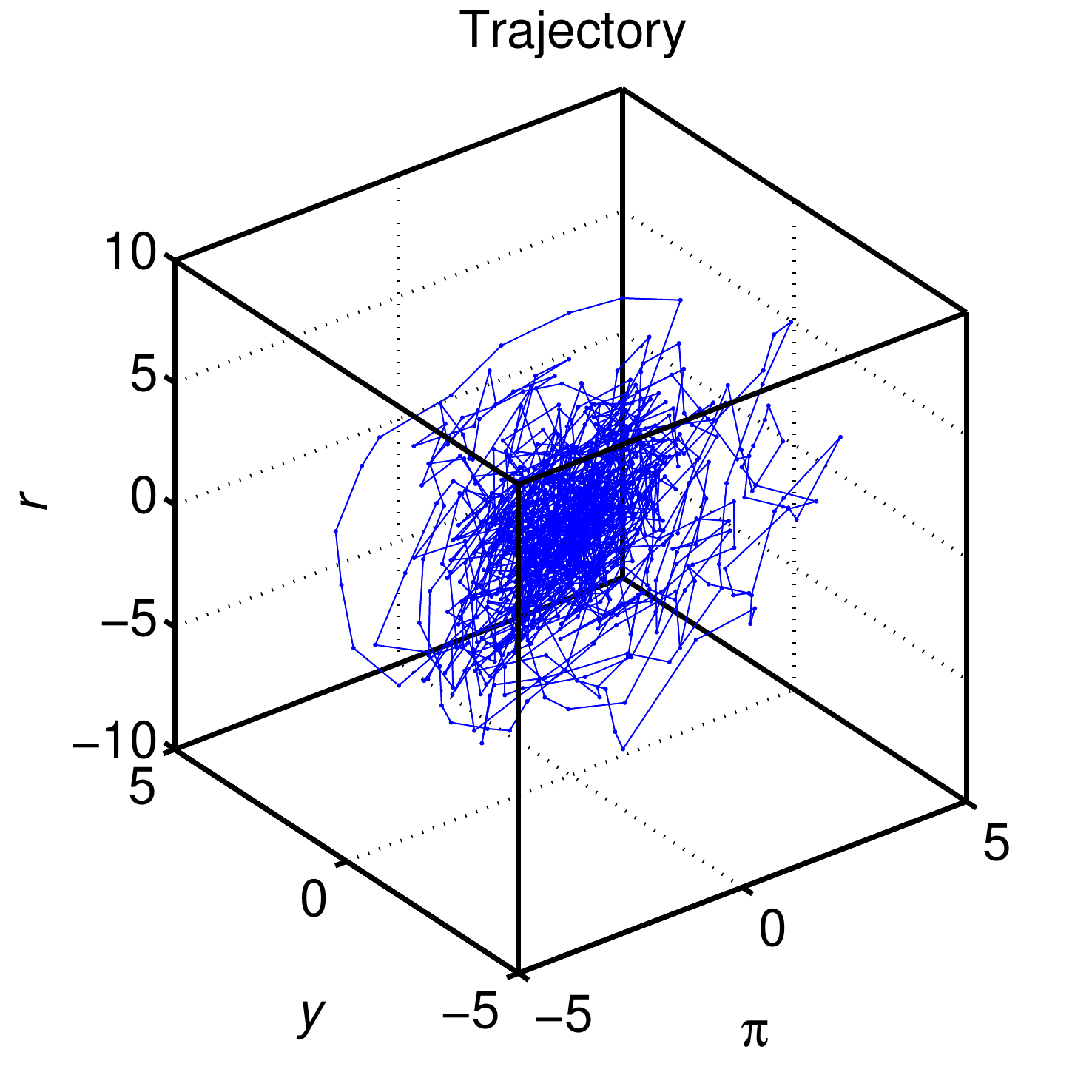}
     \includegraphics[width=.40\linewidth]{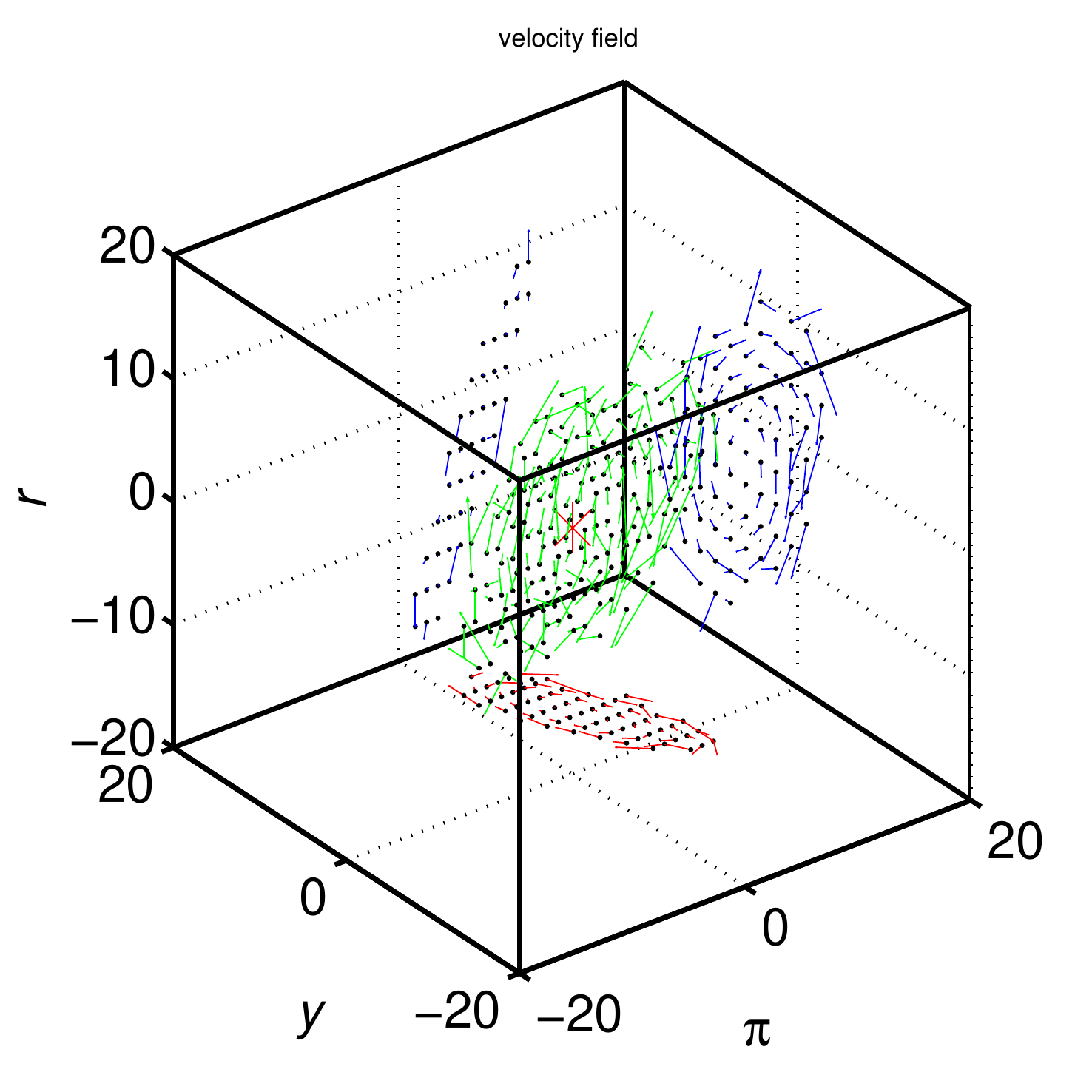}
     \includegraphics[width=.40\linewidth]{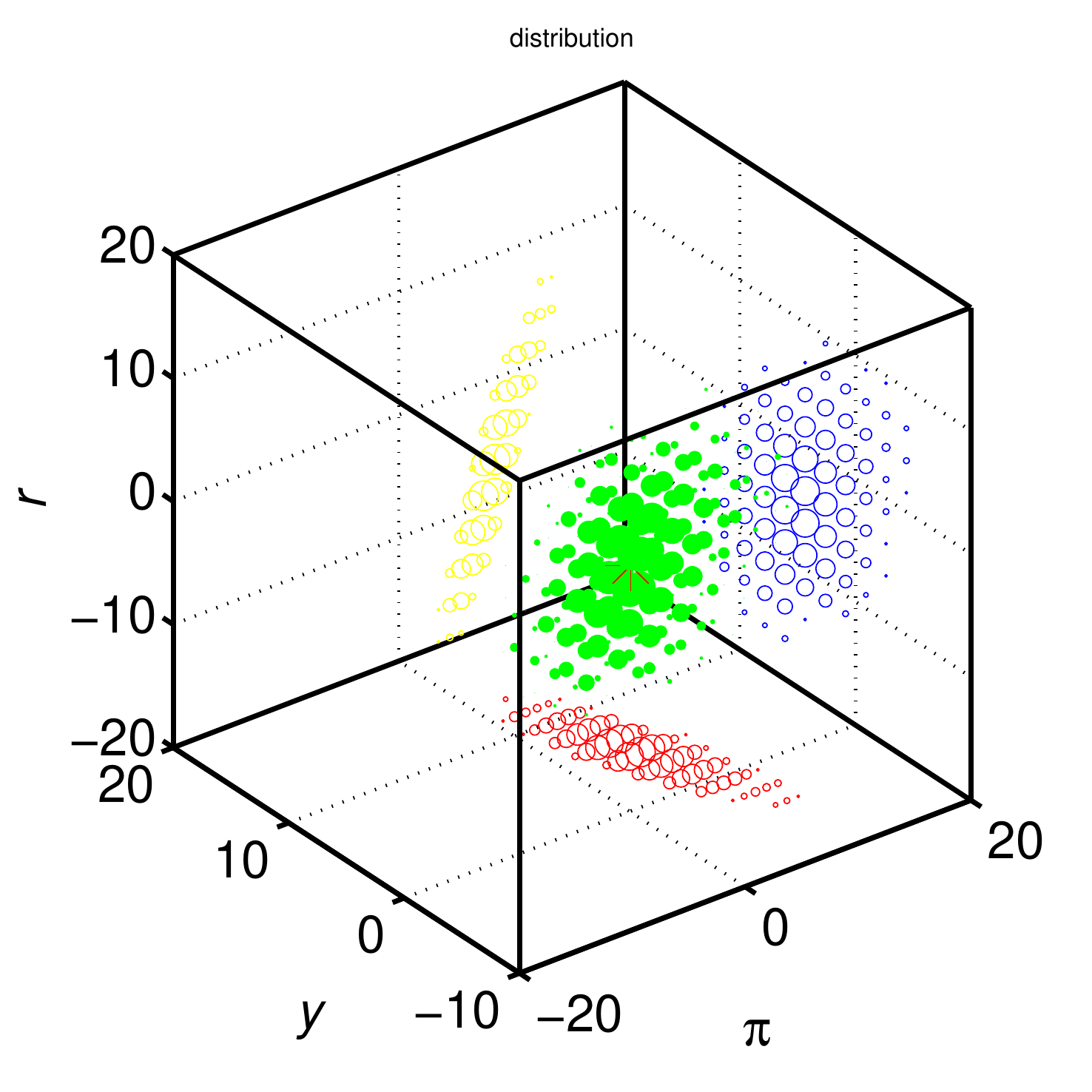}
    \end{center}
    \caption{
        \label{fig:3Ddistr}   Evolutionary body presented in ($\pi$-$y$-$r$) 3D phase diagram. Data comes from the Monte Carlo simulation basing on a representative DSGE model \cite{de2012lectures}. The simulation is 20000 periods repeated. (a) The fluctuations of three variables along time. The 100 periods of time series is randomly chosen among the simulation result.
 (b) The evolutionary trajectory of state ($\pi,y,r$) in 3D phase space. (c) The velocity field in the 3D discrete state space, and its projections onto the three 2D planes, i.e. the $\pi-y$ plane, the $y-r$ plane and the $r-\pi$ plane. The velocity vectors are plotted in 1:1 scale. In each of the 3 dimensions, the resolution of the observation 10, and then there are 10$^3$ lattices (club cells) possible observation. Each cell is employed as a testing point for distribution and velocity respectively (for more details on measurement, see section \ref{SI:velocityanddistribution} in SI). (d) The distribution of the state ($pi,y,r$) in 3D space and its projections on the three 2D phase planes. The size is related to the frequency of the state, in which larger cycle size means higher probability observed at this point. 
       }
  \end{figure}

\clearpage
\subsection{Cyclic motion in DSGE}
In order to identify the cycles in DSGE, we use a representative DSGE model, namely behaviorial macroeconomics model suggested by De Grauwe \cite{de2012lectures}. The model parameters are specified by US data \cite{gali2009monetary}. This model has inherited the main characters of classical macroeconomics dynamics | First,  when all the stochastic terms in its dynamics equations are ignored to be zero, this model will recede to classical dynamics model, its solution is $[\pi,y,r]$=$[0,0,0]$ and being in equilibrium. Second, when the system is exogenously shocked by  $\pi$, or $y$, or $r$ at time $t$ and no other noise since then, the general equilibrium will recover in several periods (for more details, see  Section \ref{sec:ge_shock}). These two points have been the contents in macroeconomics textbook.


DSGE describes the real economy, in which the noise terms can not be ignored. To obtain the regularity in DSGE, simulation have to be employed. Monte Carlo simulations are conducted to generate time series (method see Section \ref{SI:modelandsimulation}). The results can be shown directly. A sample time series simulated is shown in Fig~\ref{fig:3Ddistr} (a). A sample evolutionary trajectory is shown in Fig~\ref{fig:3Ddistr} (b) which appearing  highly disorder.

\begin{figure}
    \begin{center}
     \includegraphics[width=.30\linewidth]{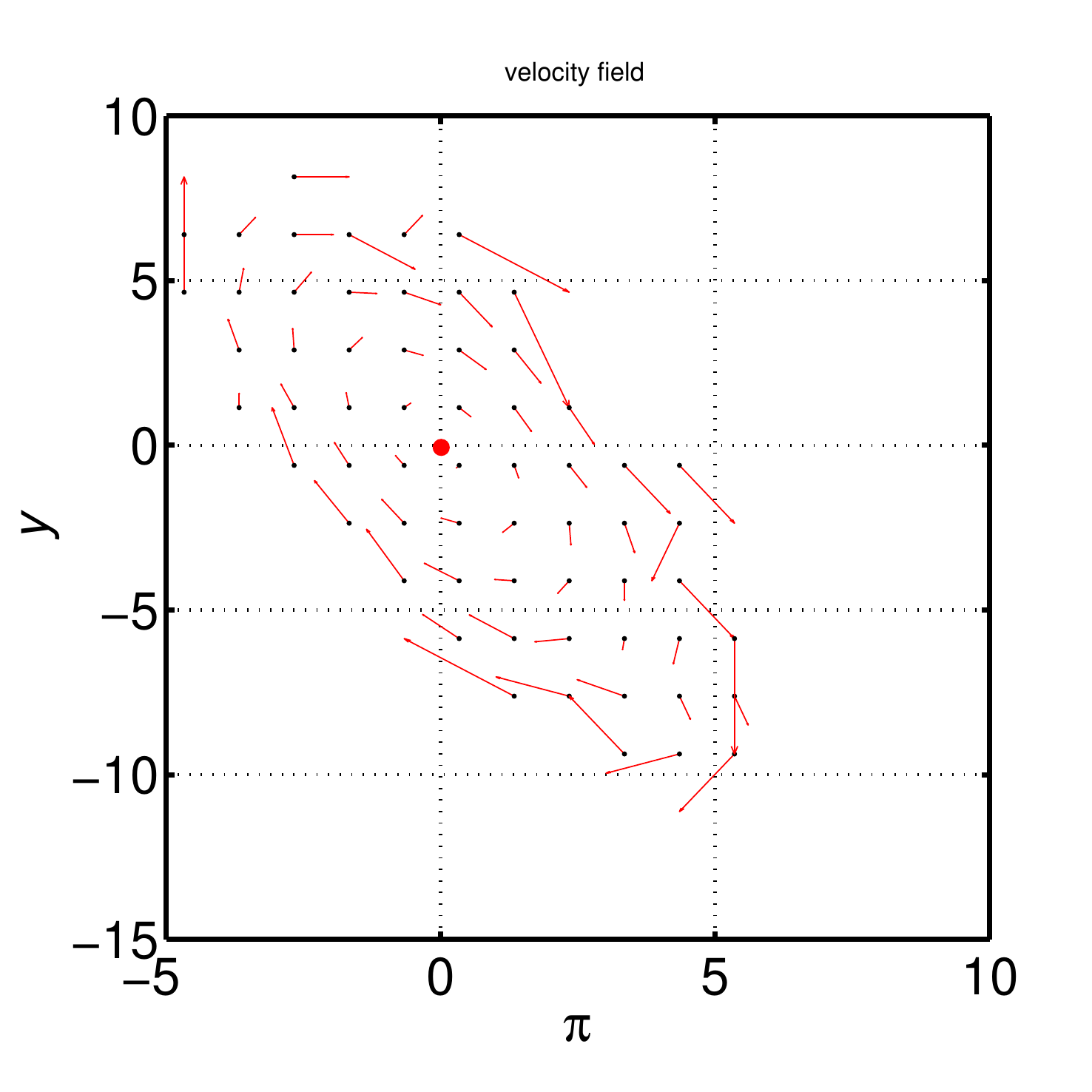}
     \includegraphics[width=.30\linewidth]{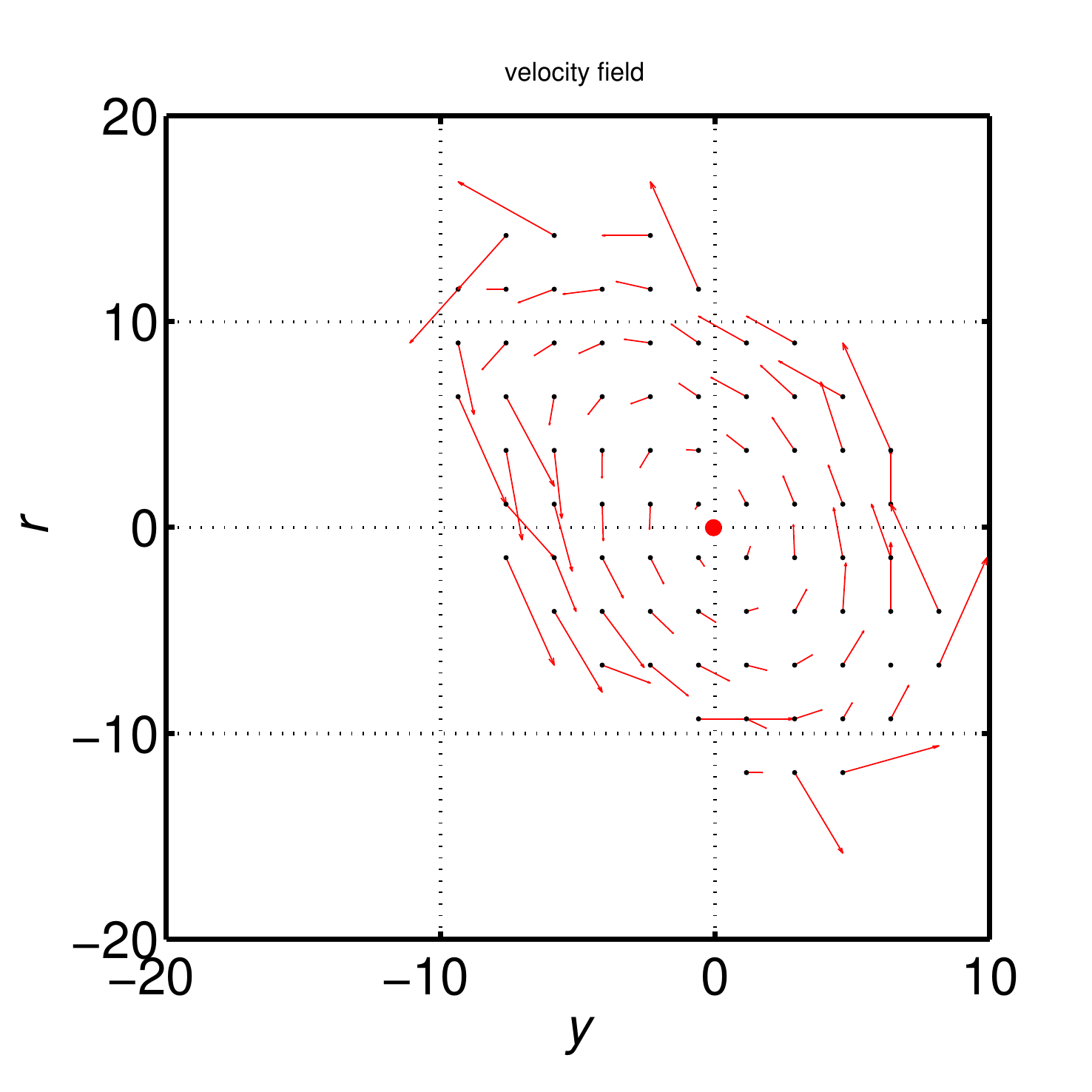}
     \includegraphics[width=.30\linewidth]{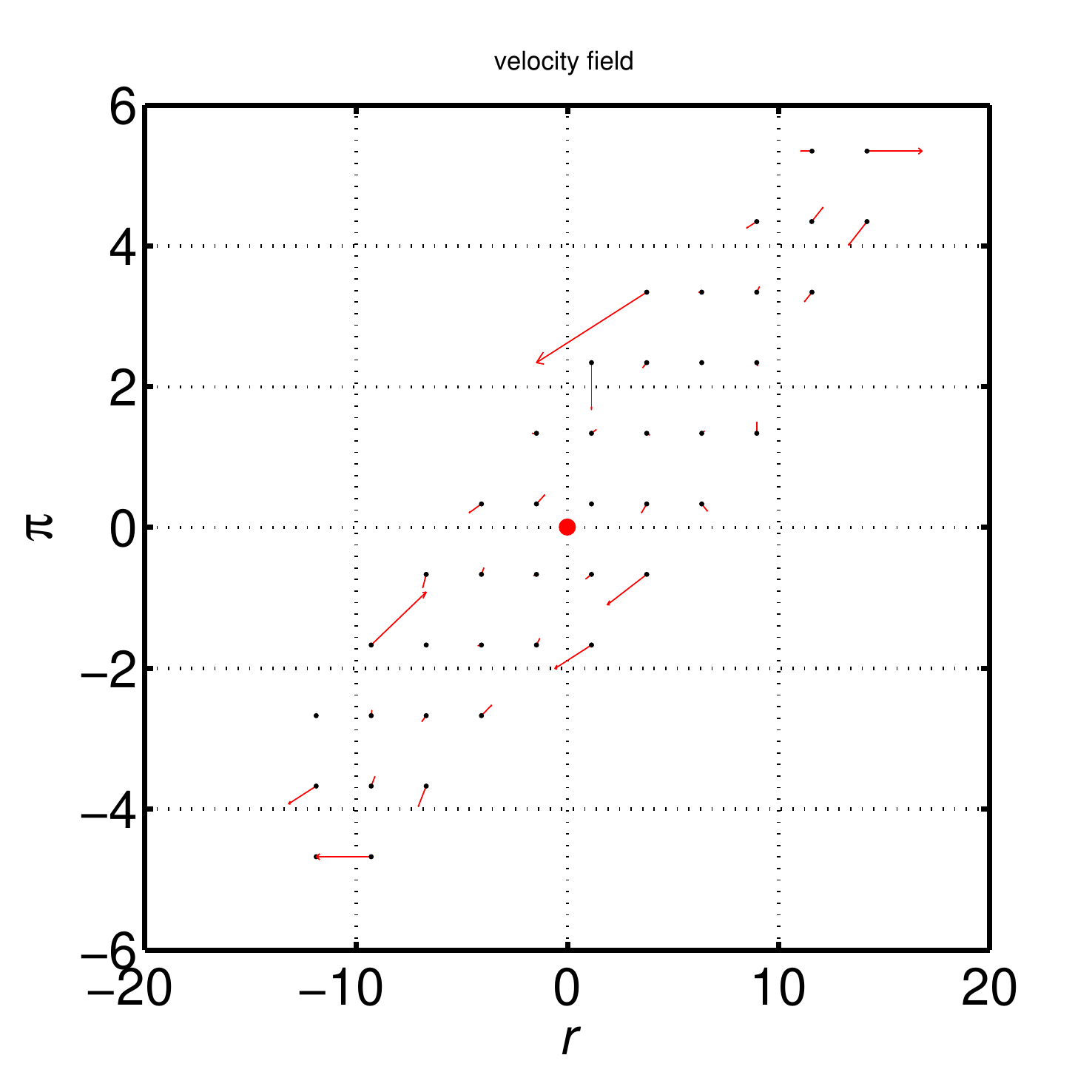}
     \includegraphics[width=.30\linewidth]{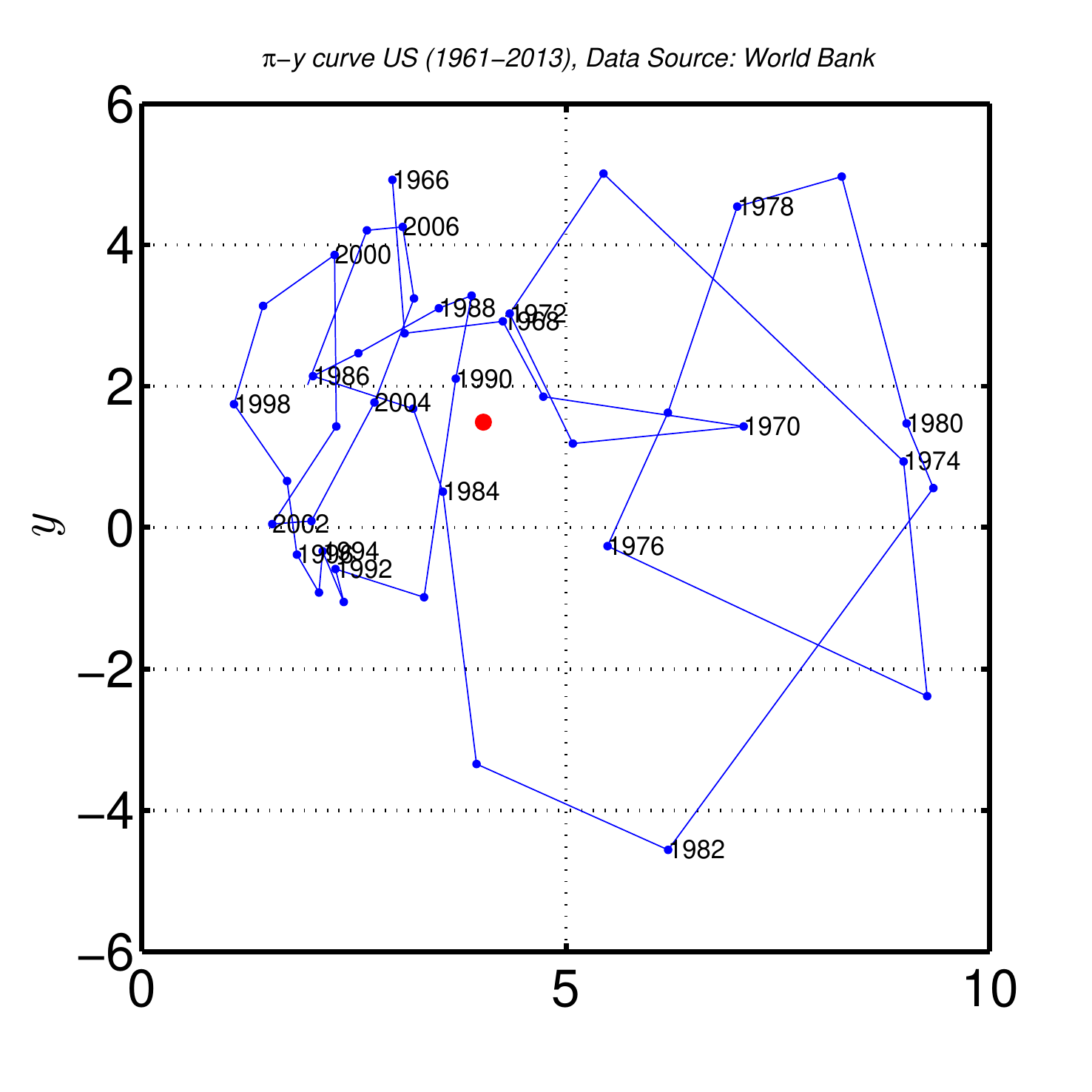}
     \includegraphics[width=.30\linewidth]{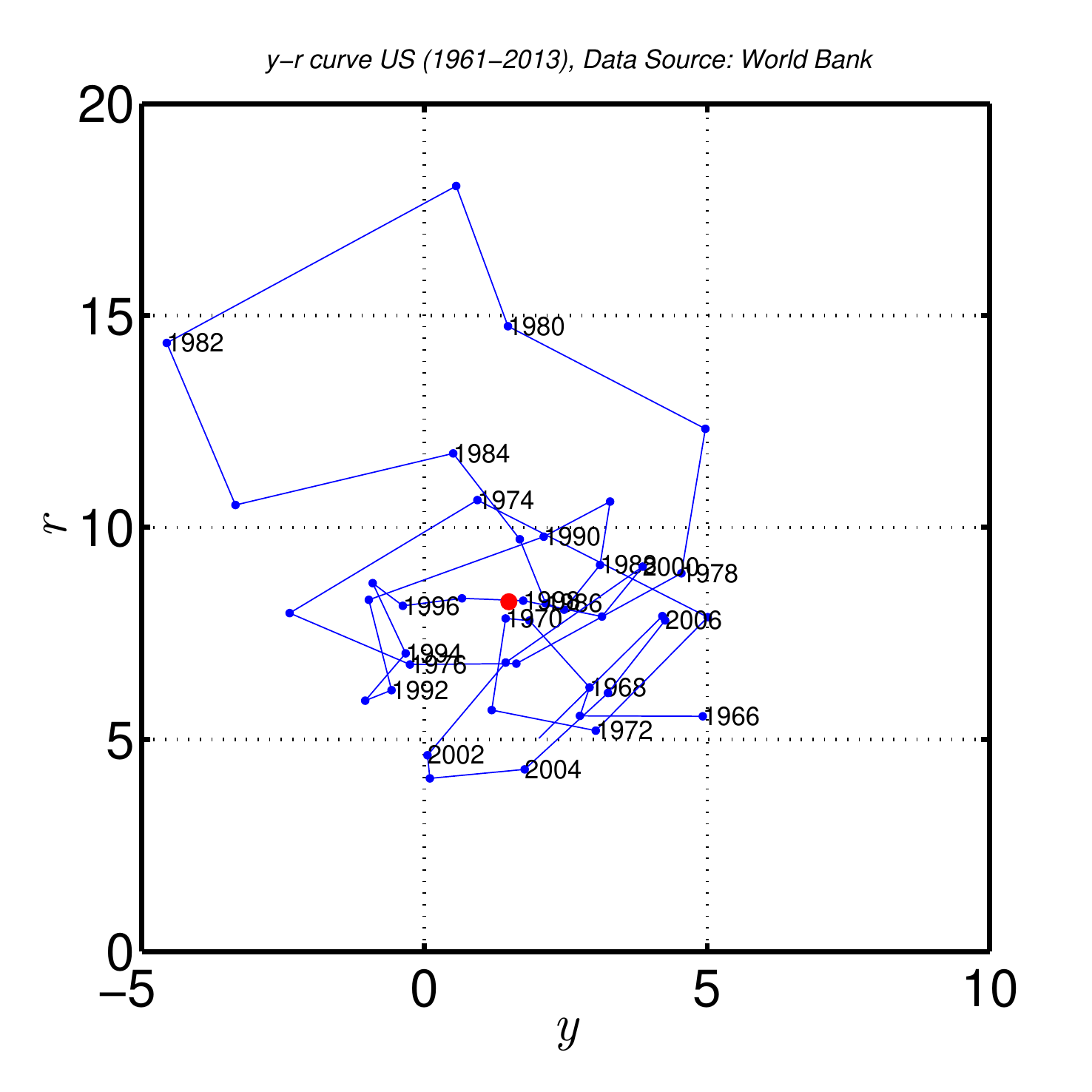}
     \includegraphics[width=.30\linewidth]{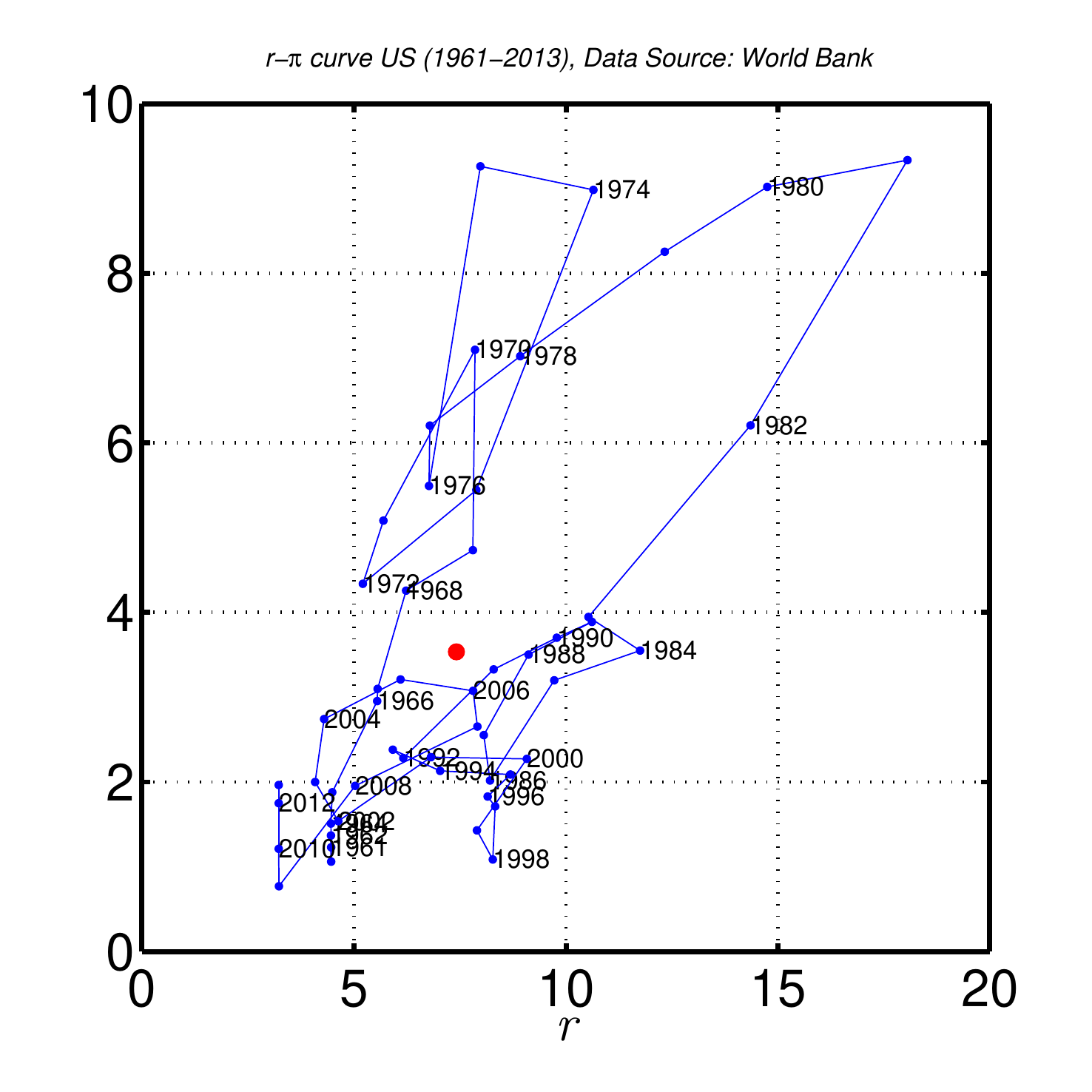}
    \end{center}
    \caption{
        \label{fig:3x3}   The up panel are the phase space projection view of the 3D velocity vector field (shown in Fig.~\ref{fig:3Ddistr}) in (a) $\pi$-$y$, (b) $y$-$r$ and (c) $r$-$\pi$ 2D space respectively.
        The low panel illustrates the empirical data trajectories in (d) $\pi$-$y$, (e) $y$-$r$ and (f) $r$-$\pi$ phase space (data source: 1961-2013 US data from World-Bank), respectively. The red dot is the  simple arithmetic average of the trajectory.
       }
  \end{figure}

Fig~\ref{fig:3Ddistr} (c) is the visualized and quantified pattern of the velocity field in $\pi$-$y$-$r$  3D  phase space. The method is shown in Section \ref{SI:velocityanddistribution}.
Fig. \ref{fig:3x3} (a)-(c) illustrate
the projection of the 3D evolutionary velocity field shown in
Fig. \ref{fig:3Ddistr} (c) to the three 2D phase planes.
It is visible that, the dynamic pattern is strongly clockwise in $\pi-y$ plane,
is strongly counterclockwise in $y-r$ plane,
and is weak or no cycles in $r-\pi$, respectively.
Clearly, the cycling can be geometrically visualized. Meanwhile, the velocity vector field quantifies the strength as well as the direction the velocity. It is clear that, at the general equilibrium state $[0,0,0]$, the velocity vector equal $[0, 0, 0]$.


We provide an interpretation for the velocity field in DSGE | Constant stochastic shock is the inherent characteristic of the DSGE model, and  also is the feature of the real world. It is hard to hold the world in the zero noise situation which is always the condition of analysis. On the contrary, the stochastic shock is constant. Therefore, the departure from equilibrium might be the normality. So the problem is not the question that how long the system will come back to the equilibrium, but the question that where the system will go, in what direction and at what speed, under the constant stochastic shocks. We answer these questions by the velocity field.


\subsection{Cyclic motion in  empirical data}
  
Fig. \ref{fig:3x3} (d)-(f) present of the trajectory of the empirical data of United State from 1960 - 2013 (For more details, see Table~\ref{tab:real} and Section \ref{SI:realdata} in SI) in the three 2D phase space. Comparing with the pattern shown in Fig. \ref{fig:3x3} (a)-(c) respectively, we can find that the cyclic trends appear similar.

We need to conform this similarity quantitatively. To this aim, we propose a two-step
angular momentum $L^{(2)}$ measurement to quantify the cyclic pattern. According to its definition, $L^{(2)} > 0$ indicates that the motion is counterclockwise,  $L^{(2)} < 0$ indicates that the motion is clockwise,  $L^{(2)}= 0$ indicates that no cycle exists. Meanwhile, cyclic motion is stronger when $|L^{(2)}|$ larger (for more details, see Section \ref{SI:nSampleingL}).

Table \ref{tab:theoL2} (right panel) lists empirical $L^{(2)}$ by
measuring on the empirical trajectories illustrated in Fig. \ref{fig:3x3} (d)-(f). The results clearly indicates that, the motion is clockwise cyclic in $\pi-y$ plane in significant ($L^{(2)} < 0$, $p$=0.0311, 2-tailed Wilcoxon test, $n$=51)
and is counterclockwise cyclic in $y-r$ plane in significant ($L^{(2)}> 0$, $p$=0.0002, 2-tailed Wilcoxon test, $n$=51),
and is not clearly cyclic in $r-\pi$ plane ($L^{(2)} \simeq 0$, $p$=0.5611, 2-tailed Wilcoxon test, $n$=51).

\subsection{Comparison of the Cyclic motion}
  
Table \ref{tab:theoL2} (left panel) lists  theoretical $L^{(2)}$ of the DSGE model. This is
 supported by three theoretical analysis. (1)  Measuring directly in the simulated trajectory. (2) Calculating directly with the simplified transition matrix of the model (for details, see Section \ref{SI:SolveL2}). Results of these two points  are illustrated in Fig. \ref{fig:TheoL2evaluted}. (3) Theoretical $L^{(2)}$ consists with the results from  simplified shocks analysis (for details, see Section \ref{sec:ge_shock}).

 Comparing the three $L^{(2)}$ components in Table \ref{tab:theoL2}, we came to the result that, the theoretical expectations meets  empirical data statistically. Because all the directions and the strengths of the cycles  measured with  $L^{(2)}$ components meet qualitatively.

One the results mentioned above, two additional results are following. (1) We have test the robustness of the existence of cycles in real data and the results are positive (for   details, see Section \ref{SI:RobustnessLn}).  
(2) The distribution of DSGE, as illustrated in Fig~\ref{fig:3Ddistr} (d), meets the empirical distribution qualitatively (for details, see Section \ref{SI:distribution}).

\begin{table}
\begin{center}
\caption{\label{tab:theoL2}Theoretical and empirical $L^{(2)}$}
\begin{tabular}{|c|c c|cccc|} 
  \hline 
 Phase &   Cycle  & Expected & Sample &&Standard& Wilcoxon \\ 
 plane &   direction & $L^{(2)}$  & size  & Mean &	error& test ($p$)   \\ 
  \hline
  $\pi$-$y$ & clockwise     &$\bar{L}_{\pi-y}^{(2)} \ll 0$	& 51  &	$-$0.8634&	0.2712&0.0311 \\ 
  $y$-$r$ &  counterclockwise   &  $\bar{L}_{y-r}^{(2)} \gg 0  $  &51 &	1.4038 &	0.4748& 0.0002\\ 
  $r$-$\pi$ & weak clockwise  & $\bar{L}_{r-\pi}^{(2)}  \geq 0 $  &	51 &	0.1858&	0.2079&	0.5611\\ 
  \hline
\end{tabular}\\
%
%
%
\end{center}
\end{table}

\section{Discussion\label{sec:discussion}}

\subsection{Equilibrium and cycle in economics}
Nash equilibrium and general equilibrium appear to be different. As mentioned in  \cite{joosten2006walras},
in the spectrum covered by the economic science, dynamic theory of general equilibrium and the evolutionary game theory seemly take rather opposite positions. However, considering the analogies
between Walrasian t$\hat{a}$tonnement processes and Darwinian dynamics, these two equilibrium concepts are at the same catalog. Meanwhile, equilibrium and cycle appear to be different too. However, 
comparing with the persistently cycling 
in Matching pennies game and Rock
Paper Scissors game \cite{XuWang2011ICCS}\cite{xu2012periodic}\cite{wang2012evolutionary}\cite{XuZhouWang2013}\cite{WangXuZhou2014socialcycling}, the cycling in stochastic general equilibrium is natural, even though it is firstly visualized and quantified in this note.
  
Our results on cycle are closely related to nonequilibrium learning in theoretical macroeconomics (for more related literature, see review \cite{fudenberg2009learning}). In macroeconomics,
debating on cycles has lasted for long time. While some macroeconomists believe that the cycles are inherent in the fundamental operation of the economy, others argue that they are a response to external (i.e. exogenous) events 
\cite{schumpeter1939business,zarnowitz1985recent,mankiw1989real}. 
In a special multi-good market experiment on general equilibrium~\cite{Plott2004globalScarf}, endogenous prices cycles have been seen. Even though,  neither in empirical data nor in DSGE model the cycles can be observed with their measurements \cite{Plott2004globalScarf}, their results are important, especially  for understanding 
the Walrasian  t$\hat{a}$tonnement processes and error and trial processes  \cite{Plott2004globalScarf}.  
A field, called as behaviorial economics 
(e.g.~\cite{gintis2009game,duffy2008experimentalmacro,DuffyAgentBased,tesfatsion2002agent,fudenberg2009learning, de2012lectures}), is developing to establish the micro
 foundation of the macro dynamic behaviors, 
 which we believe will  help us to learn equilibrium and cycle
  better. According to the 
 observations in this note, we suggest the cycle, like equilibrium, is inherent.
 
\subsection{Methodology}

Phase diagram analysis (PDA) is very common in  macroeconomics \cite{Mankiw2010,shone2002economic}.
For example, the representative relationship | Phillips curve | is a empirical result from phase diagram analysis. 
In theoretical macroeconomics, phase diagrams are often used to qualitatively characterize the interaction between the financial
and goods markets through continuous time continuous dynamic equations \cite{gali2009monetary}, but rarely used to stochastic models.
A continuous model analysis is good at trend but poor at distribution. On the contrary, a stochastic model (e.g., DSGE) analysis is good at distribution but poor at trend \cite{evans2001learning}.
Dynamic pattern in empirical data can be visualized with PDA, but rare quantified at the same time.
Identificating the pattern in DSGE is a hard task \cite{canova2007methods,consolo2009statistical,koop2013identification}.
As show in Fig. \ref{fig:3Ddistr}, our measurement can identify the trend and distribution simultaneously.  
 
Two points about the measurement we hope to emphasis | The $L^{(2)}$ measurement employed in this note is of time reversal asymmetry. As mentioned in \cite{xu2012periodic}\cite{XuZhouWang2013}, time reversal asymmetry is critically important for test out the deterministic motion in high stochastic processes. This method roots in non-equilibrium statistical physics \cite{Evans2002experimental}. Second, different from existed measurement on cycle in economics, $L^{(2)}$ can be a more natural estimator for the existence of a cycle that does not require a subjective judgment on the choice of a center. Moreover, this measurement can report the local property of dynamics field. To the best of our knowledge, these two points have not been well recognized in identifying patterns in economics dynamics \cite{blaug2010famous,fernandez2010econometrics,canova2007methods,consolo2009statistical,
koop2013identification,shone2002economic,lang2008economists}.

\subsection{Future topic}

One of the big unanswered questions in macroeconomics is that: how to get a
tractable micro to macro model. Regrettably, difference from our
previous works in experimental games in which micro (e.g., conditional
response \cite{WangXuZhou2014socialcycling}) to macro can be found, in this work we cannot find this tractable relationship neither in model nor in data. We hope to see experiments, either in field or in laboratory, to reconstruct the cycles in general equilibrium.

We hope the cycles observed in the equilibrium, both Nash equilibrium and general equilibrium, could be a stimulator on further research, especially on fundamental concepts in economics. For example,  the fine structure of the IS-LM curve, the Goodwin cycles and so on \cite{Mankiw2010}. 
We believe that, with rigorously mathematical analysis, the dynamics equation model and the stochastic model (e.g., DSGE) can be merged and traced in phase space quantitatively instead of schematically. Because, in phase space, the entanglement of deviations from equilibrium can be measured definitely. 
 We hope these methods and results can be helpful to macroeconomic engineering, as well as to understanding equilibrium concept in economic science. \\

~~\\

\section{Method and material}
\subsection{Model and simulation\label{SI:modelandsimulation}}
The model comes from  \emph{The Lectures on Behavioral Macroeconomics} by Paul De Grauwe \cite{de2012lectures}, which can be called as New-Keynesian macroeconomic model  or as a dynamic stochastic general equilibrium (DSGE) Model. The model have the same framework as the general AS-AD model in textbooks  \cite{Mankiw2010}\cite{gali2009monetary}), and is regarded as a representative model for macroeconomics \cite{Linnemann2014}\cite{gatti2013lectures}. As the main propose of this note is to illustrate the cyclic entanglement of the deviations in general equilibrium and the methods, we use the algorithm of the behavioral model (see the appendix of the first chapter in  \cite{de2012lectures})  to generate the time series, but do not involve into the model.

The main variables of the time series, presented in Fig. \ref{fig:3Ddistr}, are output ($y_t$), inflation ($\pi_t$) and nominal interest rate ($r_t$). In our simulation, we do not change any parameter chose by the author. Before using this algorithm on this study, we replicated the results shown in the first chapter where the algorithm had been used by the author \cite{de2012lectures} and obtained same results shown in that chapter.

\subsection{Empirical data\label{SI:realdata}}
We use U.S. data to test whether the cycles exist in the phase space,
and comparing with the De Grauwe model of behaviorial macroeconomics \cite{de2012lectures}.
The data comes from the web site of World Bank accessed in Sept. 25, 2014.
\begin{table}
\begin{center}
\caption{\label{tab:real}Source of empirical data (United State, World Bank )}
\begin{tabular}{|c|c|c|c|}
  \hline
  Variable   symbol & Variable Name & series  code  & Year \\
  \hline
$\pi$ & Inflation, GDP deflator (annual \%)	   & NY.GDP.DEFL.KD.ZG & 1961-2013\\
  $i$ & Real interest rate (\%) & FR.INR.RINR & 1961-2013 \\
 $g_r$ & GDP growth (annual \%)	 & NY.GDP.MKTP.KD.ZG & 1961-2013 \\
  \hline
\end{tabular}
\end{center}
\end{table}
\subsubsection{Inflation rate | $\pi$}
The source data of $\pi$ can be obtained according to Table \ref{tab:real} directly.

\subsubsection{Nominal interest rates | $r$}
The relation between real and nominal interest rates and the expected inflation rate is given by the Fisher equation
        $1+r = (1+i) (1+\pi)$,
where $r$ is the nominal interest rate, $i$ is the real interest rate and $\pi$ is the expected inflation rate. Because of $i\ll 1$ and $\pi\ll1$ and then $i\pi$ is very small, as an approximate, we use
\begin{equation}\label{eq:Fisher_equation}
 r=i+\pi
\end{equation}
in this study. The source data of $i$ and $\pi$ can be obtained according to Table \ref{tab:real}.

\subsubsection{Output gap | $y$}

We use a moving average method to evaluate the potential GDP rate $\left(g_p\right)$ for the output gap $\left( y \right)$.
The calculation are explained  following.
Suppose within the year interval $[\alpha,\beta]$,
$T$ presents the number of years used to calculate the moving average.
For potential GDP growth rate for the year $t$, we use  a simple moving average  which is the unweighted mean of the real GDP growth rate $\left(g_p\right)$ between year [$t-T, t+T$]. The potential GDP growth rate at year $t$ is defined as
      \begin{equation}\label{eq:potentialgrowth_g}
       {g_{p}}(t) =  \frac{1}{2T+1}{\sum_{\tau=t-T}^{t+T}g_r(\tau)},
      \end{equation}
 in which, $g_r(\tau)$ is the real GDP growth rate at year $\tau$. Specially, the potential GDP growth rate for the year before $ \alpha+T $ is defined as
       ${g_{p}}(t)\mid_{t<\alpha+T} =  {g_{p}}(\alpha+T)$,
while for the year after $\beta-T$ as
      ${g_{p}}(t)\mid_{t>\beta-T} =  {g_{p}}(\beta-T)$. Then,      
the potential GDP at year, denoted as $t$ ${G_{p}}(t)$, can be defined as
      \begin{equation}\label{eq:potentialgdp_y}
       {G_{p}}(t) = G_\alpha\prod_{\tau=\alpha+1}^{t}\left(1+{g_{p}}(\tau)\right),
      \end{equation}
in which, $G_\alpha$ is the initial GDP at year $\alpha$. Then the percentage of the output gap at year $t$ is defined as
 \begin{equation}\label{eq:outputgap_y}
       y(t) = \frac{{G_{a}(t)} - {G_{p}}(t)}{{G_{p}}(t)} \times 100,
      \end{equation} \\
in which, $G_{a}(t)$ is the actual GDP (constant 2005 US\$,  coded as NY.GDP.MKTP.KD in World Bank database) at year $t$.
In our study case, we choose $T=5$, $\alpha=1960$, $\beta=2013$. Accordingly, the output gap $y(t)$ can be obtained and shown in sub-figure in Fig. \ref{fig:3x3}. Changing $T$ from 3 to 10, we do not observed significant difference on the main results in this note. The potential GDP in both end, $g_r$ betwwen 1960-1965 and 2008-2013, are approximately as $g_r (1965)$ and $g_r (2008)$, respectively. When cutting off the samples (1960-1965 and 2008-2013), we do not observe significant difference on main results.

\begin{figure}
    \begin{center}
     \includegraphics[width=.30\linewidth]{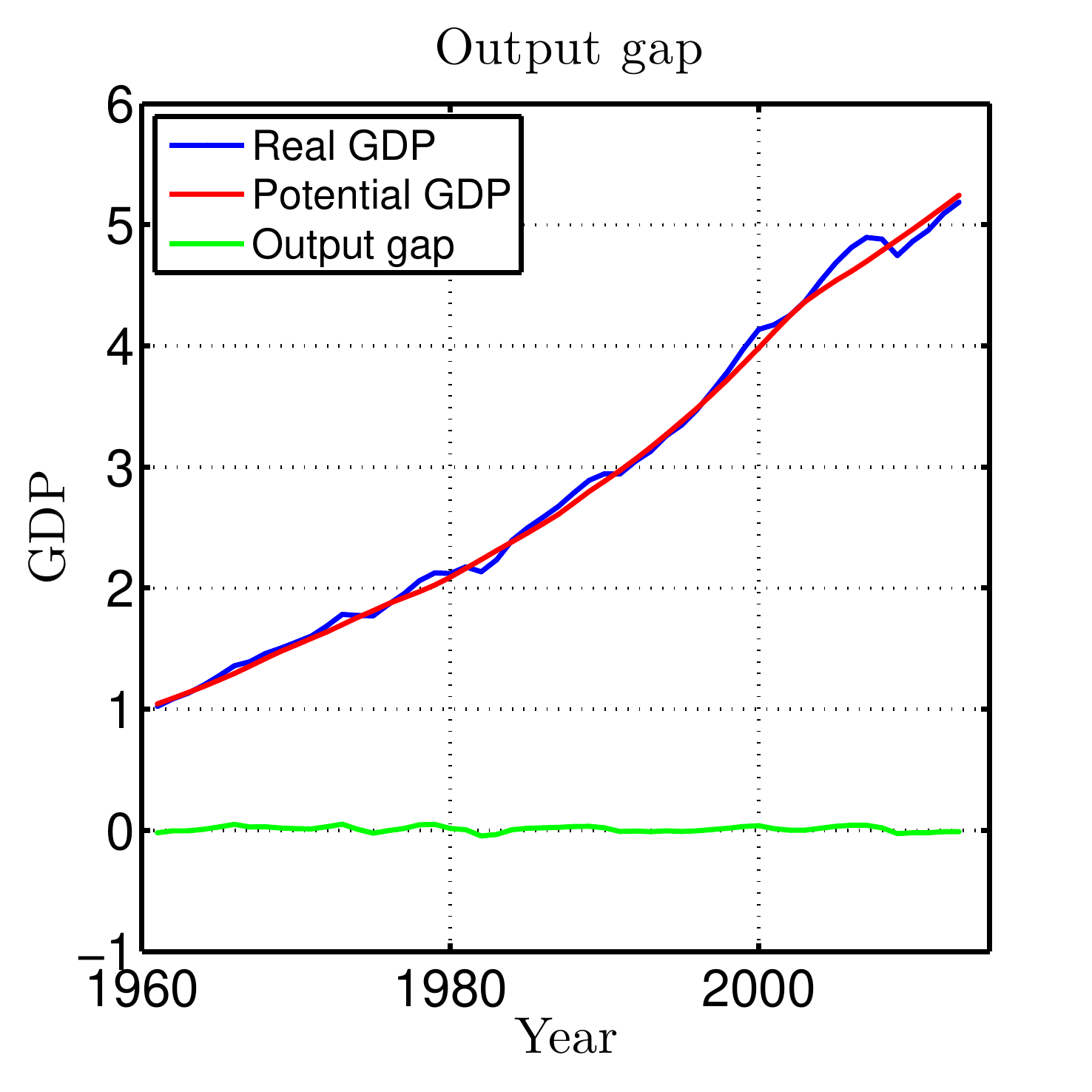}
     \includegraphics[width=.30\linewidth]{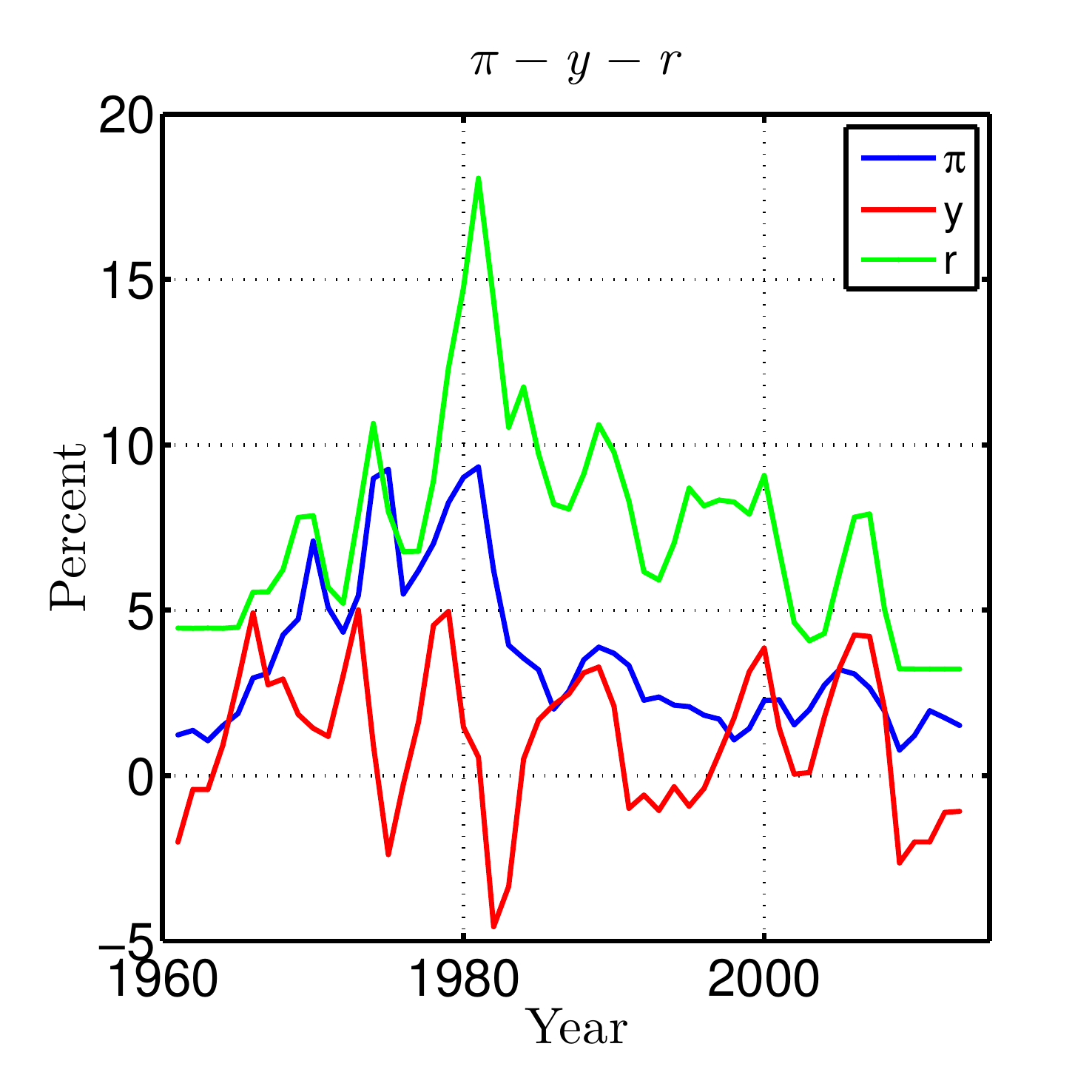}
     \includegraphics[width=.30\linewidth]{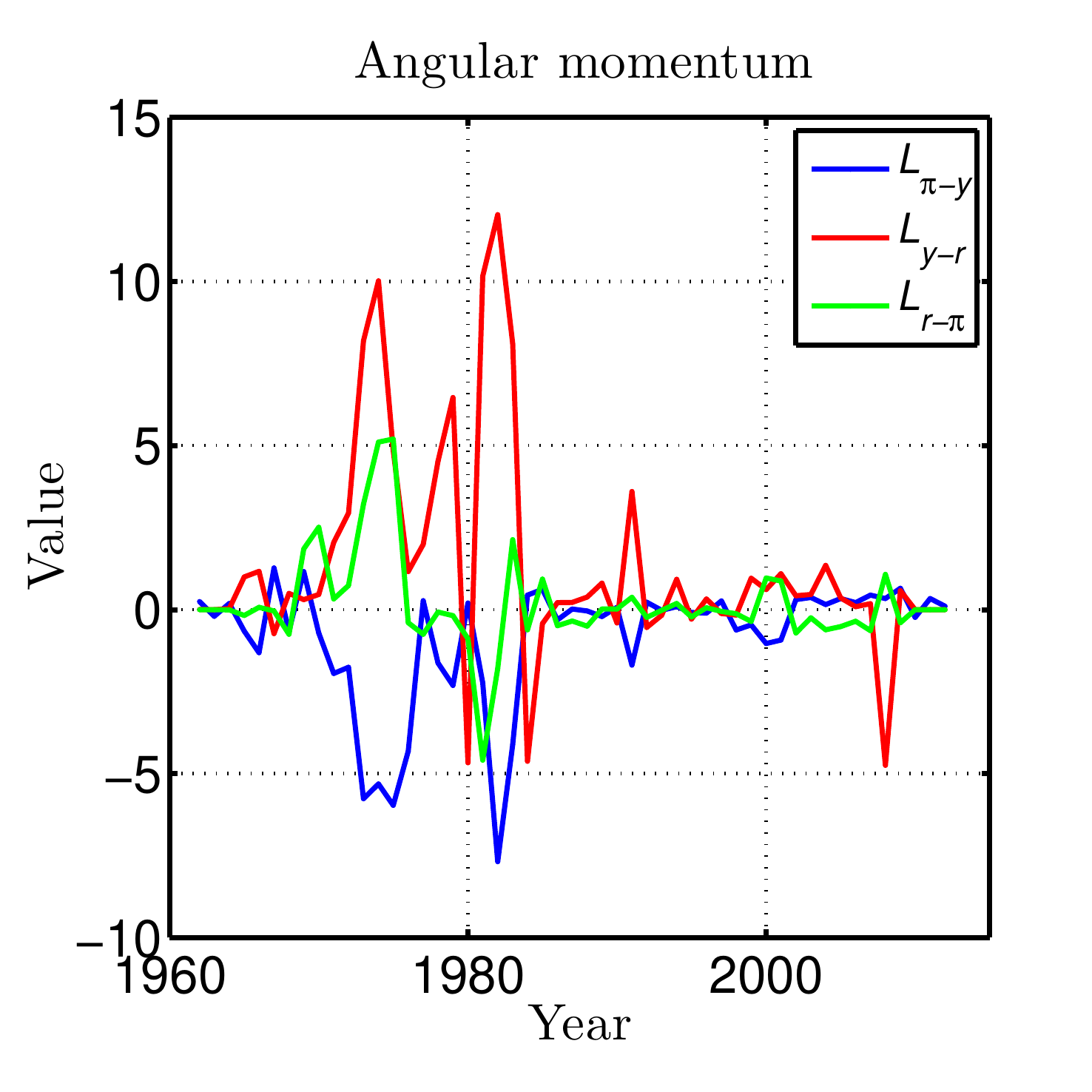}
    \end{center}
    \caption{
        \label{fig:3x3Add}   The up panel are the phase space projection view of the 3D velocity vector field   (shown in Fig.~\ref{fig:3Ddistr}) in $\pi$-$y$, $y$-$r$ and $r$-$\pi$ 2D space respectively.
        The middle panel illustrates the empirical data trajectories in $\pi$-$y$, $y$-$r$ and $r$-$\pi$ phase space (data source: 1961-2013 US data from World-Bank), respectively. The red dots are the center (simple arithmetic average over all of the vectors) of the trajectory. The bottom panel illustrates the output gap (in green), time series of the three variables (see legend) and the $L^{(2)}$ in the three phase planes (numerical results is shown in Table \ref{tab:theoL2}), respectively.
       }
  \end{figure}

Fig. \ref{fig:3x3Add}, from left to right, illustrates the real grow rate, the time series of the three variables, and the three components of the empirical $L^{(2)}$ measured, respectively.

\subsection{Measurement}
 
To explore the  patterns in general equilibrium, we extend the measurements used in our previous research
\cite{XuWang2011ICCS}\cite{wang2012evolutionary} \cite{XuZhouWang2013}\cite{WangXuZhou2014socialcycling} in this note.

\subsubsection{$n$-sampling angular momentum\label{SI:nSampleingL}}

\begin{figure}
    \begin{center}
     \includegraphics[width=.35\linewidth]{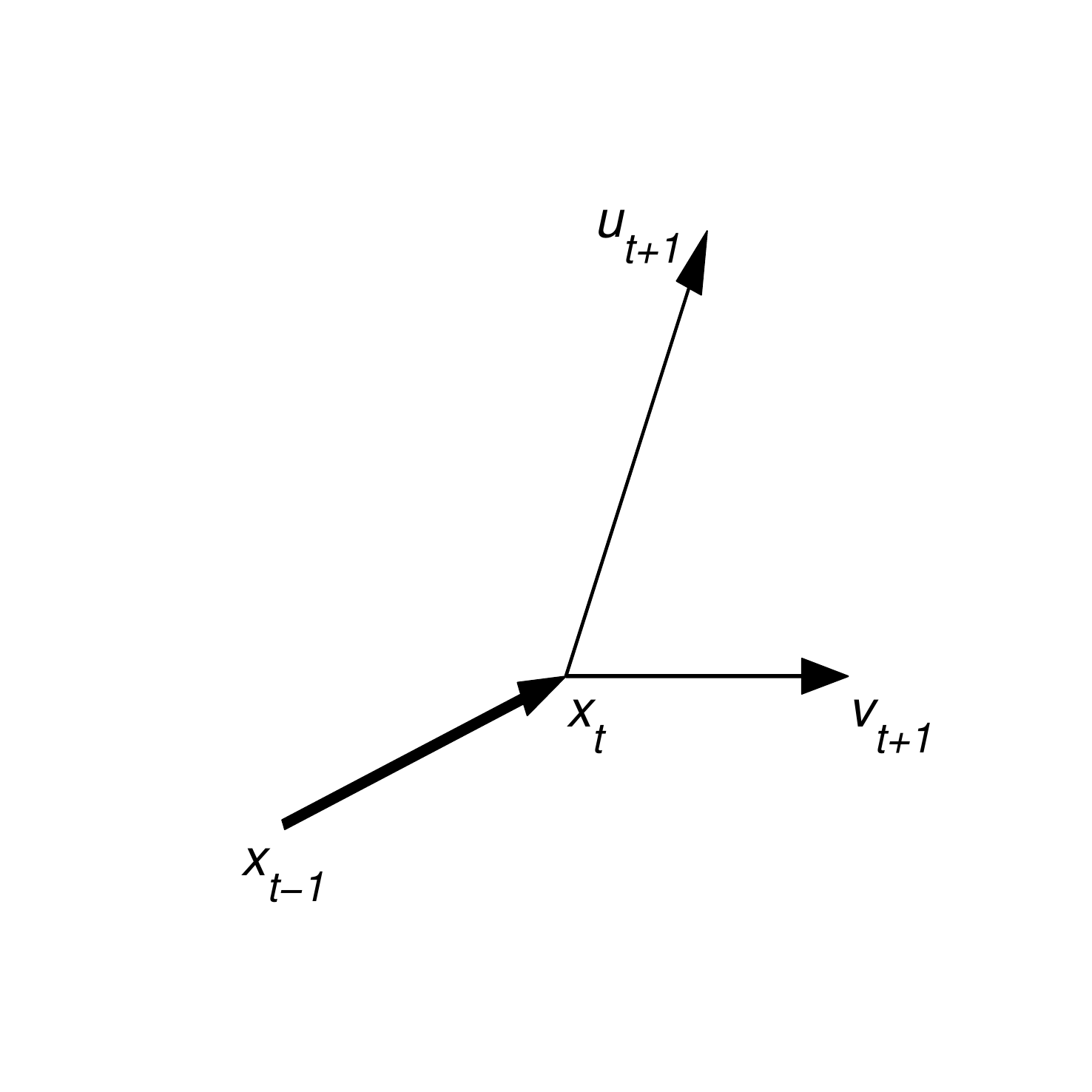}
    \end{center}
    \caption{
        \label{fig:L2_schematic_diagram}   Schematic diagram for $L^{(2)}$ measurement formulated as Eq.~\ref{eq:Ln}. $L^{(2)}$ reports the direction and the amplitude of the next transit ($x_{t} \rightarrow x_{t+1}$) in respect to current transit ($x_{t-1}\!\!\rightarrow\!\!x_t$).  In 2D condition, if $x_{t+1} = u_{t+1}$, the $L^{(2)}$ is positive. Alternatively, If $x_{t+1} = v_{t+1}$, the $L^{(2)}$ is negative.
  Accordingly, in this study case, we defined that, accumulated $L^{(2)}$ should be positive when the motion is counterclockwise and negative when clockwise, meanwhile, cyclic motion is stronger when $|L^{(2)}|$ larger. A metaphor for $L^{(2)}$ is: If a cyclic motion is homogeneously deterministic, $L^{(2)}$ will deviate from zero definitively. Notice that this measurement is a mathematical abstract for cyclic motion. Two points need to be emphasised briefly: (1) Indexed by $L^{(2)}$, the cyclic motion is not necessary surrounding any general equilibrium points (attractors); (2) This angular momentum measurement can also be denoted as $L(x_{t+1}\!\! \leftarrow\!\! x_{t} | x_{t}\!\!\leftarrow\!\!x_{t-1})$ and regarded as a Bayesian approach.
       }
  \end{figure}

We propose a a measurement, called as the mean angular momentum and denoted as $L$, to quantify the cyclic motions.
In the phase space $S$, an evolutionary trajectory ($x_1, x_2, ..., x_n$) forms during $n$ time periods ($t_1, t_2, ..., t_n$).  In respect to the center of the trajectory $\sum_{i=1}^{n} x_i/n$,  $L^{(n)}$ is the average value of the cross product of the two vectors of each successive transitions. Explicitly, $L^{(n)}$ is defined as
\begin{equation}\label{eq:Ln}
  L^{(n)}_S = \frac{1}{n} \sum_{i=1}^{n}  x_i \times x_{i+1}.
\end{equation}
In this study case, $S$ represent the $\pi$-$y$-$r$ 3D space. Like $x_t$, $L^{(n)}_S$ is a vector having three components, denoted as ($L^{(n)}_\pi, L^{(n)}_y, L^{(n)}_r$) or ($L^{(n)}_{y-r}, L^{(n)}_{r-\pi}, L^{(n)}_{\pi-y}$), denoting the cycling motion in $y$-$r$, $r$-$\pi$ and $\pi$-$y$ in 2D space.
In a given long trajectory (time series) having its length $l$ in phase space $S$, given a fixed natural number $n$, we can obtain $l/n$ samples of $L^{(n)}_S$, and then the probability spectrum of the $L^{(n)}_S$ can be obtained. In \cite{wang2012evolutionary}, we have used this measurement to illustrate the cycles in experimental games, in which the $n$ is set as the total length of a experimental session (e.g., $n$=150) in \cite{Binmore2001}.

In main text, we use $n$=2 to show the statistic results. Fig \ref{fig:L2_schematic_diagram} is the schematic diagram and definition of the $L^{(2)}$ measurement. $L^{(2)}$ straightforwardly reports the instantaneously cyclic motions. As a vector, $L^{(2)}(t)=\frac{1}{2}(x_{t+1}-x_{t}) \times (x_{t}-x_{t-1}) $ reports the direction and the amplitude of the transition of the nest transition ($x_t$ to $x_{t+1}$) referring to current transition ($x_{t-1}$ to $x_t$). In a small step deterministic processes, referring to each transition, its next transition has a deterministic direction and its angular momentum is deterministic and should deviate from zero definitively. Projected in a given phase plane, the observed component of $L^{(2)}$ is  positive when the motion being counterclockwise, and negative when clockwise.

Table \ref{tab:theoL2} demonstrates the theoretical expectations (hypothesis) on $L^{(2)}$ according to the theoretical velocity field pattern illustrated in Fig. \ref{fig:3Ddistr}. To test the robustness, we have test also $n$=$3,4,5,,...,12$ for $L^{(n)}$, and the results are shown in Table \ref{tab:robust}.

\begin{table}
\begin{center}
\caption{\label{tab:robust} Robustness test}
\begin{tabular}{crrrrrr}
\hline
	 $n$ &  ~~~~ $\bar{L}_{\pi-y}^{(n)}$ ~~~~ & ~~~~ $p$  ~~~~ & $\bar{L}_{y-r}^{(n)}$  &  ~~~~ $p$  ~~~~ & $\bar{L}_{r-\pi}^{(n)}$ &  ~~~~ $p$  ~~~~ \\
\hline
	 2 & $-$0.8634 & 0.0311 & 1.4038 & 0.0002 & 0.1858 & 0.5611 \\
	 3 & $-$2.8083 & 0.0024 & 4.4615 & 0.0000 & 0.3800 & 0.9116 \\
	 4 & $-$5.4011 & 0.0004 & 8.6940 & 0.0000 & 0.3296 & 0.6870 \\
	 5 & $-$8.2188 & 0.0001 & 13.4340 & 0.0000 & $-$0.0246 & 0.7505 \\
	 6 & $-$11.1249 & 0.0000 & 18.0349 & 0.0000 & $-$0.5237 & 0.8407 \\
	 7 & $-$14.0079 & 0.0000 & 22.4741 & 0.0000 & $-$1.1396 & 0.7144 \\
	 8 & $-$16.9161 & 0.0000 & 26.6510 & 0.0000 & $-$1.8652 & 0.5310 \\
	 9 & $-$19.9469 & 0.0000 & 30.6053 & 0.0000 & $-$2.5685 & 0.5134 \\
	 10 & $-$23.1497 & 0.0000 & 34.5579 & 0.0000 & $-$3.4236 & 0.3401 \\
	 11 & $-$26.4547 & 0.0000 & 38.5523 & 0.0000 & $-$4.4856 & 0.1710 \\
	 12 & $-$29.9195 & 0.0000 & 42.6605 & 0.0000 & $-$5.7208 & 0.0860 \\
  \hline
\end{tabular}
\end{center}
\end{table}

\subsubsection{Velocity vector field and distribution\label{SI:velocityanddistribution}}
Velocity vector field is a diagrammable presentation of the state depended motion. Measurement for velocity field has been well defined and illustrated in \cite{XuWang2011ICCS} and \cite{XuWang2011}.

\subsection{Additional material}

\subsubsection{Distribution in empirical data vs simulation \label{SI:distribution}}
Statistic results on distribution, in US empirical data, is that the positive dependence on $r-\pi$ in significant ($p$=0.000, OLE, $n$=53), which support the theoretical expectations too. In  $y-r$ ($p$=0.368, OLE, $n$=53) nor $\pi-y$ ($p$=0.301, OLE, $n$=53), we do not find significant result. In the simulated time series generated by the DSGE model, randomly sampling of the same length gives the same result. This consists with the simulated distribution patterns shown in the diagrams shown in the first row in Fig. \ref{fig:3x3}.

\subsubsection{Robustness test on cycles \label{SI:RobustnessLn}}

  To test the robustness of the results of cycles identified by $L^{(2)}$, we take the sampling interval $n$=3, 4, 5, 6, 7, 8, 9, 10, 11 and 12 (year) to calculate the $L^{(n)}$, respectively. We observed none of the theoretical expectation is violated statistically. On the contrary, the \emph{all} observed $L^{(n)}_{\pi-y}$ meets the expectations of clockwise cycles, and \emph{all} $L^{(n)}_{y-r}$ meet the expectations of clockwise cycles, meanwhile, weak and no cycles (in short run) in $r-\pi$ plane.
The mean observations and the statistic result (the Wilcoxon matched-pairs
signed-ranks test by comparing $L^{(n)}$ with 0, because if a system in randomness,
$L^{(n)}$ should be zero statistically)
are shown in Table~\ref{tab:robust}).

\subsubsection{Theoretical evaluation on cycle \label{SI:SolveL2}}

For understanding the inherent cyclic pattern better, an analytical method is expected. We can try to simplified the model to evaluate $L^{(2)}$ analytically.

 Recall that in \cite{de2010scientific}\cite{de2012lectures},  the research strategy consists in comparing the dynamics of this behavioral model with the same structural model under rational expectations which can be interpreted as a stylized DSGE-model, and can be written in matrix notation, referring to Eq.(15) in \cite{de2010scientific}, as follows:
\begin{equation}\label{eq:agg3x3Eq}
 \left(
   \begin{array}{ccc}
     1 & -b_2 & 0 \\
     0 & 1 & -a_2 \\
     -c_1 & -c_2 & 1 \\
   \end{array}
 \right)
   \left(
     \begin{array}{c}
       \pi_t \\
       y_t \\
       r_t \\
     \end{array}
   \right)
  =  \left(
   \begin{array}{ccc}
      b_1 & 0   & 0 \\
     -a_2 & a_1 & 0 \\
        0 & 0   & 0 \\
   \end{array}
 \right)
   \left(
     \begin{array}{c}
       E_t \pi_{t+1}\\
       E_t y_{t+1}\\
       E_t r_{t+1}\\
     \end{array}
   \right)
 +
\left(
   \begin{array}{ccc}
     1 - b_1 & 0   & 0 \\
     0 & 1-a_1 & 0 \\
        0 & 0   & c_3 \\
   \end{array}
 \right)
   \left(
     \begin{array}{c}
       \pi_{t-1}\\
       y_{t-1}\\
       r_{t-1}\\
     \end{array}
   \right)
 +
 \left(
     \begin{array}{c}
       \eta_t \\
       \varepsilon_t \\
       \upsilon_t \\
     \end{array}
   \right).
 \end{equation}
Theoretically, this equation can be solved when the coefficient matrix in the l.h.s being of full rank, i.e. its determinant is not equal to zero.  Commonly, solution can be obtained \cite{de2012lectures,jang2012identification} by the Binder and Pesaran \cite{binder1995multivariate} procedure. In this note, we do not fall into the brute force iteration procedure, but in a simplified condition,  calculate $L^{(2)}$ explicitly to illustrate the inherent cycles.

\begin{figure}
    \begin{center}
          \includegraphics[width=.30\linewidth]{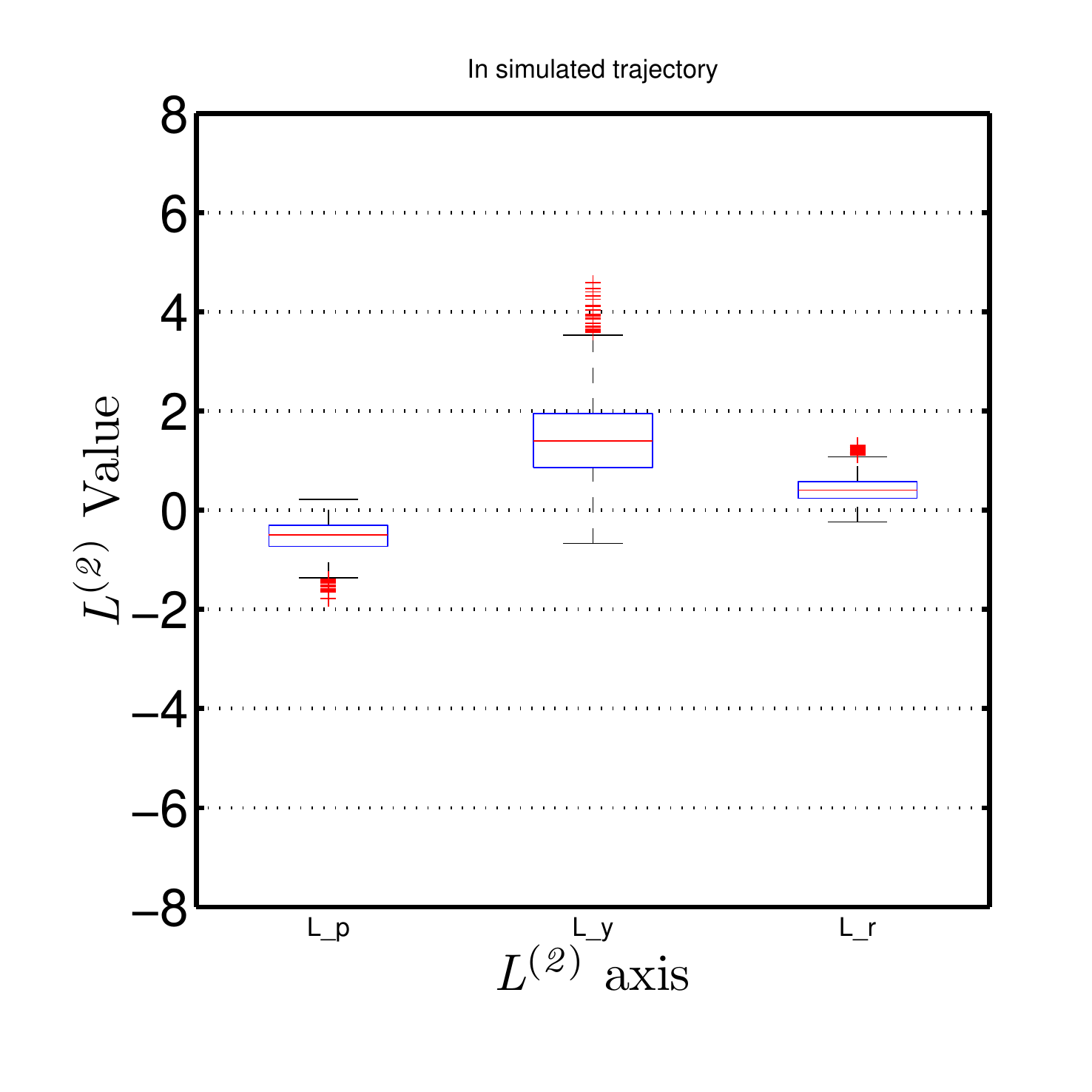}
          \includegraphics[width=.30\linewidth]{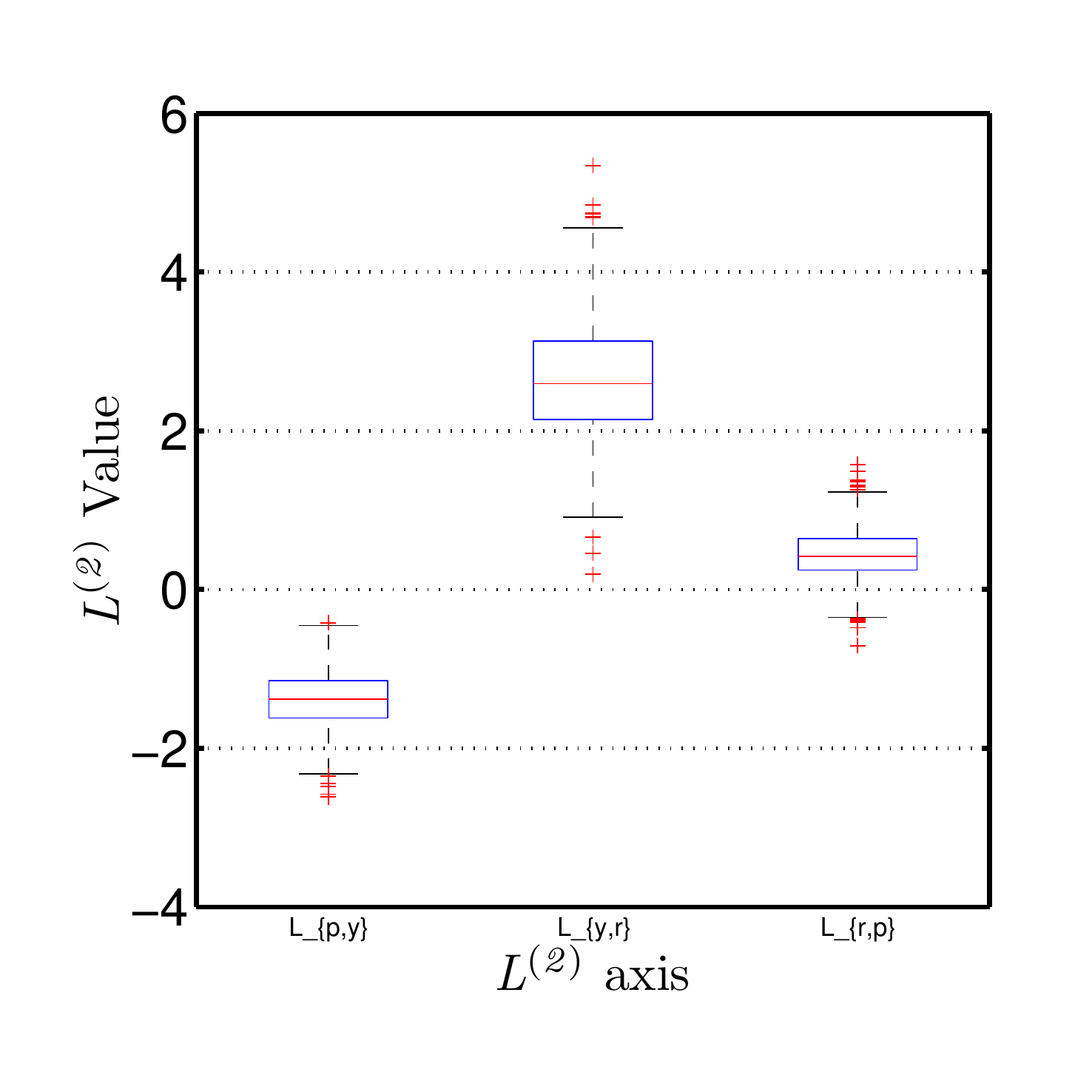}
    \end{center}
    \caption{
        \label{fig:TheoL2evaluted} Theoretical evaluation of $L^{(2)}$.  (1) The left panel is the  quartiles box  of distribution of the observed $L^{(2)}$ measured in the trajectories by 1000 repeated simulations and each simulation includes 2000 rounds repeated. For details of the simulation, see Section \ref{SI:modelandsimulation}.  (2) The right panel is the quartiles box  of distribution of the observed $L^{(2)}$ by simplified numerical evaluation (For more details, see Eq. \ref{eq:munRes} and Section \ref{SI:SolveL2}). These results are summarized as the theoretical $L^{(2)}$
       }
  \end{figure}

Under consideration of the heuristics for the forecasts regarding the output gap and inflation expectations, the forward
looking term $ (E_t \pi_{t+1},
       E_t y_{t+1},
       E_t r_{t+1})^T$
    is substituted by the equivalent expressions for the discrete
    choice mechanism given.
    Using extrapolative rule, according to Eq.(1.5) in \cite{de2012lectures}, $E_t y_{t+1}$ = $y_{t-1}$; similarly, according to Eq.(1.15) in \cite{de2012lectures}, $E_t \pi_{t+1}$ = $\pi_{t-1}$.
    It follows that the model becomes purely
    backward-looking, and can be written as a recursive formula. Thus Eq. \ref{eq:agg3x3Eq} turn out to be
    \begin{equation}\label{eq:agg3x3Eq2}
 \left(
   \begin{array}{ccc}
     1 & -b_2 & 0 \\
     0 & 1 & -a_2 \\
     -c_1 & -c_2 & 1 \\
   \end{array}
 \right)
   \left(
     \begin{array}{c}
       \pi_t \\
       y_t \\
       r_t \\
     \end{array}
   \right)
  =
\left(
   \begin{array}{ccc}
     1   & 0   & 0 \\
     -a_2 & 1  & 0 \\
        0 & 0   & c_3 \\
   \end{array}
 \right)
   \left(
     \begin{array}{c}
       \pi_{t-1}\\
       y_{t-1}\\
       r_{t-1}\\
     \end{array}
   \right)
 +
 \left(
     \begin{array}{c}
       \eta_t \\
       \varepsilon_t \\
       \upsilon_t \\
     \end{array}
   \right).
 \end{equation}
This equation can  be solved by backward-induction \cite{jang2012identification}. If ignore the noise term, we have,
\begin{equation}\label{eq:agg3x3Eq7}
  \left(
     \begin{array}{c}
       \pi_t \\
       y_t \\
       r_t \\
     \end{array}
   \right)
  =
C \left(\begin{array}{ccc} a_{2}\, b_{2} + a_{2}\, c_{2} - 1 & - b_{2} & - a_{2}\, b_{2}\, c_{3}\\ - a_{2}\, \left(c_{1} - 1\right) & -1 & - a_{2}\, c_{3}\\ a_{2}\, \left(c_{2} + b_{2}\, c_{1}\right) - c_{1} &  - c_{2} - b_{2}\, c_{1} & - c_{3} \end{array}\right)
   \left(
     \begin{array}{c}
       \pi_{t-1}\\
       y_{t-1}\\
       r_{t-1}\\
     \end{array}
   \right)  =
F
   \left(
     \begin{array}{c}
       \pi_{t-1}\\
       y_{t-1}\\
       r_{t-1}\\
     \end{array}
   \right),
 \end{equation} 
in which $F$ indicates the monthly action. When the constant $C  \equiv a_{2}\, c_{2} + a_{2}\, b_{2}\, c_{1} - 1 \neq 0$, which has been pointed out as Eq.(1.22) in \cite{de2012lectures} and as Eq.(1.22) in \cite{de2010scientific}.
The model was calibrated in such a way that the time units can be considered to be months (see p.12 in \cite{de2012lectures}), and we use $F^{12}$ as the action to calculate the yearly $L^{(2)}$ | Because the empirical data which was reported yearly from 1960-2013. Then, using the parameters $a_1, a_2, b_1, b_2, c_1, c_2, c_3$ in the models (see p. 36 in \cite{de2012lectures}), the monthly action is \\
\begin{center}
$F = \left(
	\begin{array}{ccc}
	 0.9955 & 0.0448 & -0.0045 \\
	 -0.0897 & 0.8969 & -0.0897 \\
	 1.4484 & 0.5157 & 0.4484 \\
	\end{array}
\right)$
\end{center}
and, if no noise exists, the yearly action is
\begin{equation}\label{f12}
  F^{12} =\left(
	\begin{array}{ccc}
	 0.4360 & 0.1408 & -0.0296 \\
	 -1.2129 & -0.1841 & 0.0283 \\
	 0.2559 & 0.3022 & -0.0738 \\
	\end{array}
\right)
\end{equation}
Then, the expected $L^{(2)}$, the angular momentum in the successive transit
    from $x_{t-1}$ to $x_{t}$ to $x_{t+1}$ can be expressed as
\begin{eqnarray}
  L^{(2)} &=& \frac{1}{2}(x_{t+1} - x_{t}) \times (x_{t} - x_{t-1}) \\
          &=&  \frac{1}{2} F y \times y,
\end{eqnarray}
in which $y=(F - 1) x_{t-1} = x_{t} - x_{t-1}$ and $y$ is employed as
the benchmark from which direction and amplitude the vector $X_{t+1}$ will go. Substitute $Y$ with $F^{12}$, by 53  repeated random sampling (normally distributed noise with standard deviation being 1  in respect to the equilibrium [0,0,0]) the location $X_{t-1}$,  and the shock at $t$ and $t+1$ being normally distributed noise with standard deviation being 0.5 as suggested in \cite{de2012lectures}. We conduct such calculation 1000 times, the distribution of the average $L^{(2)}$ are shown in Fig. \ref{fig:TheoL2evaluted} (right panel). Over all, the mean values of $L^{(2)}$ are approximately as
\begin{equation}\label{eq:munRes}
   [L^{(2)}_{\pi-y},L^{(2)}_{y-r},L^{(2)}_{r-\pi}]  = [ -1.392,    2.637,    0.4361]
\end{equation}
This results is illustrated in the right panel in Fig. \ref{fig:TheoL2evaluted}. Accordingly, the expectations are approximately expressed as
\begin{center}
 $|L^{(2)}_{\pi-y}| > |L^{(2)}_{y-r}| > |L^{(2)}_{r-\pi}|$
\end{center}
and
\begin{equation}\label{eq:uneq}
   L^{(2)}_{y-r} \gg L^{(2)}_{r-\pi} \geq 0 \gg L^{(2)}_{\pi-y}
\end{equation}
form which the theoretical prediction can be shown in Table \ref{tab:theoL2}.

\subsubsection{Cycle and single shock  \label{sec:ge_shocks}}

To illustrate the existence of cycles in DSGE is
due to the constantly noise, theoretically, we can control the noise (the term in the right most
in Eq.~\ref{eq:agg3x3Eq2}).
We can see that, the cyclic motion of DSGE could appear even when an exogenous shock at
 the equilibrium.

We give an example of an exogenous $\pi$ shock to illustrate the cycles. We assume that the system is in equilibrium  $[\pi,y,r]$ = $[0,0,0]$ at $t-1$, then a positive $+10$ exogenous shock is added at $\pi$ at $t$, and no exogenous shock is added since then. Such system will evolute according to its own dynamic equations \ref{eq:agg3x3Eq7}. 

\begin{figure}
    \begin{center}
     \includegraphics[width=.221\linewidth]{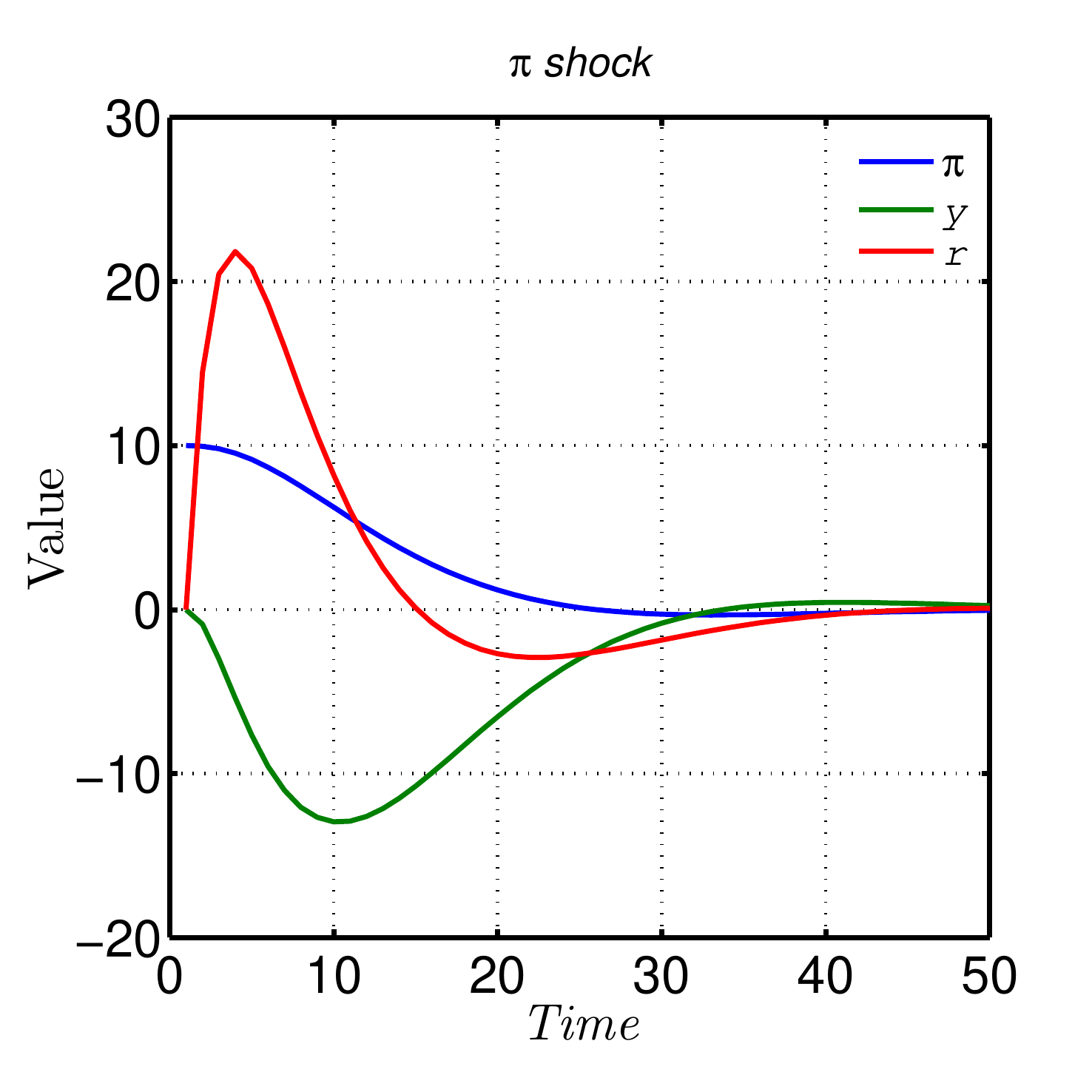}
     \includegraphics[width=.221\linewidth]{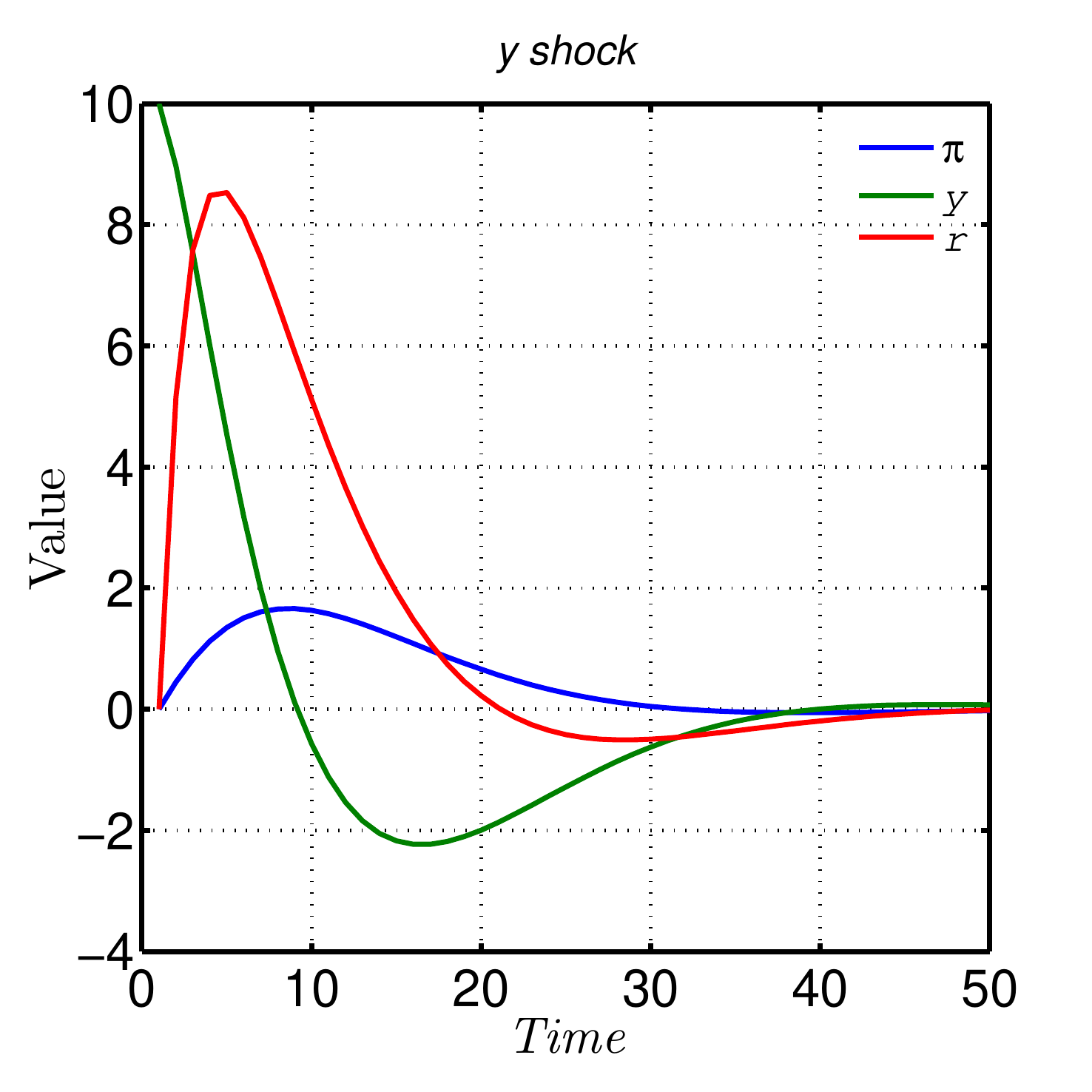}
     \includegraphics[width=.221\linewidth]{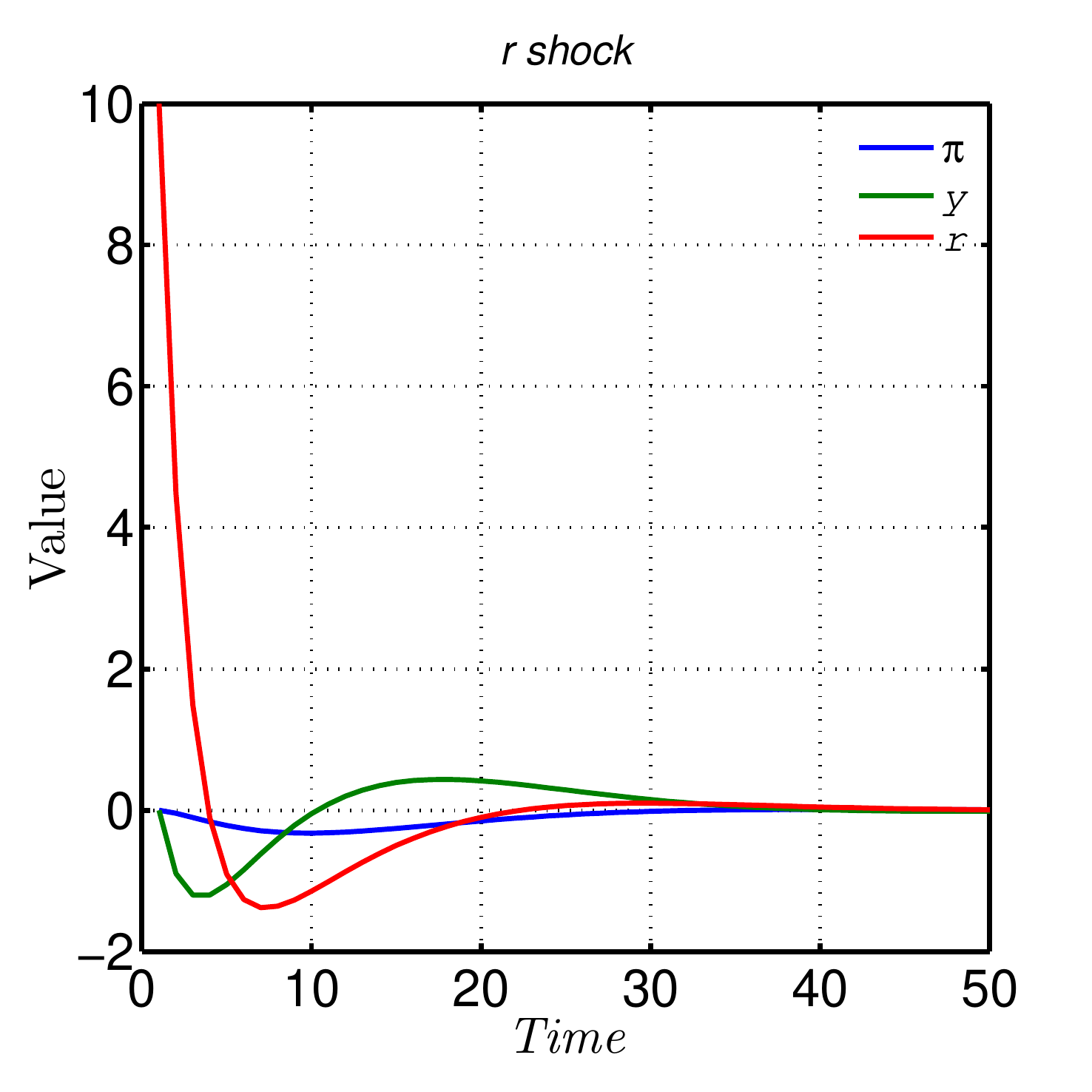} \\
     \includegraphics[width=.221\linewidth]{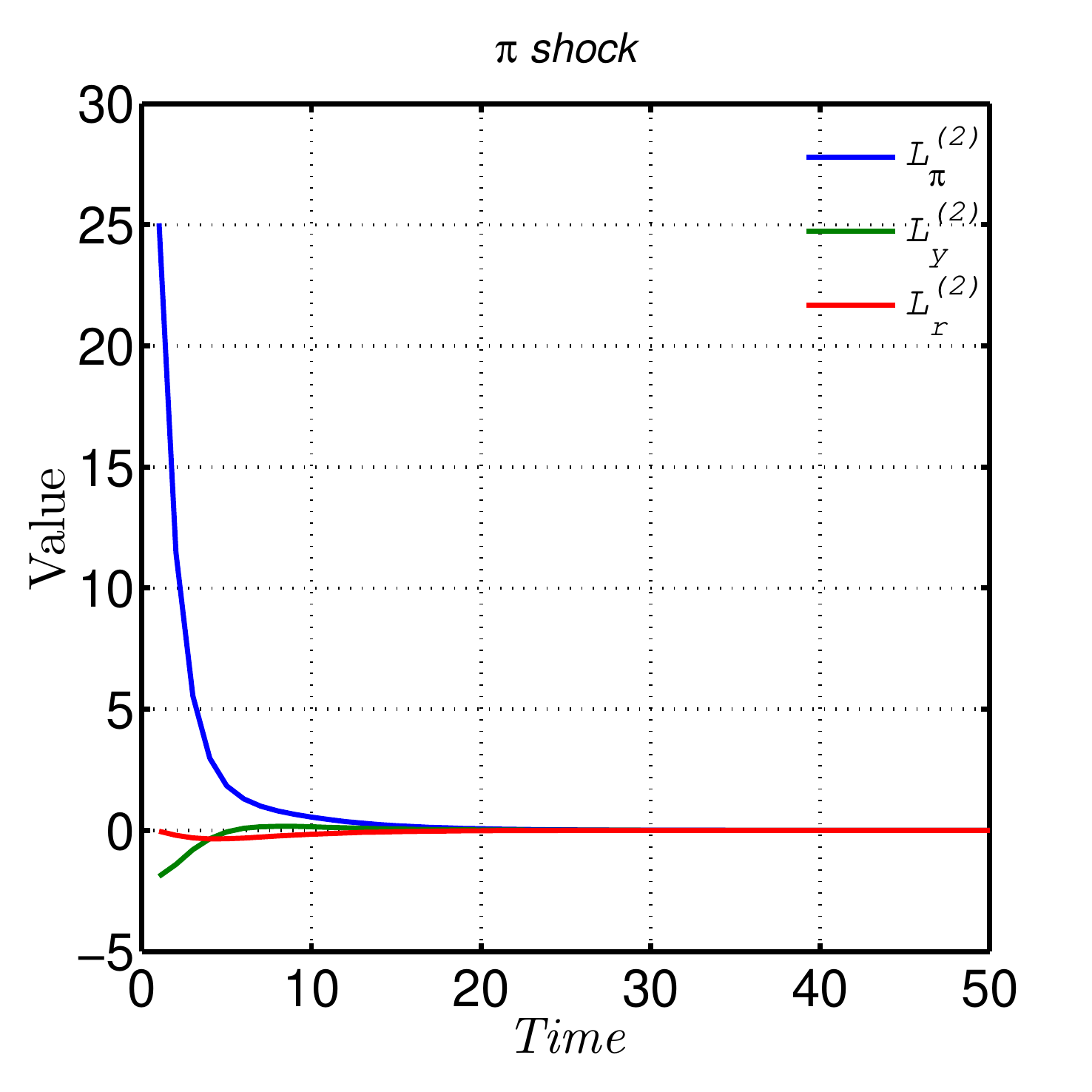}
     \includegraphics[width=.221\linewidth]{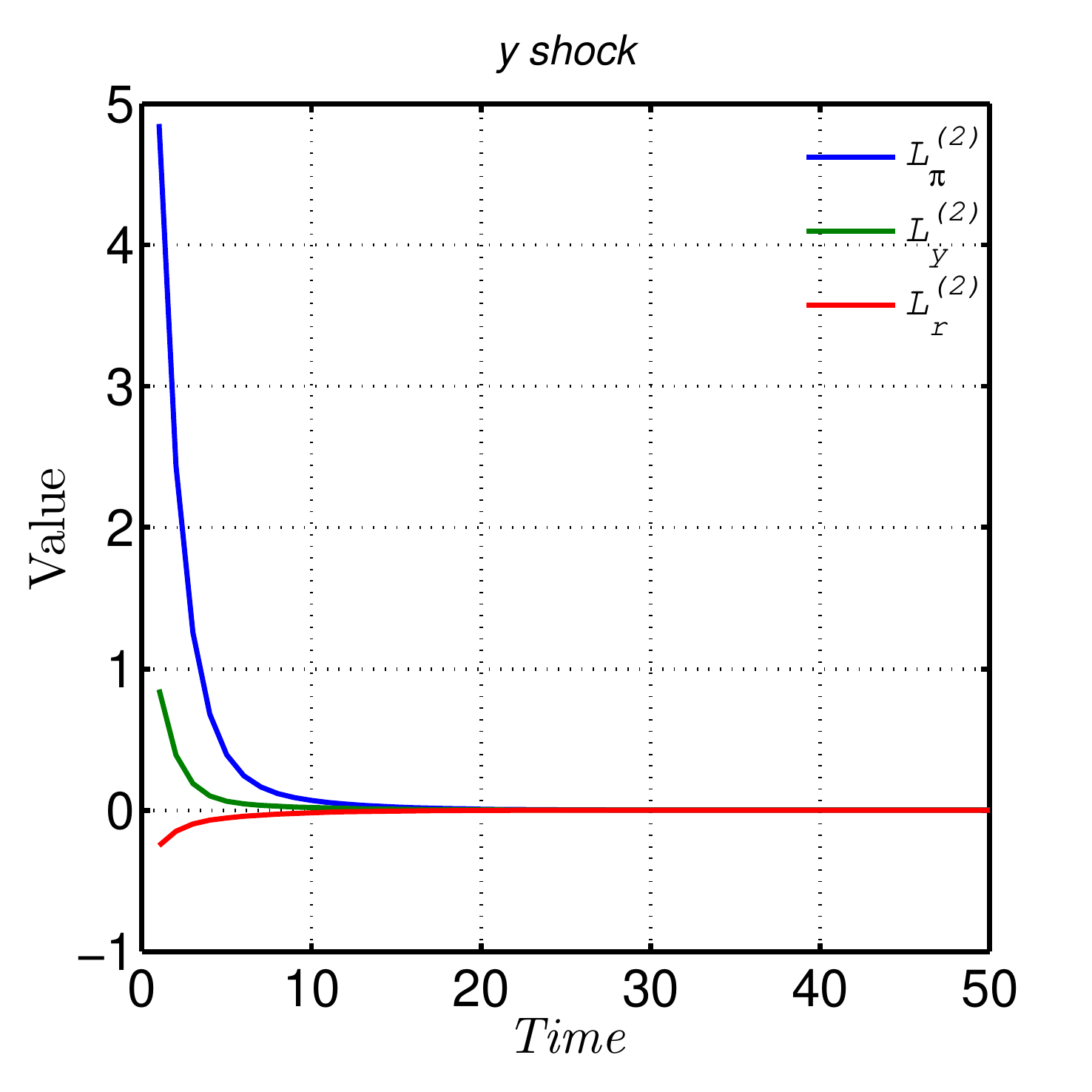}
     \includegraphics[width=.221\linewidth]{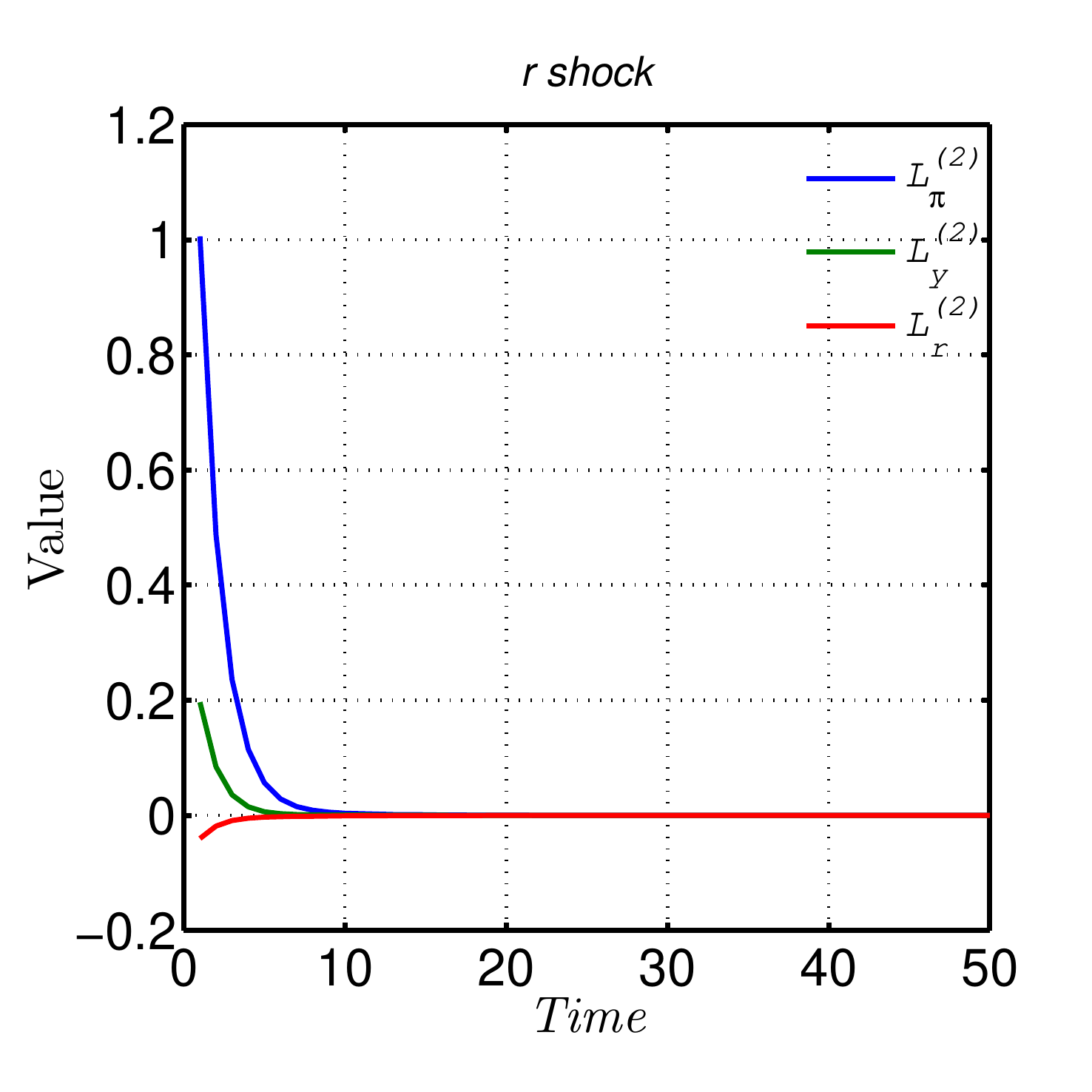} \\
     \includegraphics[width=.221\linewidth]{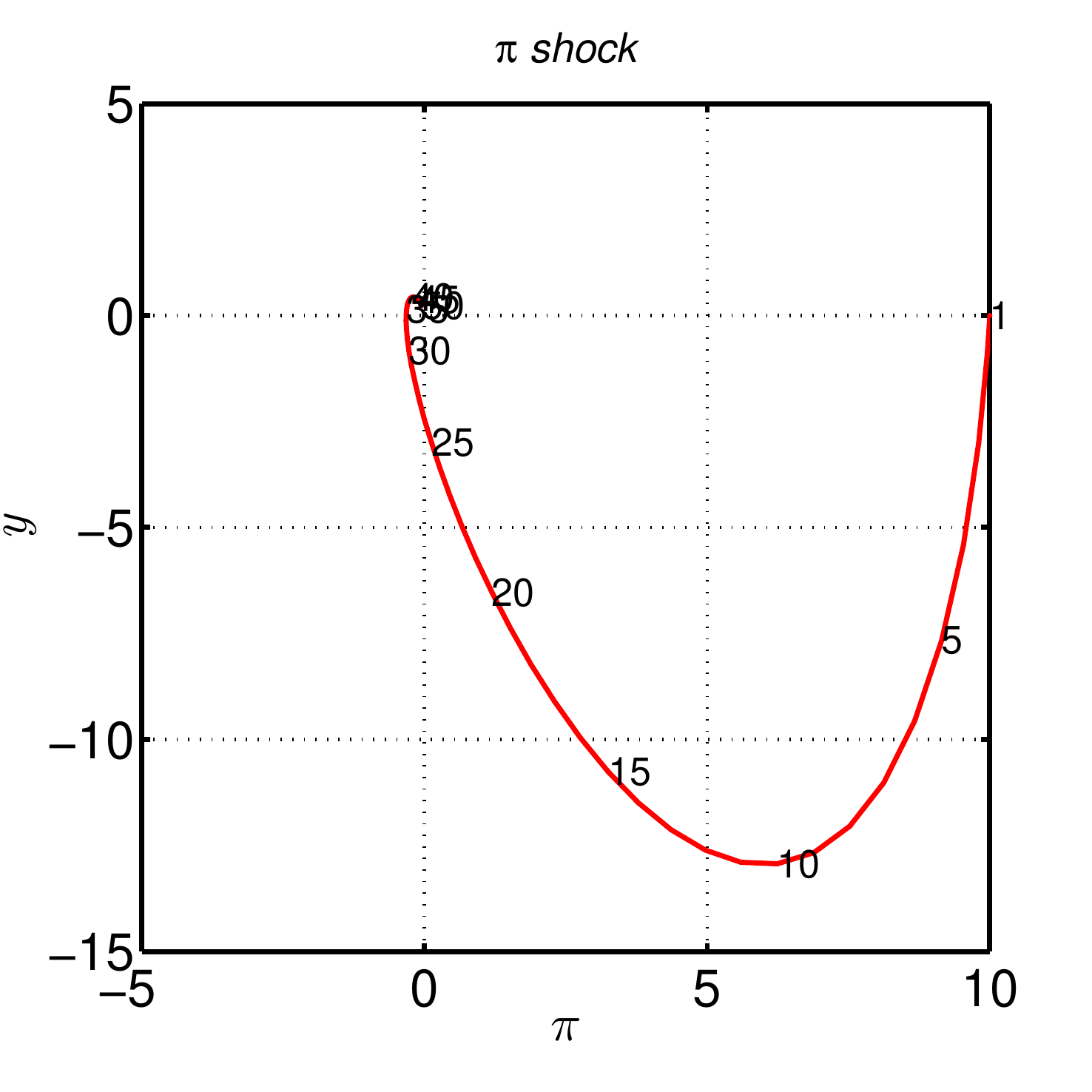}
     \includegraphics[width=.221\linewidth]{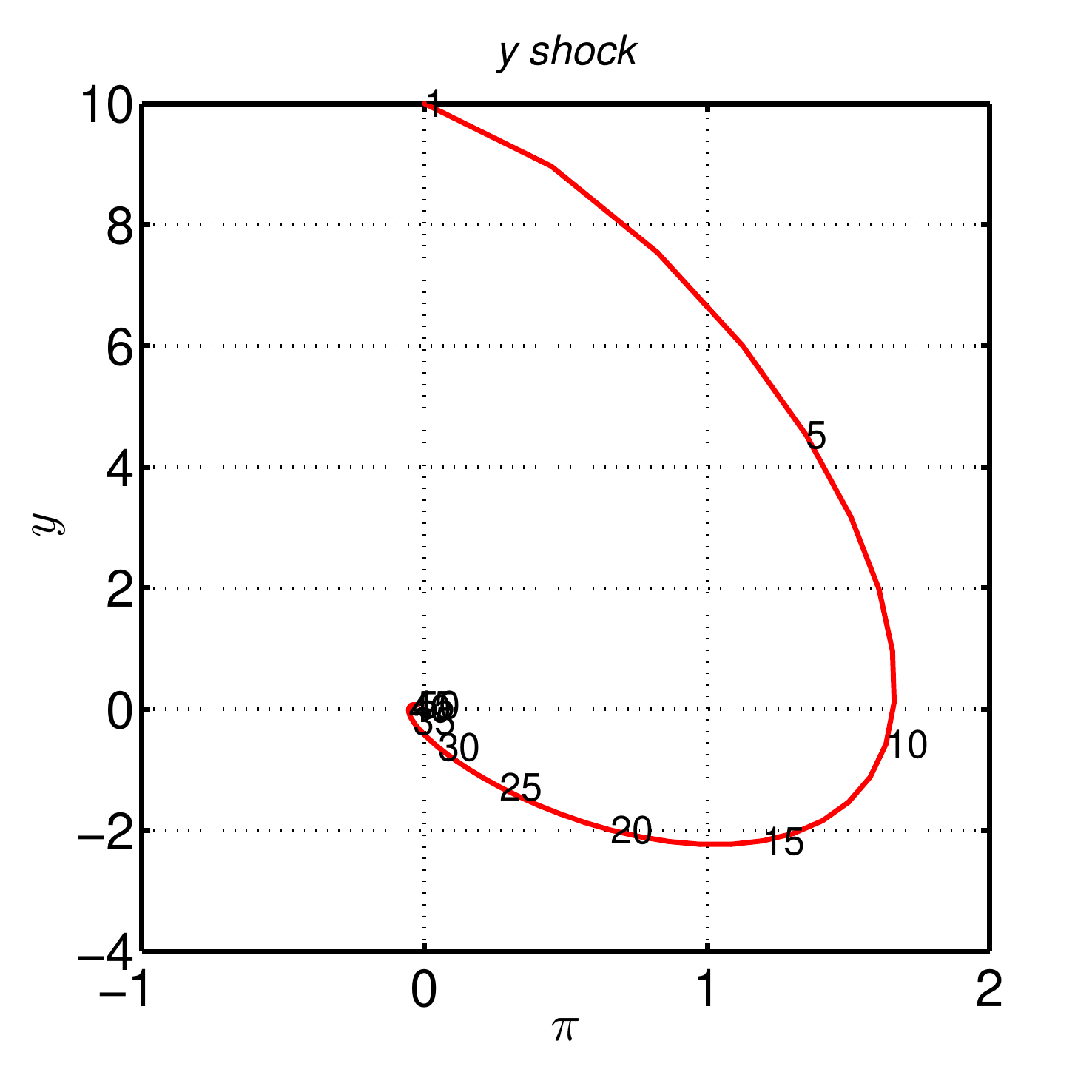}
     \includegraphics[width=.221\linewidth]{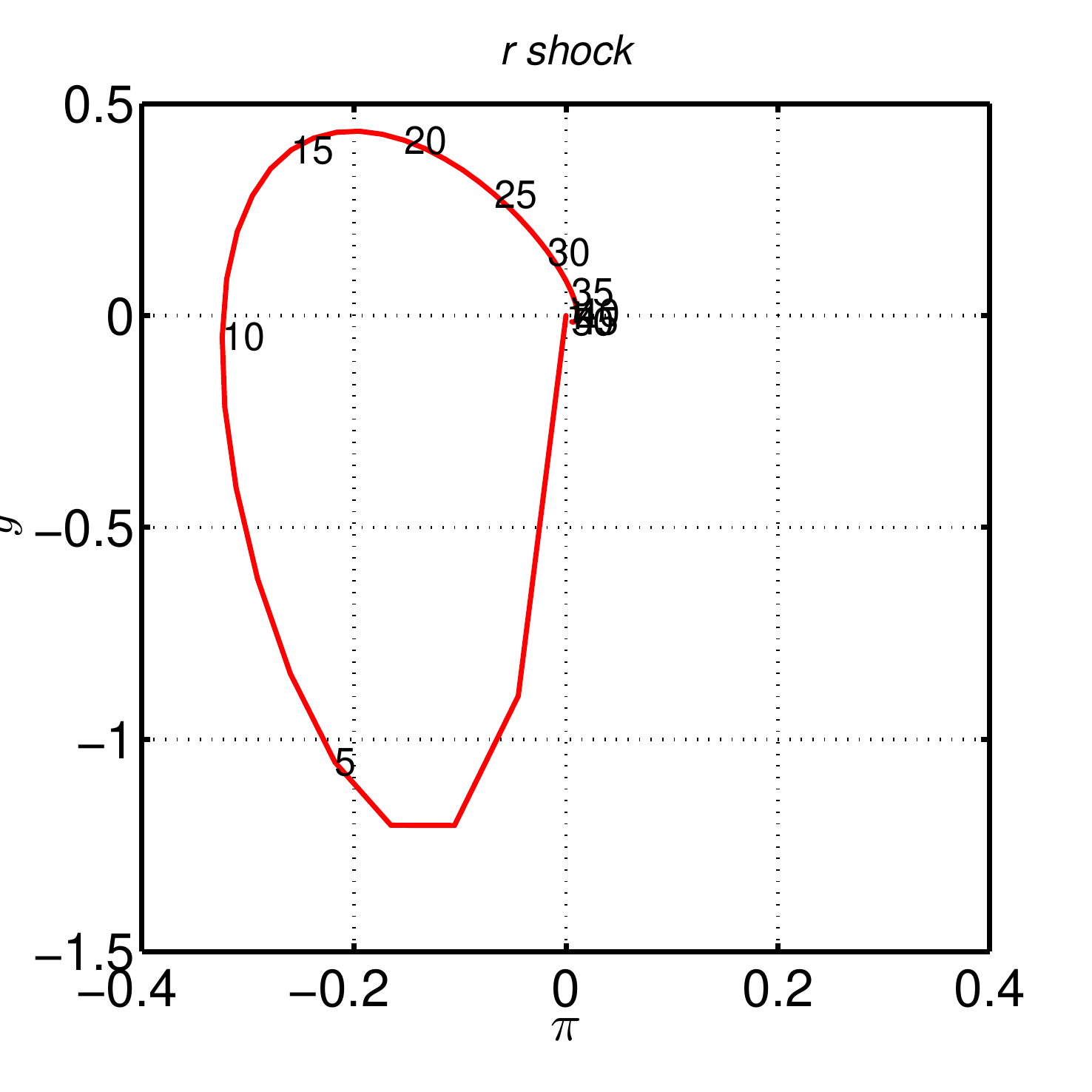} \\
     \includegraphics[width=.221\linewidth]{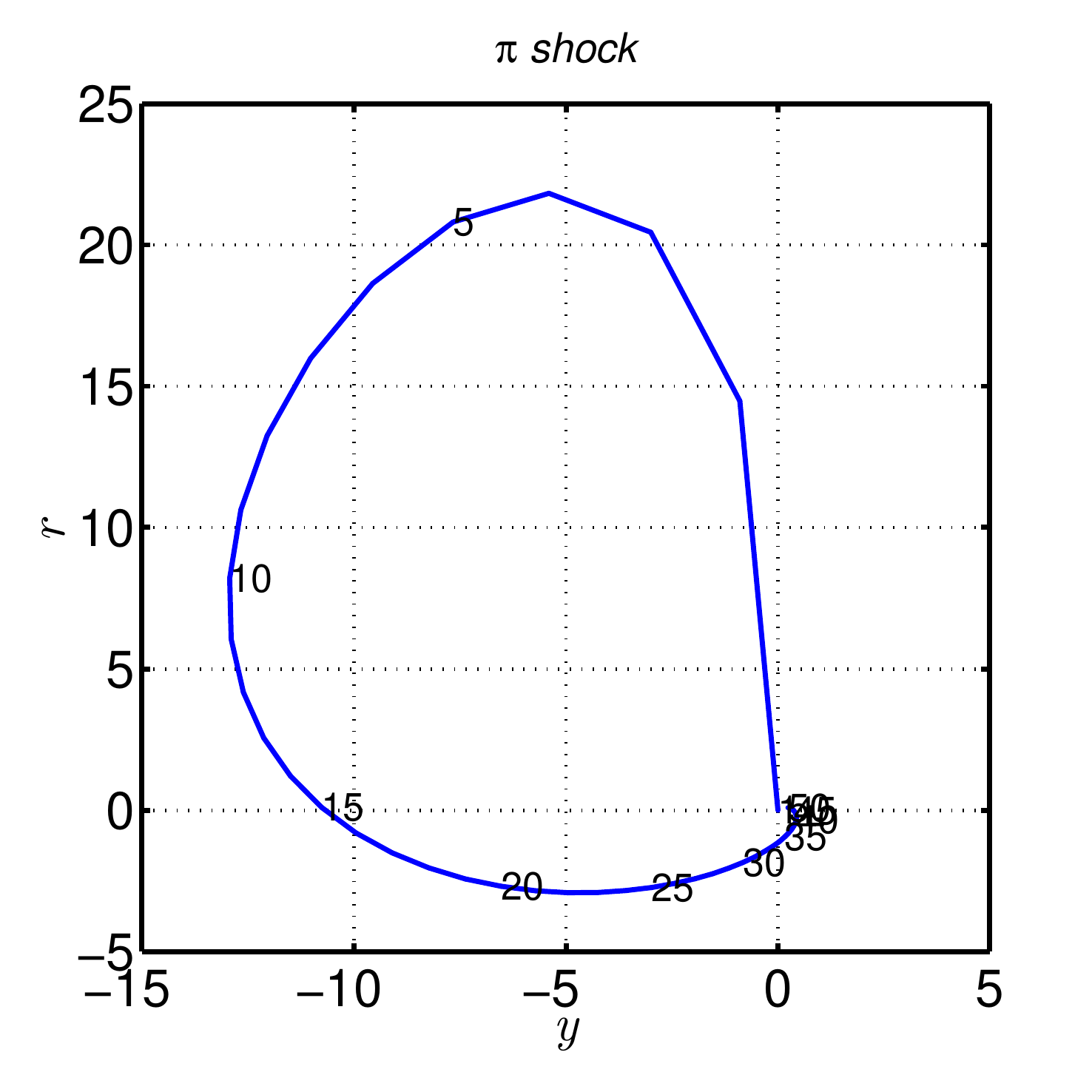}
     \includegraphics[width=.221\linewidth]{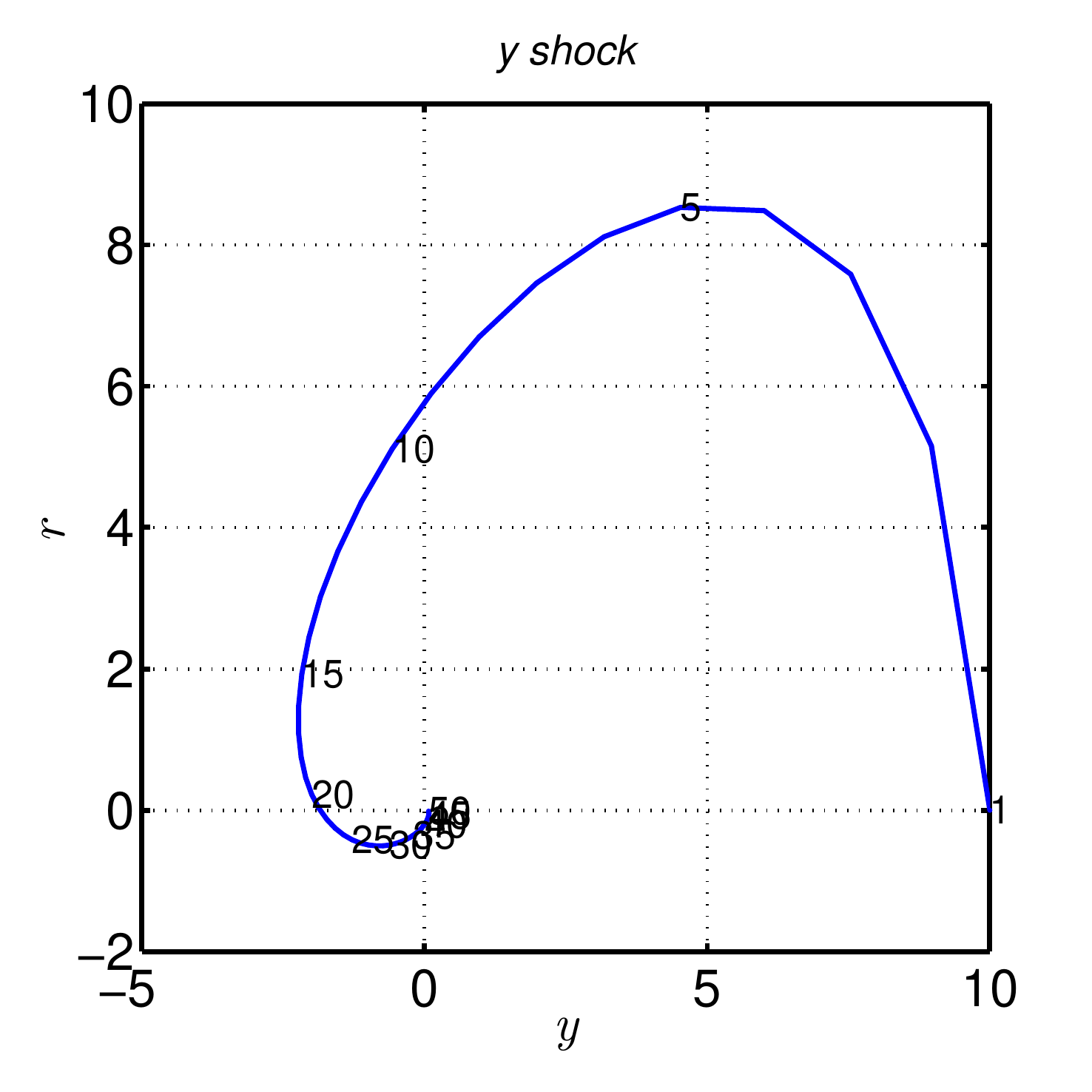}
     \includegraphics[width=.221\linewidth]{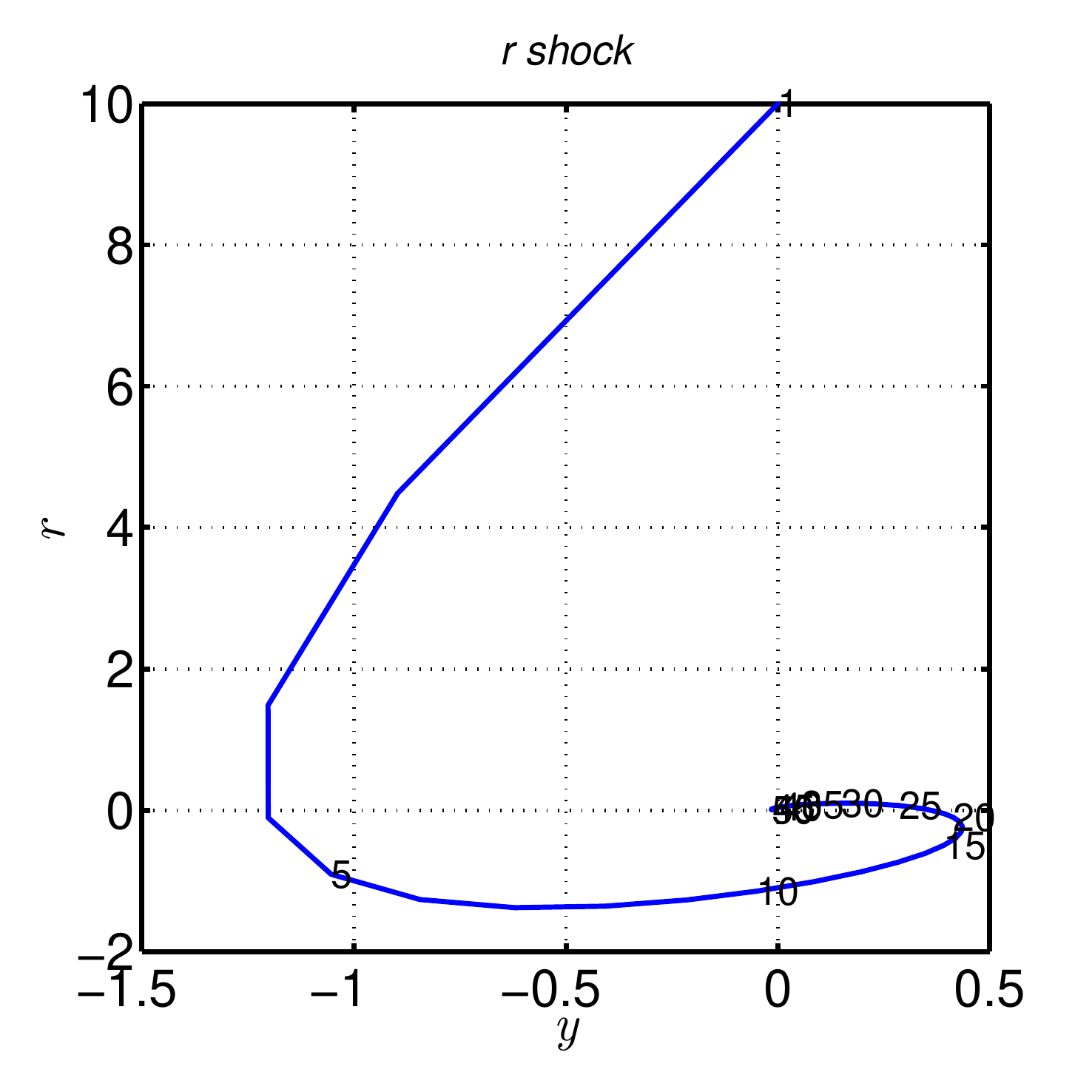} \\
     \includegraphics[width=.221\linewidth]{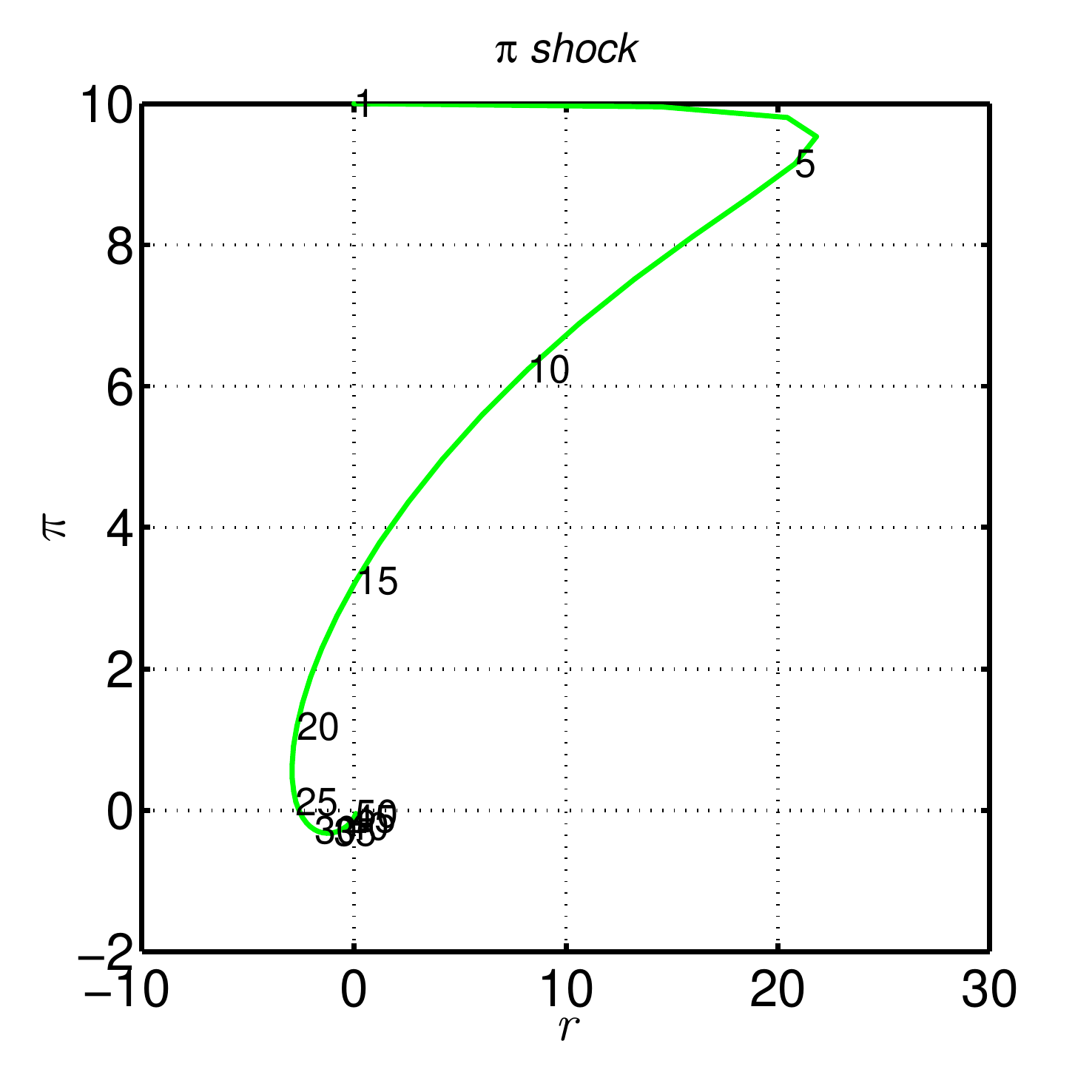}
     \includegraphics[width=.221\linewidth]{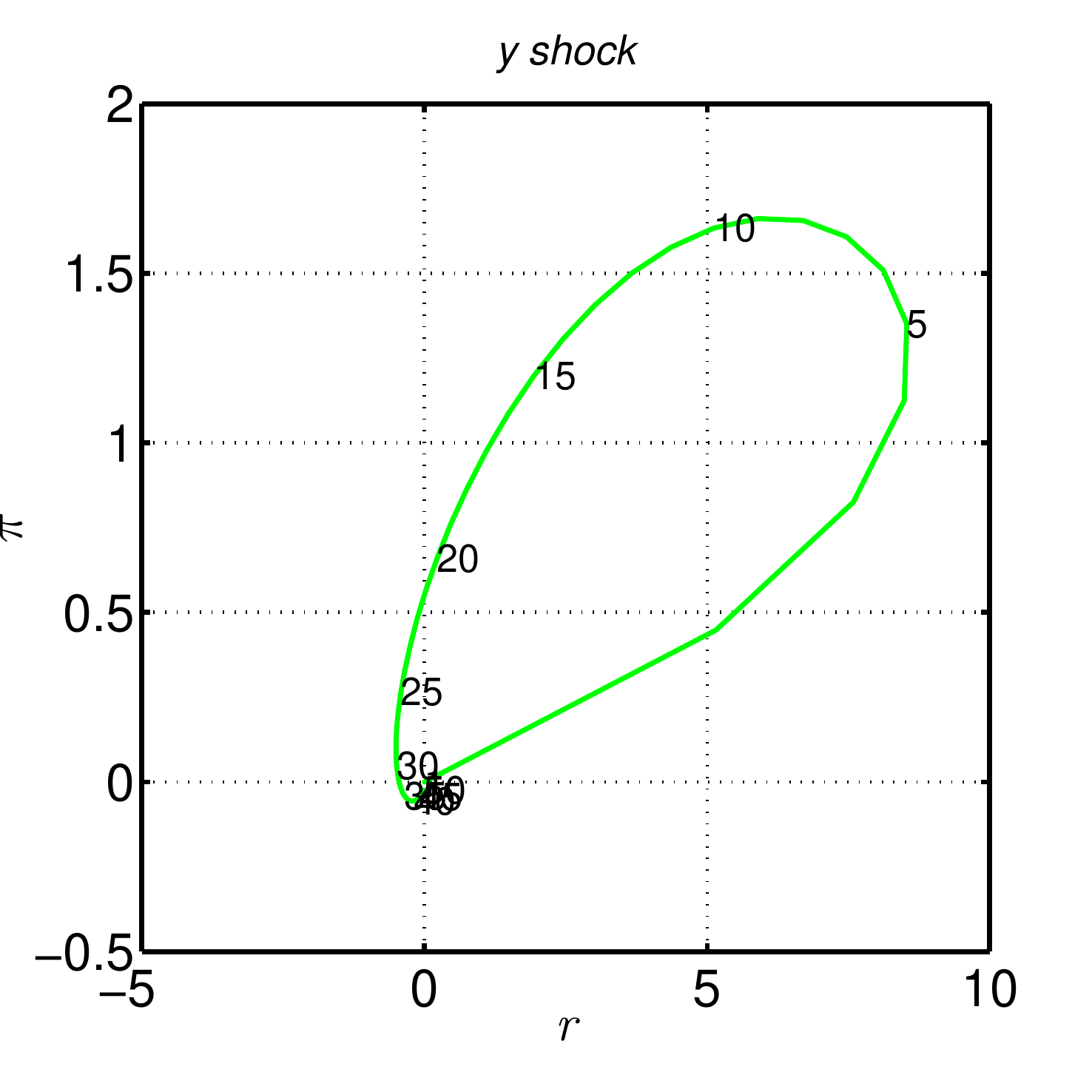}
     \includegraphics[width=.221\linewidth]{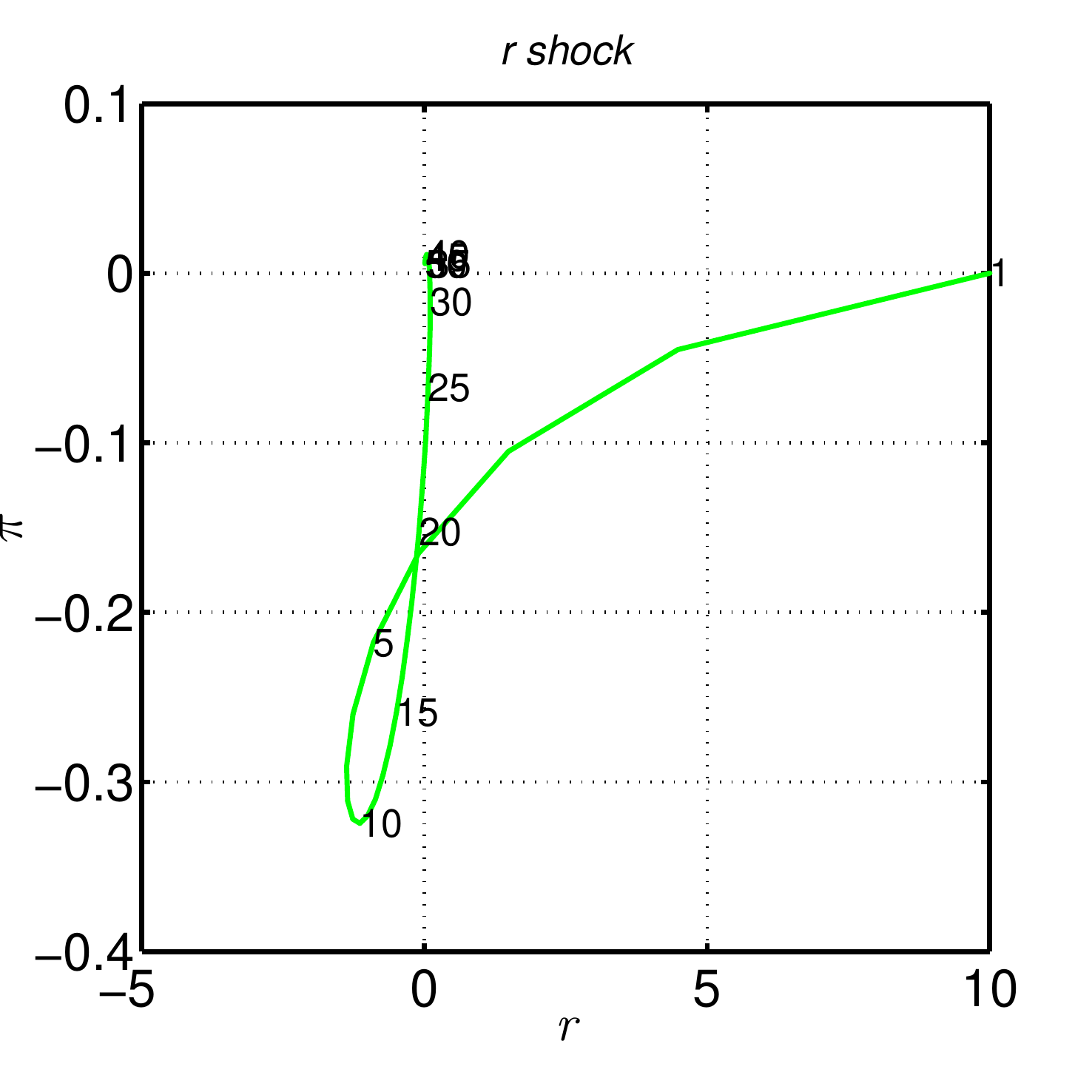}
    \end{center}
    \caption{
        \label{fig:Cycle_and_shock_positive}   The up panel are the presentation of the positive shock and recovering of the general equilibrium. The middle panel are evolutionary trajectory in $\pi-y$ phase diagram, in which we can see the trajectory is of clockwise constantly. The low panel are related $L^{(2)}$ observed, in which we can see  $L^{(2)}_{\pi} > 0$ always (means counter clockwise cycle in $y-r$ phase plane), $L^{(2)}_{r} > 0$ always (means clockwise cycle in $\pi-y$ phase plane).
       }
  \end{figure}

As illustrated in Fig. \ref{fig:Cycle_and_shock_positive} (the top left most sub figure), the system will recover its equilibrium after several tens monthly rounds. In this recovering processes,  $L^{(2)}$ and the cyclic patterns in the three 2D space can be quantified and visualized.
Similarly, we can add the exogenous shock at $y$ or at $r$ respectively. We can also add negative exogenous shocks at $\pi$, $y$ or $r$ respectively. In these six examples, we can analysis the $\pi-y-r$ entanglements, the $L^{(2)}$ measured  and the evolutionary trajectory in the three phase planes.

\begin{figure}
    \begin{center}
     \includegraphics[width=.221\linewidth]{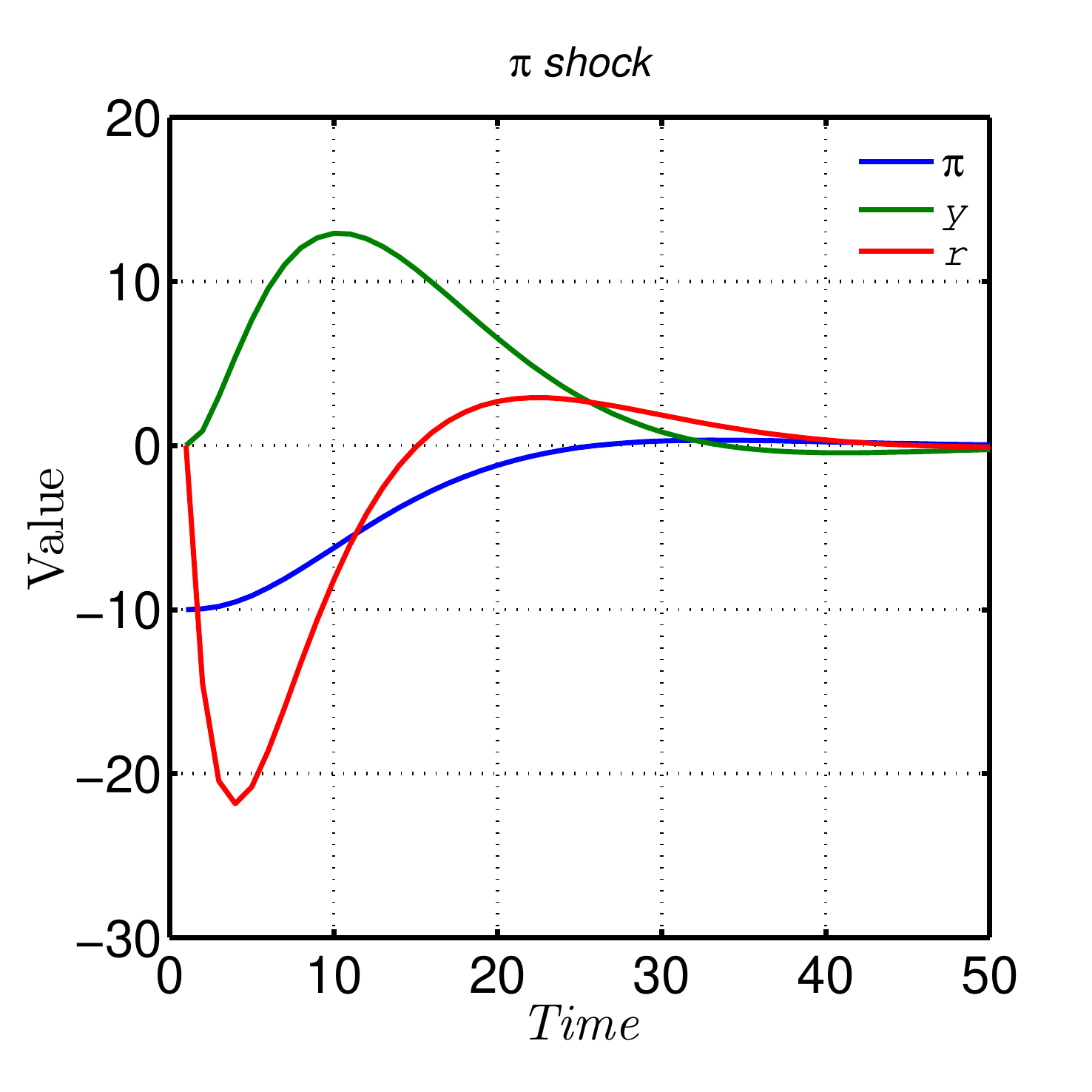}
     \includegraphics[width=.221\linewidth]{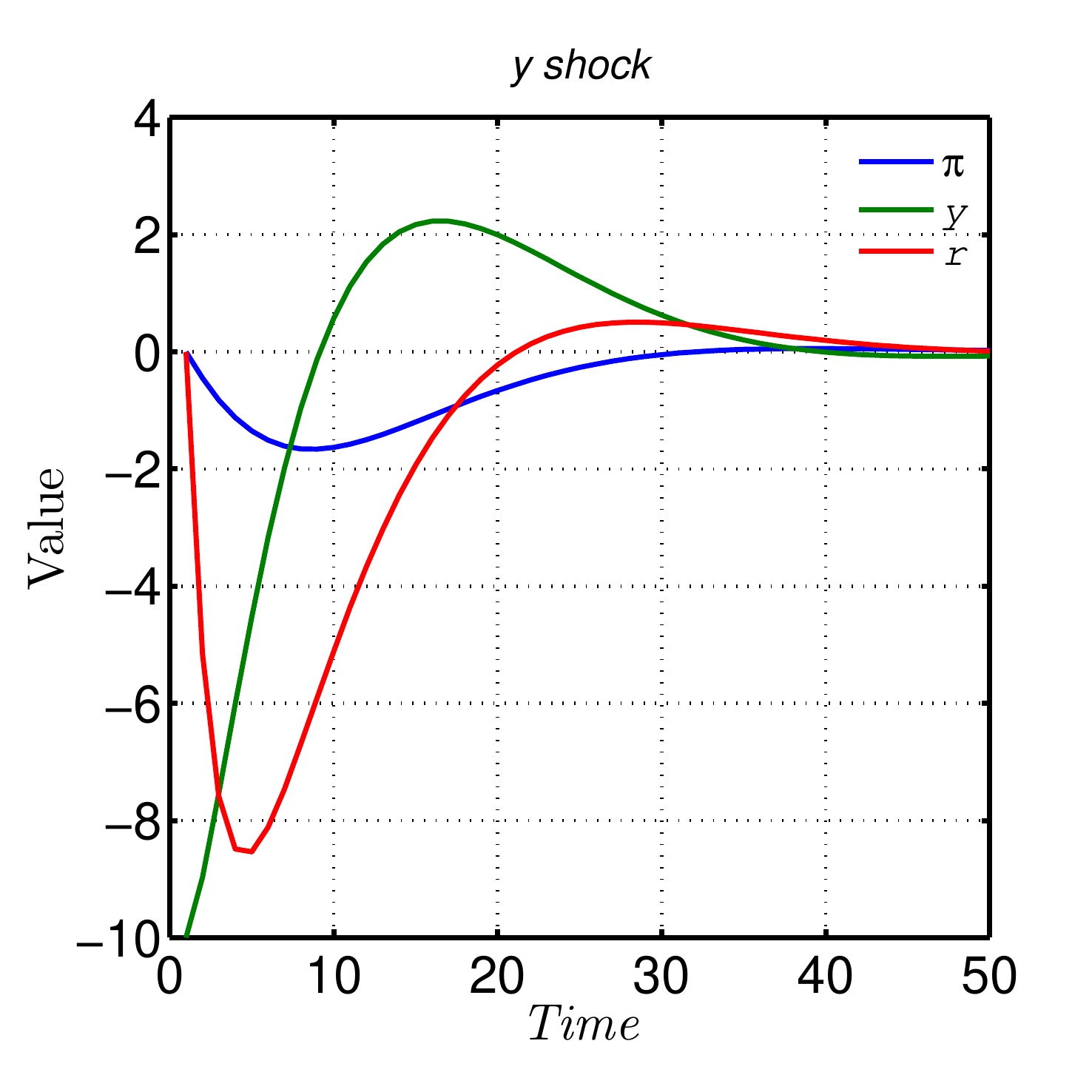}
     \includegraphics[width=.221\linewidth]{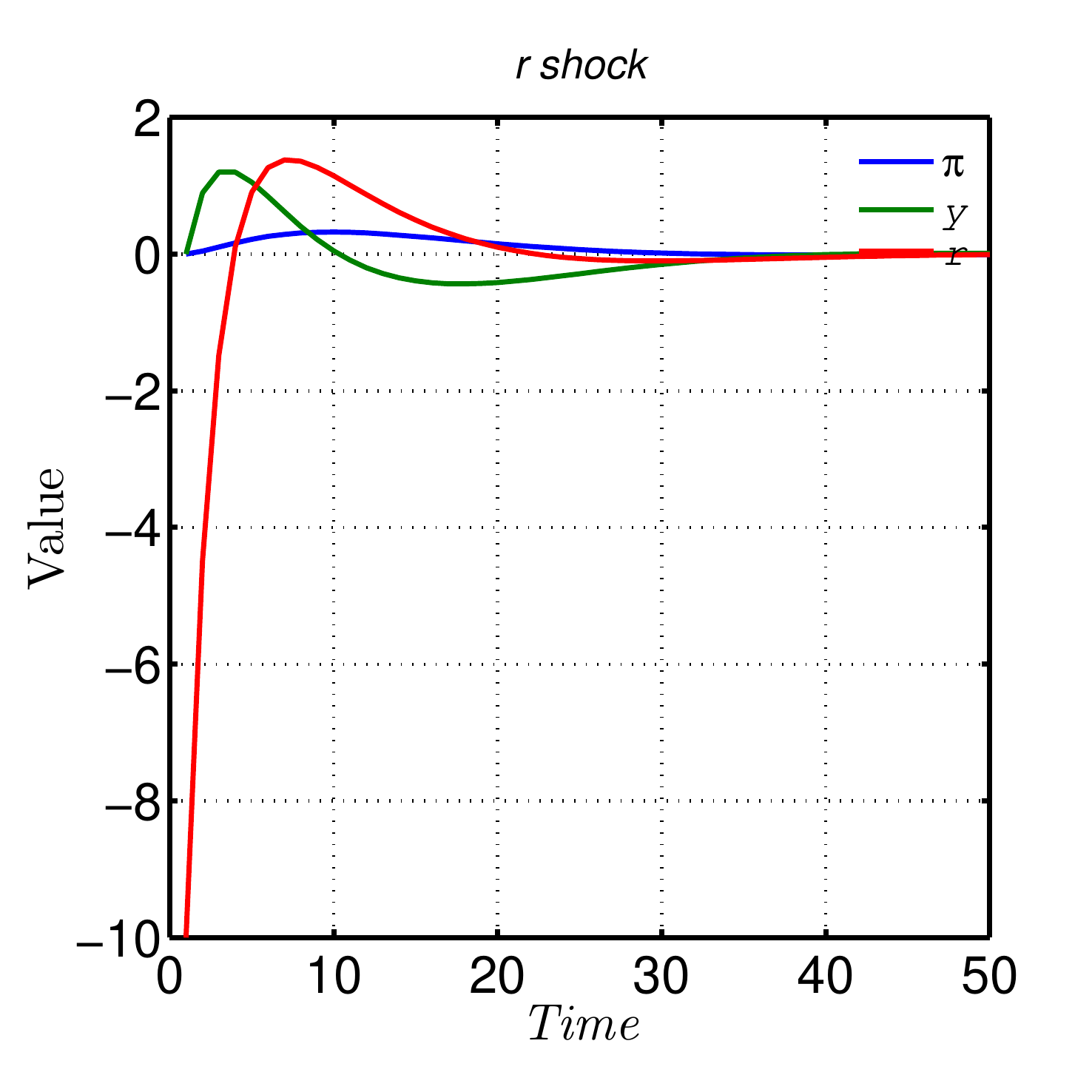} \\
     \includegraphics[width=.221\linewidth]{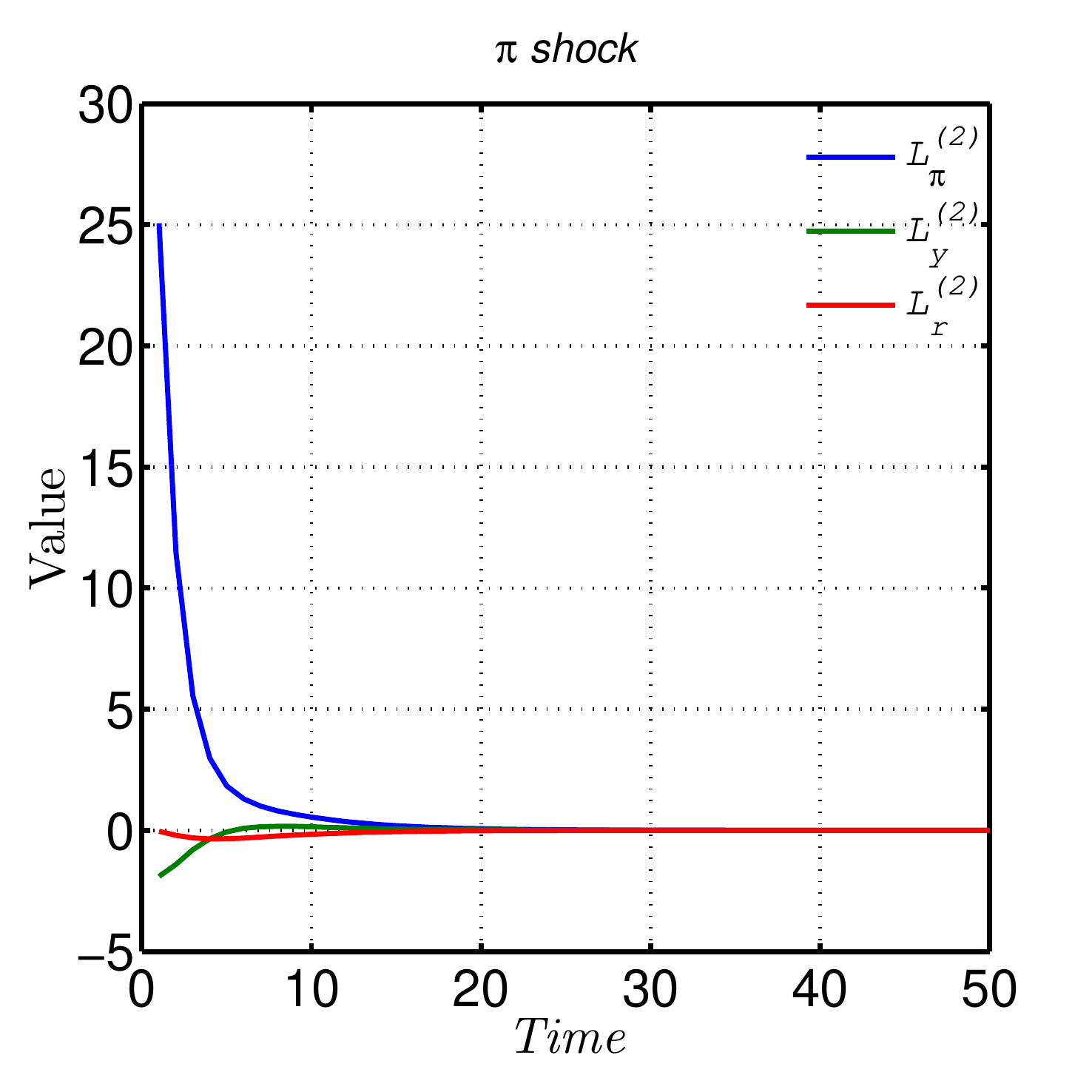}
     \includegraphics[width=.221\linewidth]{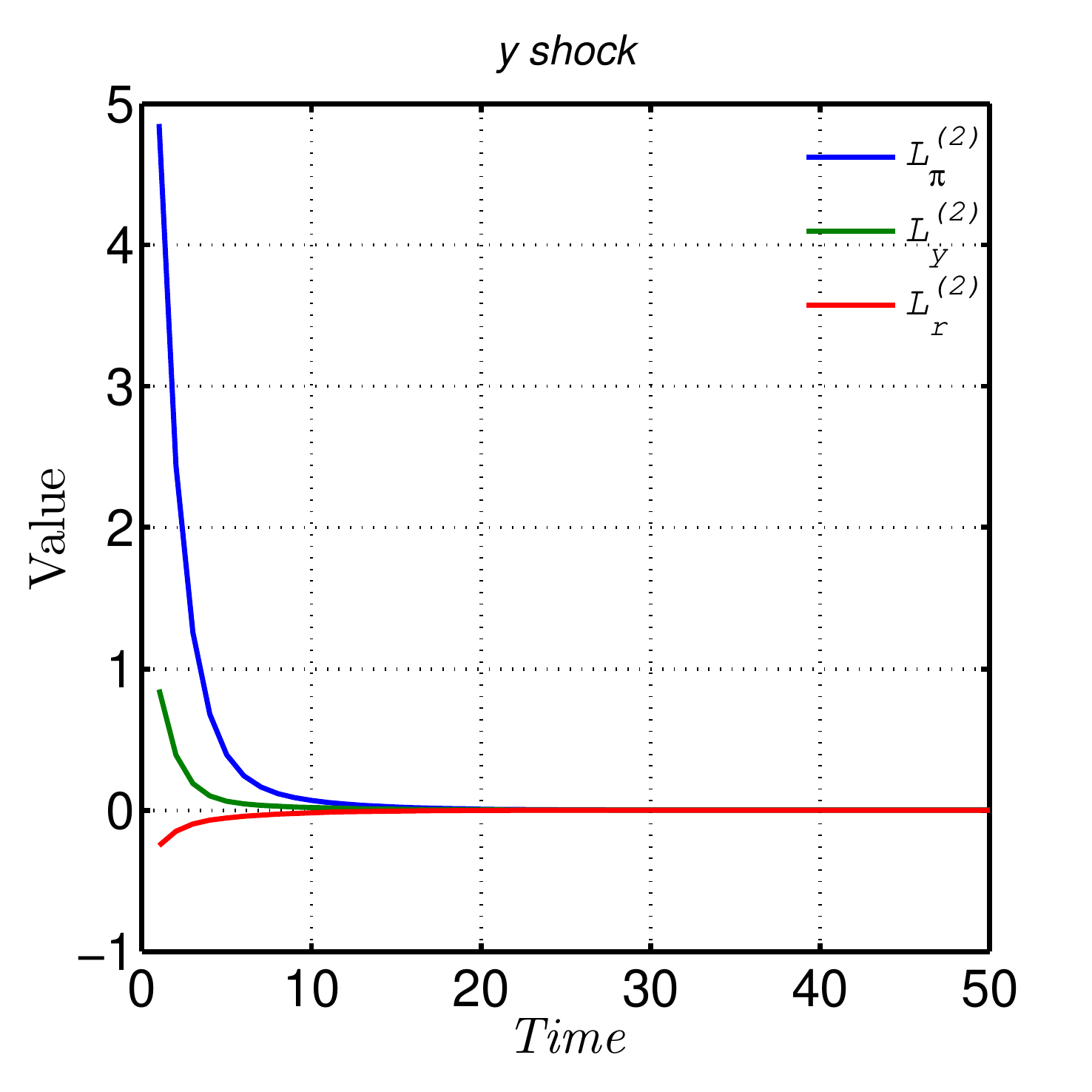}
     \includegraphics[width=.221\linewidth]{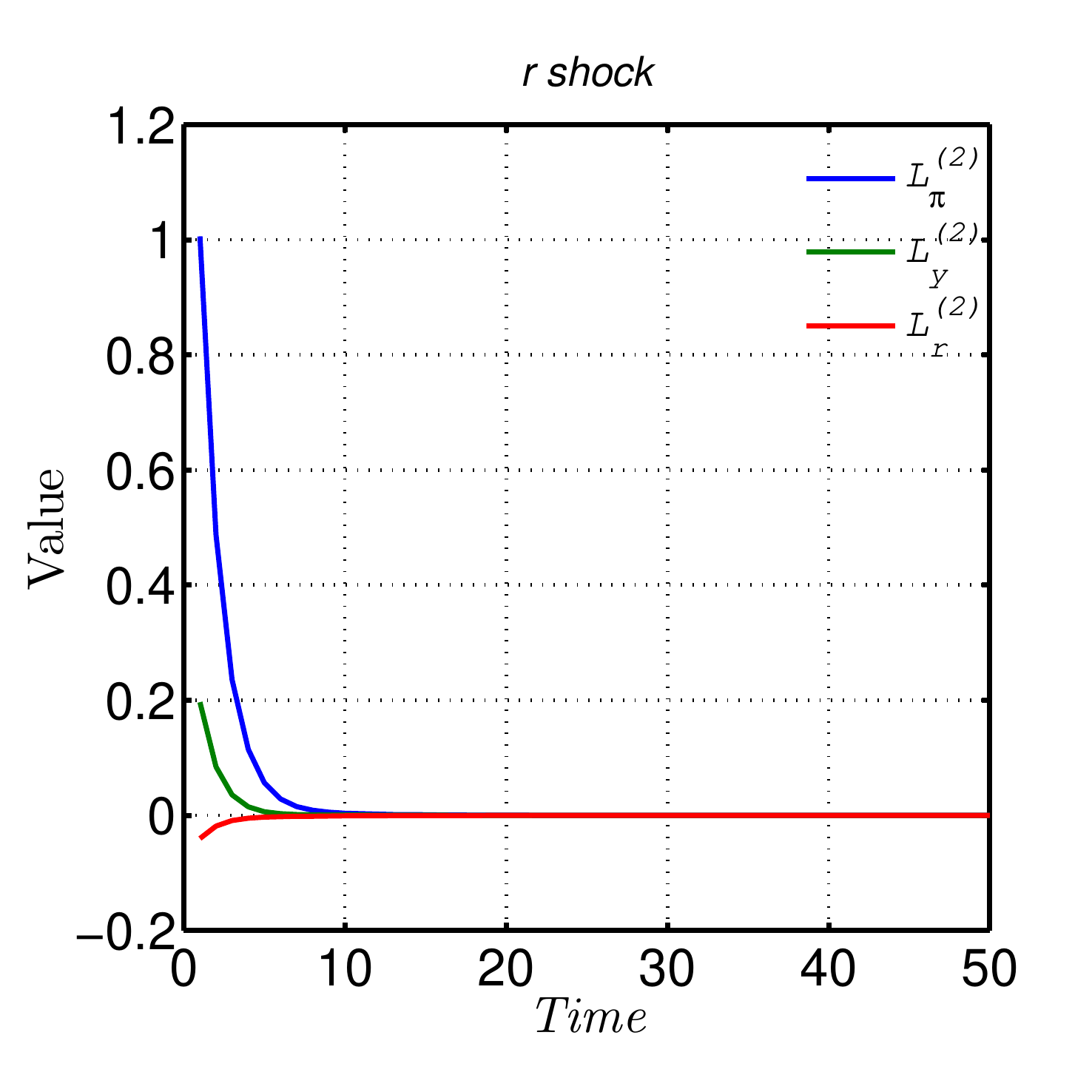} \\
     \includegraphics[width=.221\linewidth]{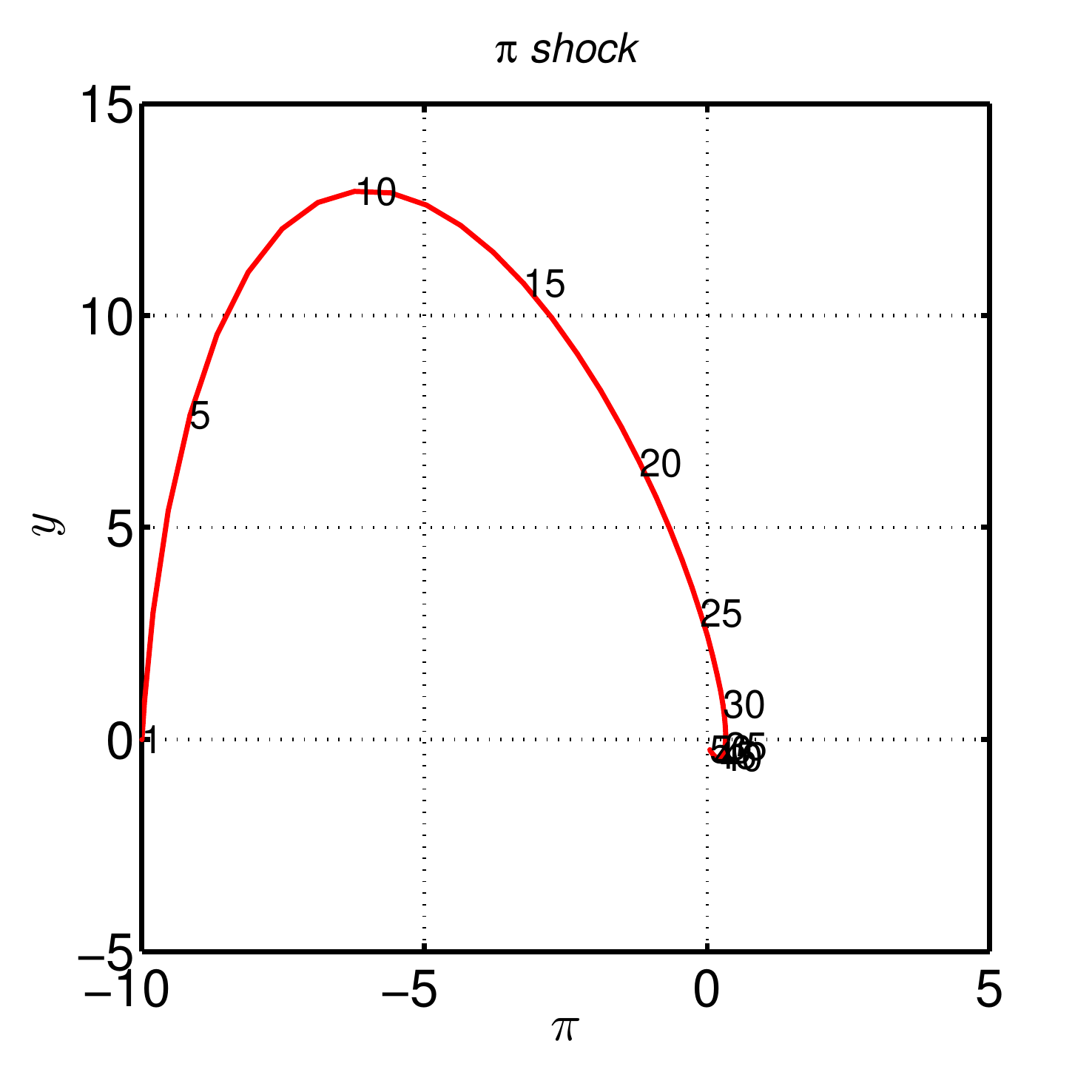}
     \includegraphics[width=.221\linewidth]{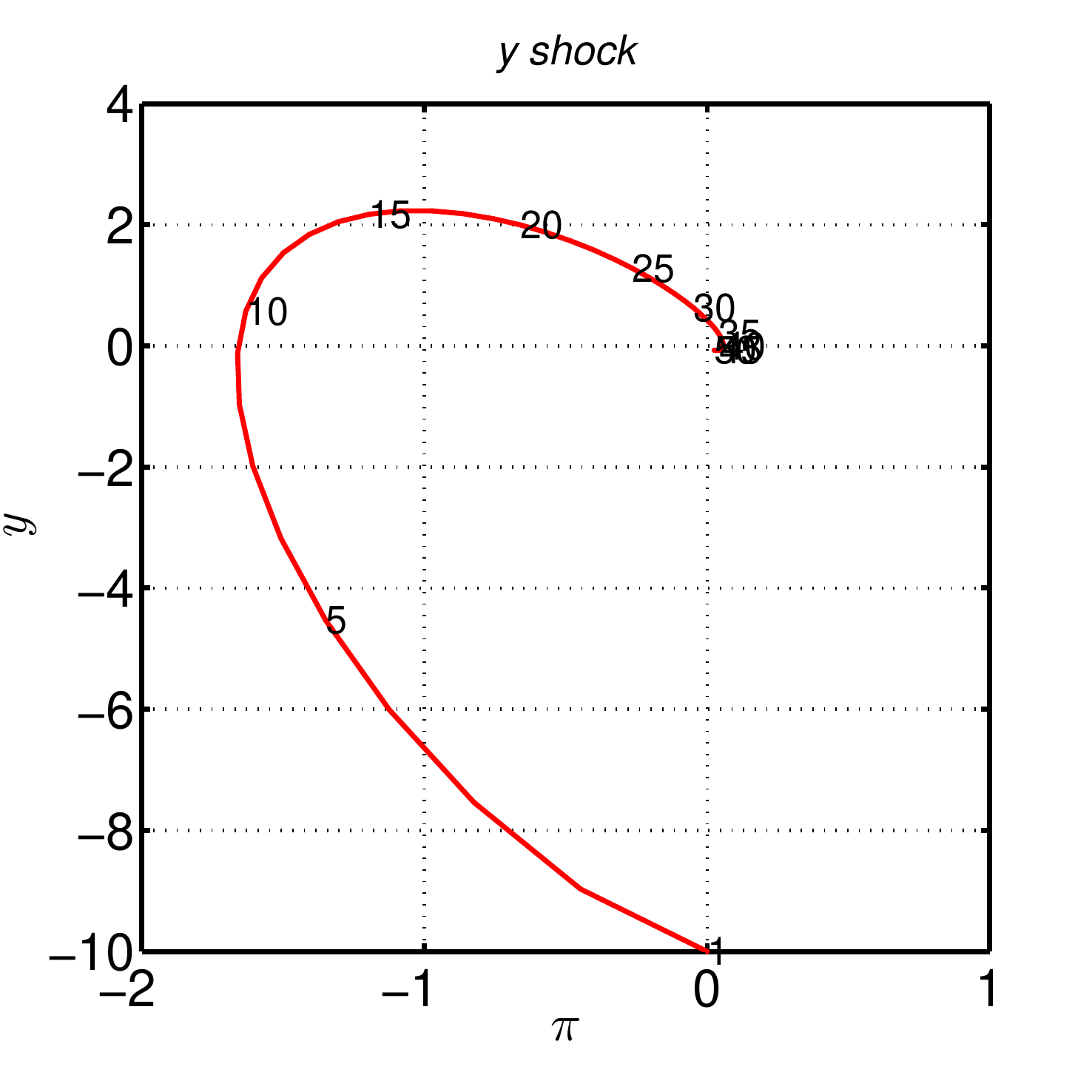}
     \includegraphics[width=.221\linewidth]{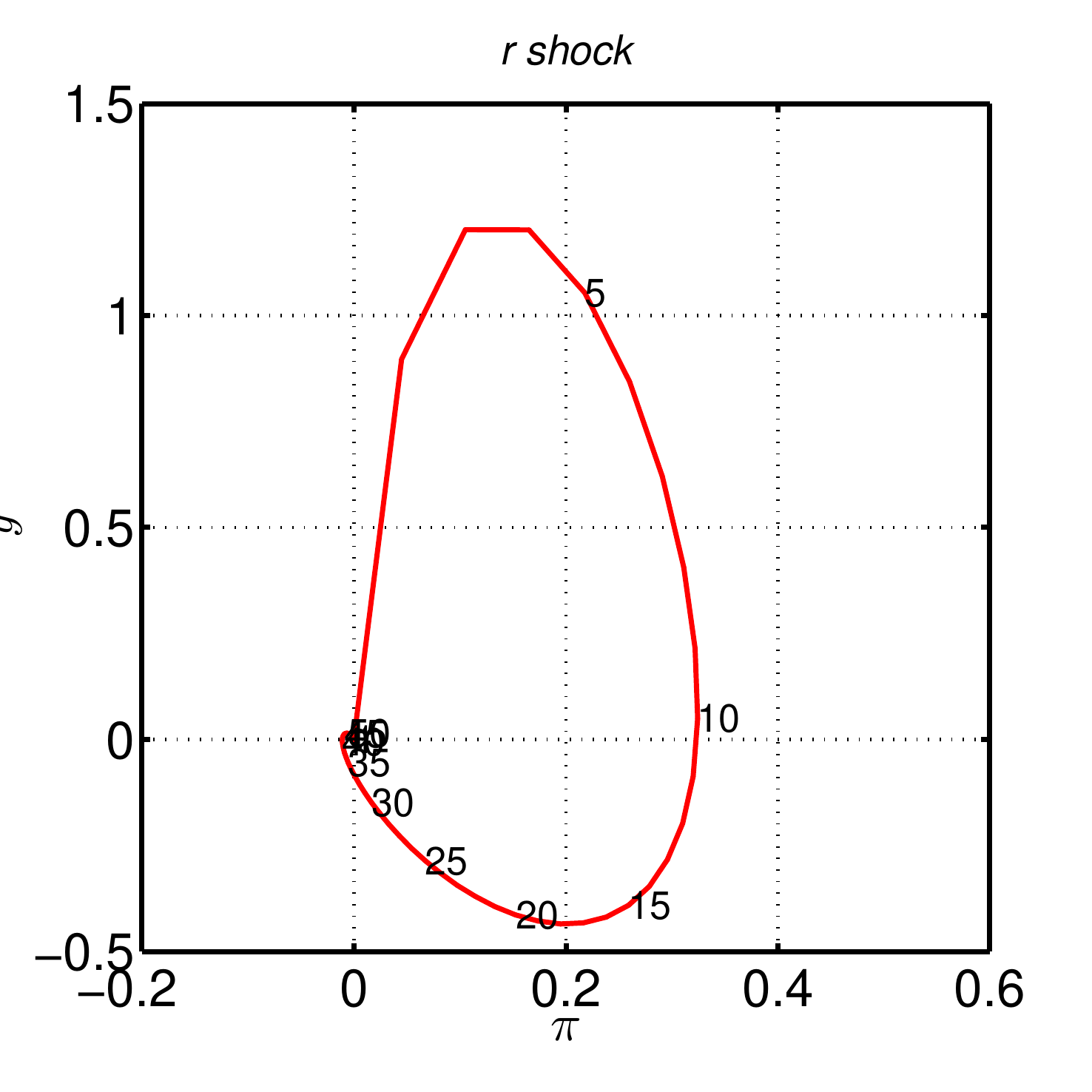} \\
     \includegraphics[width=.221\linewidth]{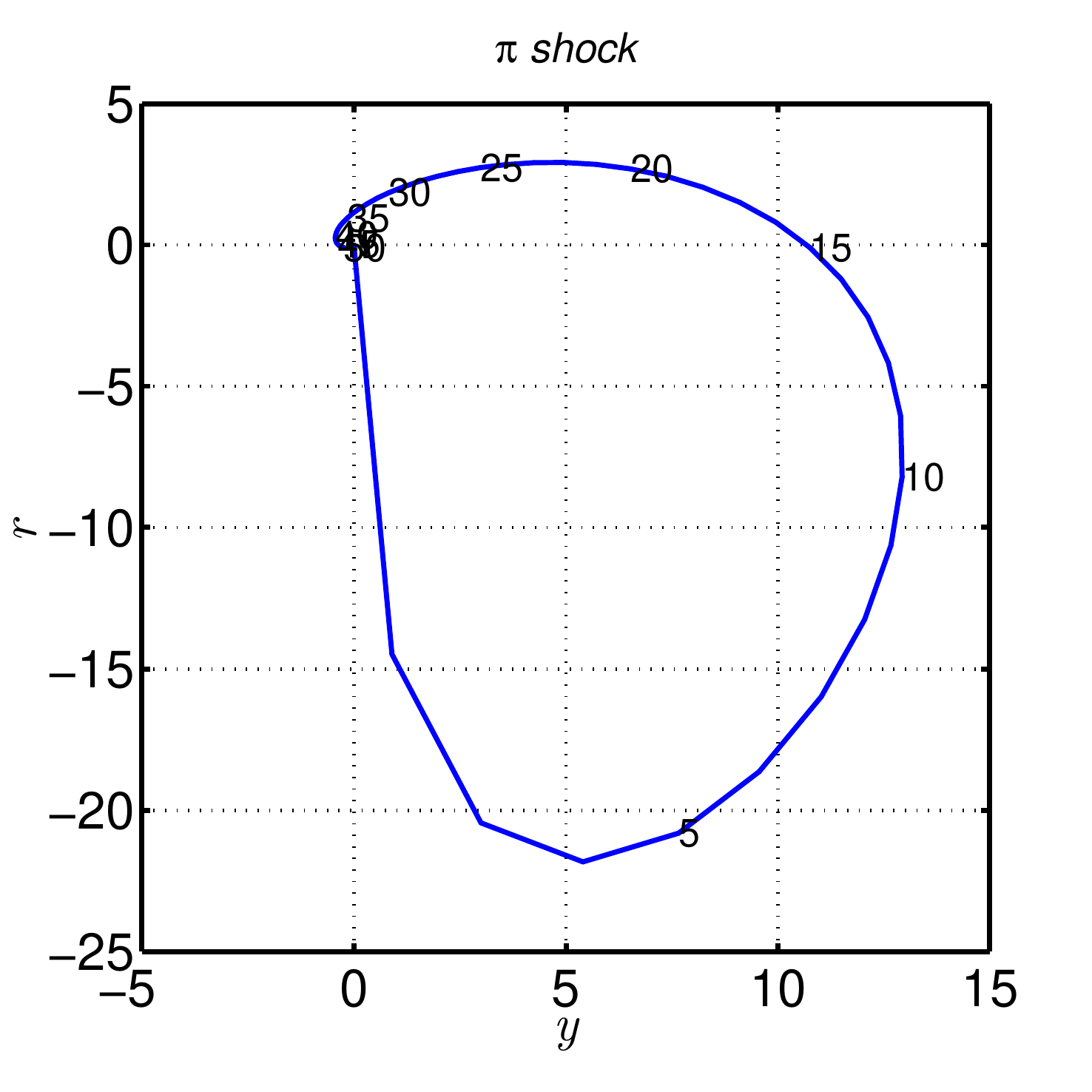}
     \includegraphics[width=.221\linewidth]{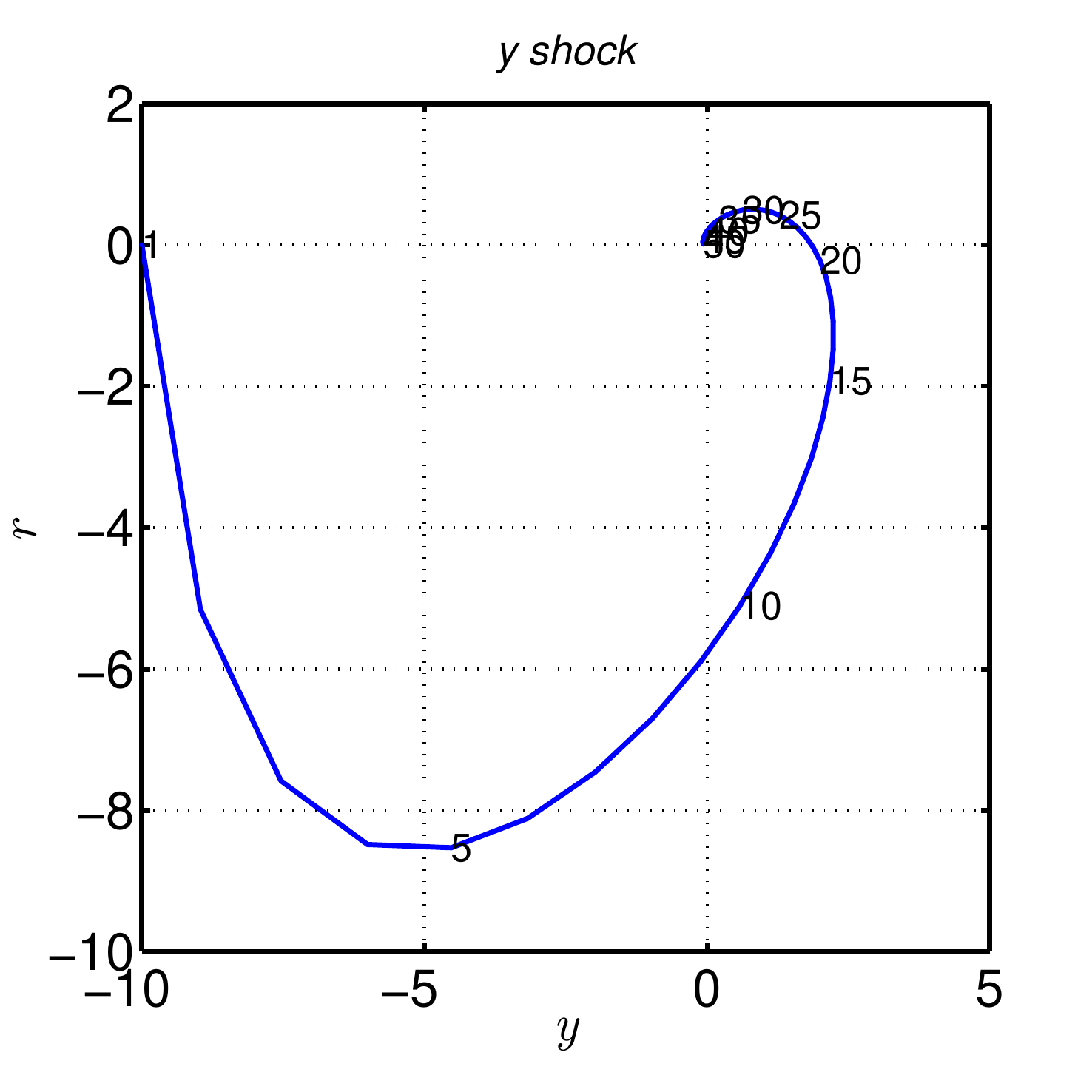}
     \includegraphics[width=.221\linewidth]{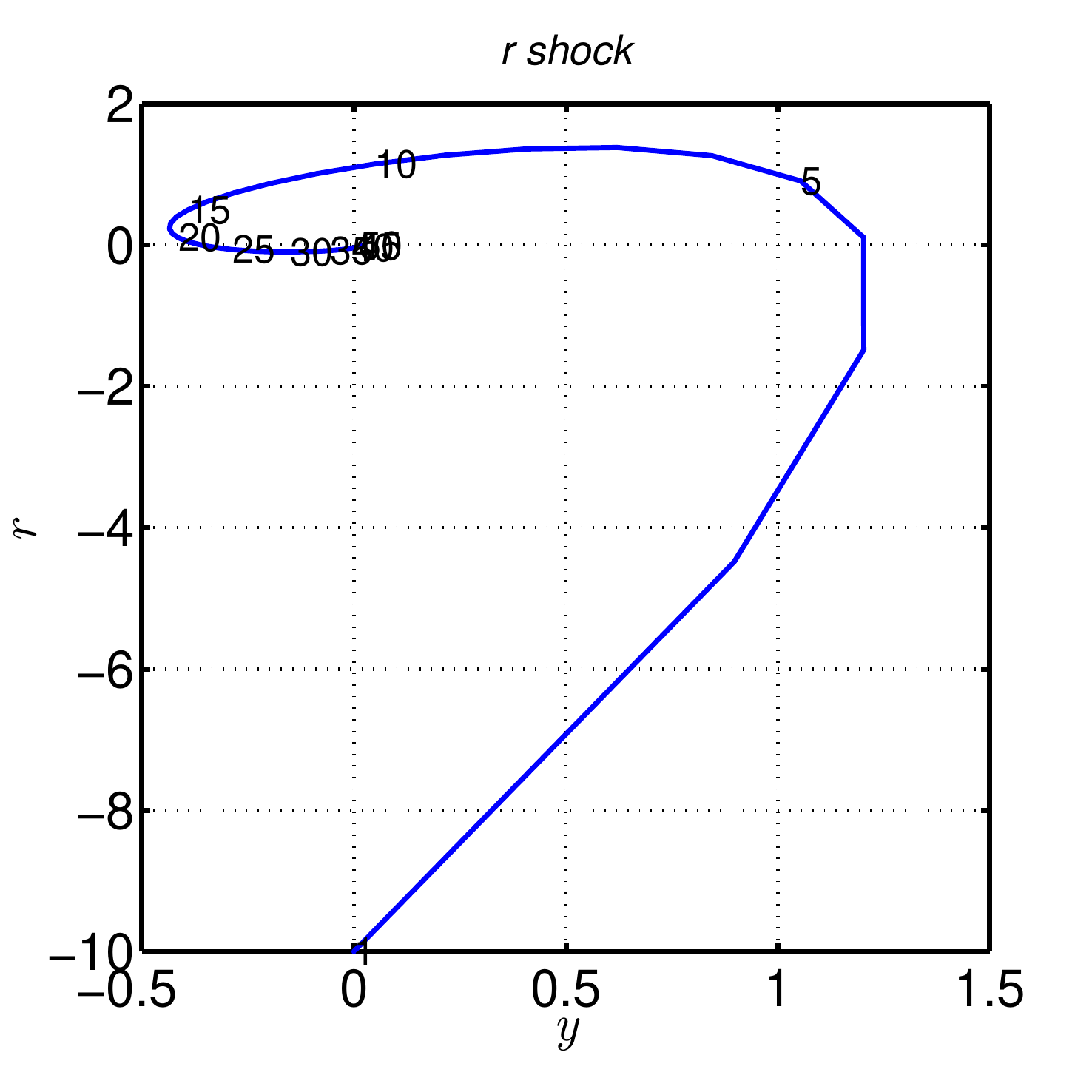} \\
     \includegraphics[width=.221\linewidth]{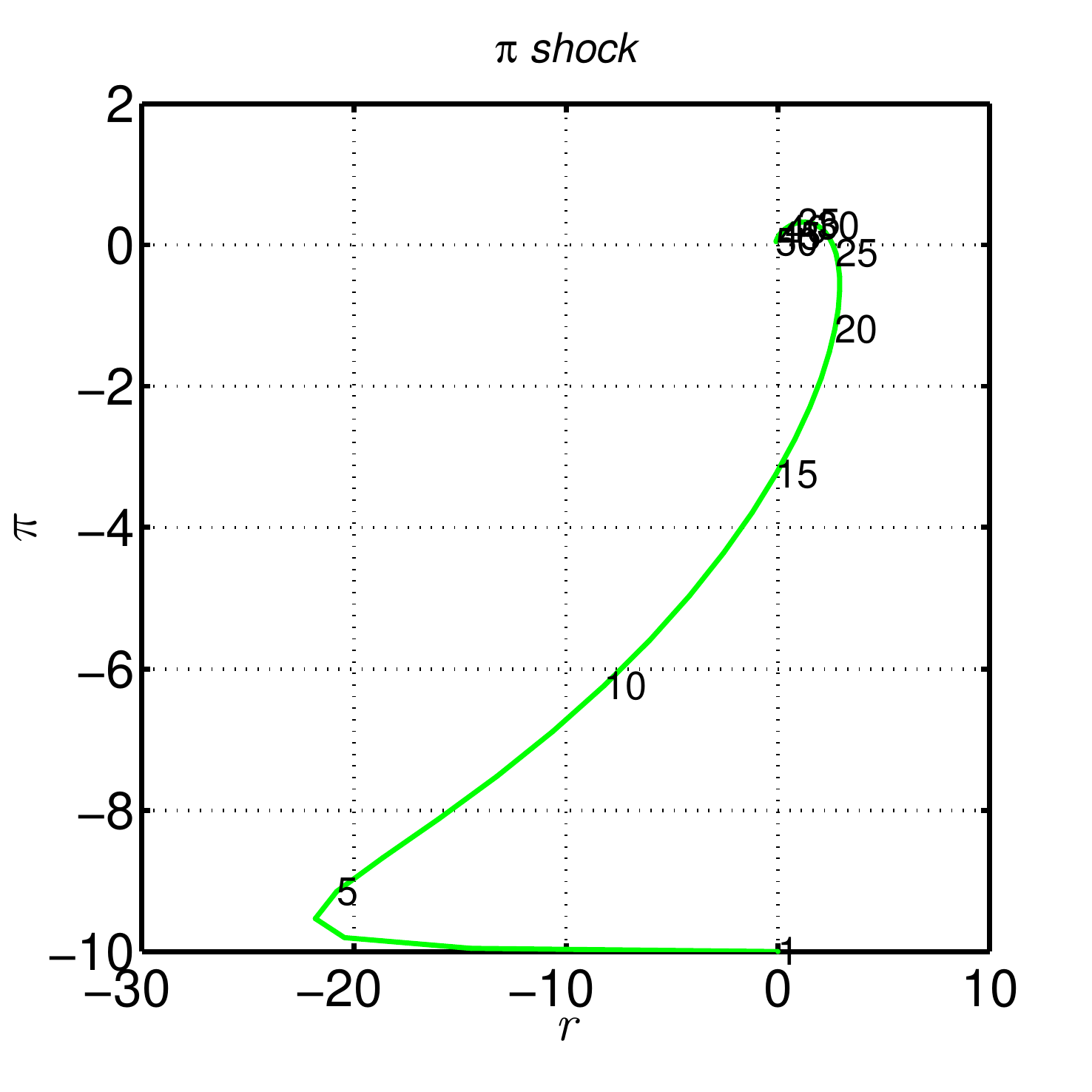}
     \includegraphics[width=.221\linewidth]{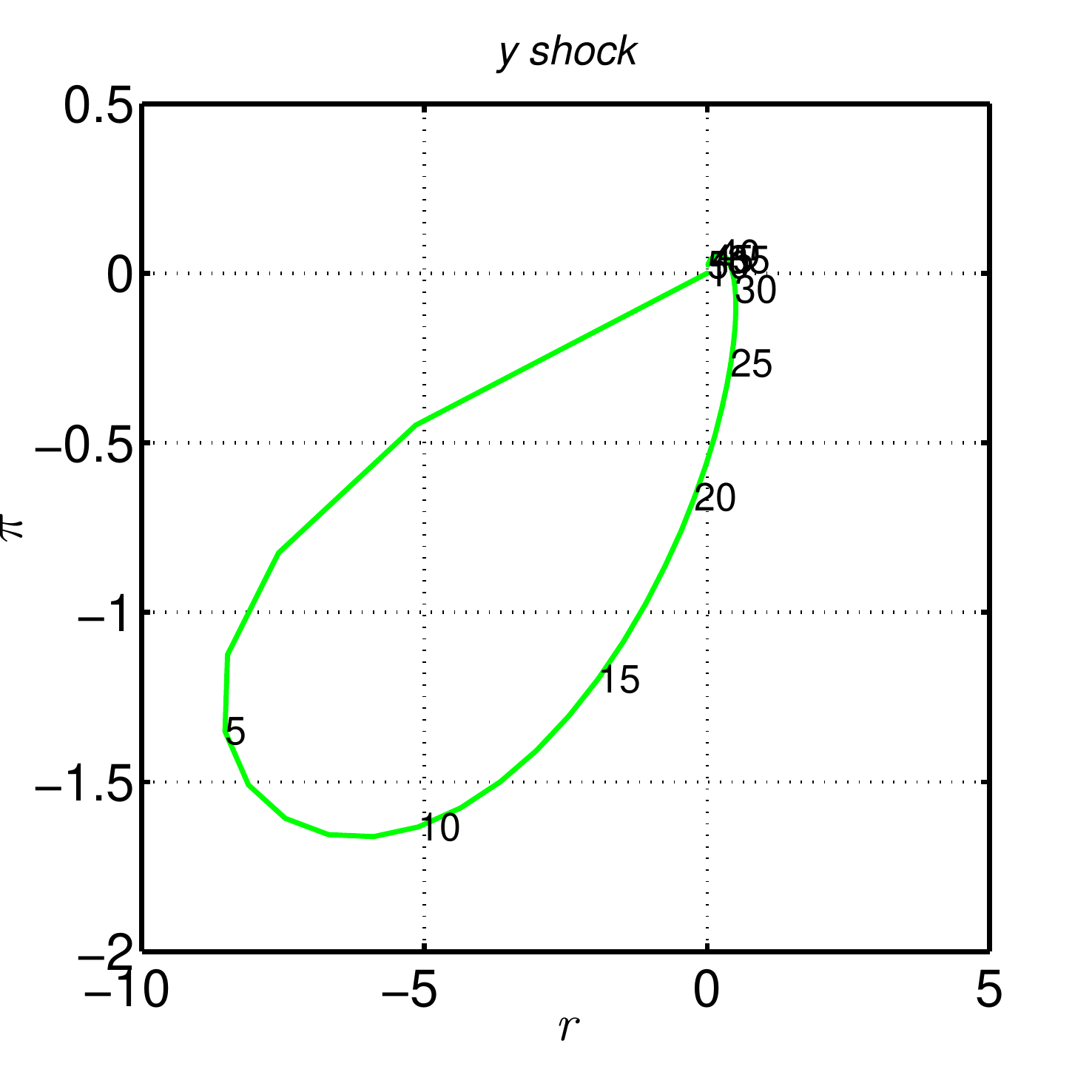}
     \includegraphics[width=.221\linewidth]{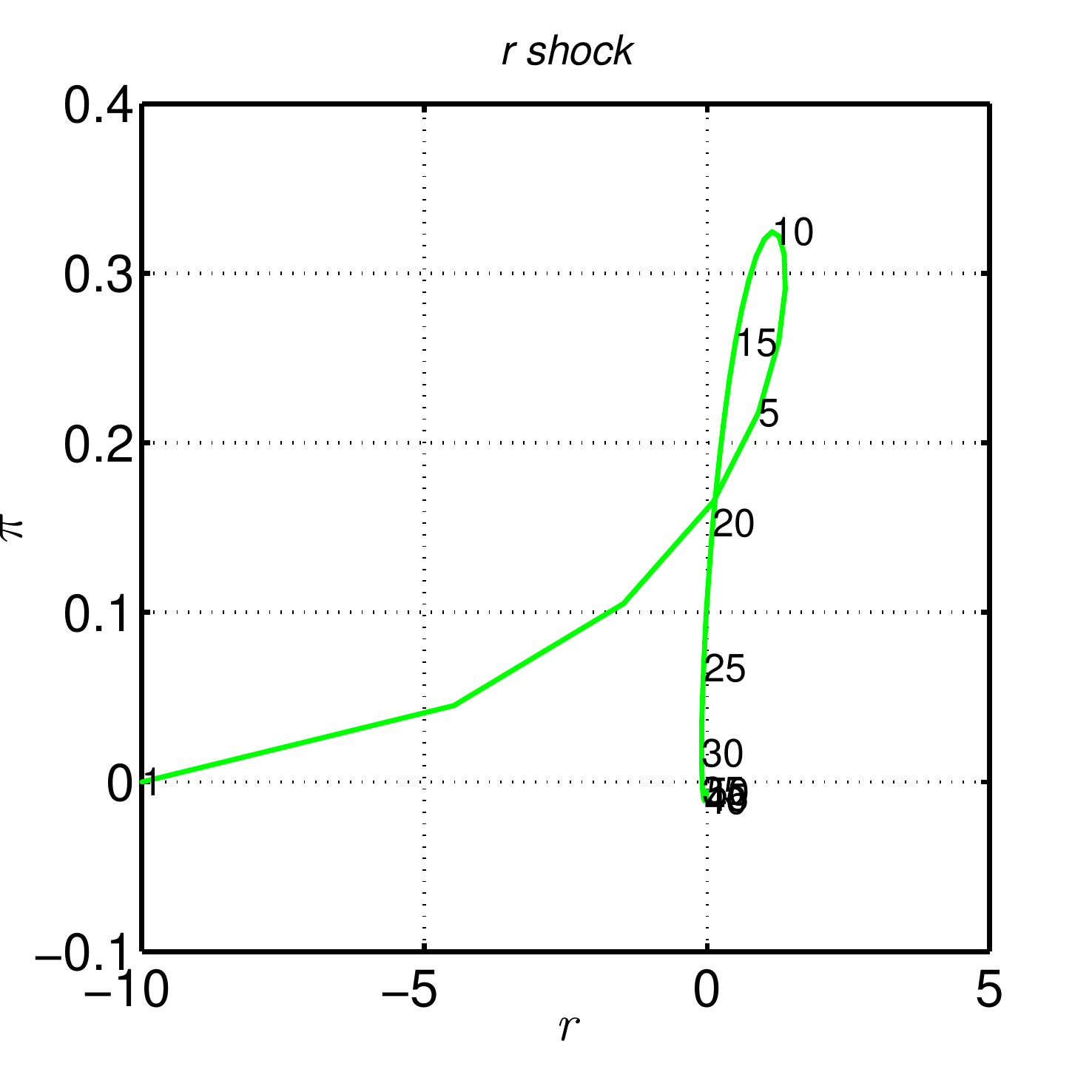}
    \end{center}
    \caption{
        \label{fig:Cycle_and_shock_negative}   The up panel are the presentation of the negative shock and recovering of the general equilibrium. The middle panel are evolutionary trajectory in $\pi-y$ phase diagram, in which we can see the trajectory is of clockwise constantly. The low panel are related $L^{(2)}$ observed, in which we can see  $L^{(2)}_{\pi} > 0$ always (means counter clockwise cycle in $y-r$ phase plane), $L^{(2)}_{r} > 0$ always (means clockwise cycle in $\pi-y$ phase plane).
       }
  \end{figure}

Fig. \ref{fig:Cycle_and_shock_positive} and \ref{fig:Cycle_and_shock_negative} show the results. It is clear that, the system, described by Eq. \ref{eq:agg3x3Eq7}, will return to its equilibrium [0,0,0] quickly (about 50 periods). Importantly, over all the six examples in Fig. \ref{fig:Cycle_and_shock_positive} and \ref{fig:Cycle_and_shock_negative}, we can have the results following:
\begin{description}
  \item[(1) On $L^{(2)}_{r}$] |  all of the $L^{(2)}_{r}$ in the six shocks are negative, means the cycles are definitively clockwise  in $\pi-y$ phase plane;
  \item[(2) On $L^{(2)}_{\pi}$] | all of the $L^{(2)}_{\pi}$ in the six shocks are positive, means the cycles are definitively counterclockwise in $y-r$ phase plane;
  \item[(3) On $L^{(2)}_{y}$] |  the $L^{(2)}_{y}$ in the six shocks are not definitively, means the cycles are not confident in $\pi-r$ phase plane;
\end{description}
These results consist with the velocity field pattern simulated shown in Fig \ref{fig:3x3}, the analysis results shown in Eq. \ref{eq:uneq} as well as the simulated results shown in Fig. \ref{fig:TheoL2evaluted}. \\

\subsubsection{Cycle and noise amplitude \label{sec:ge_shock}}

\begin{figure}
    \begin{center}
     \includegraphics[width=.30\linewidth]{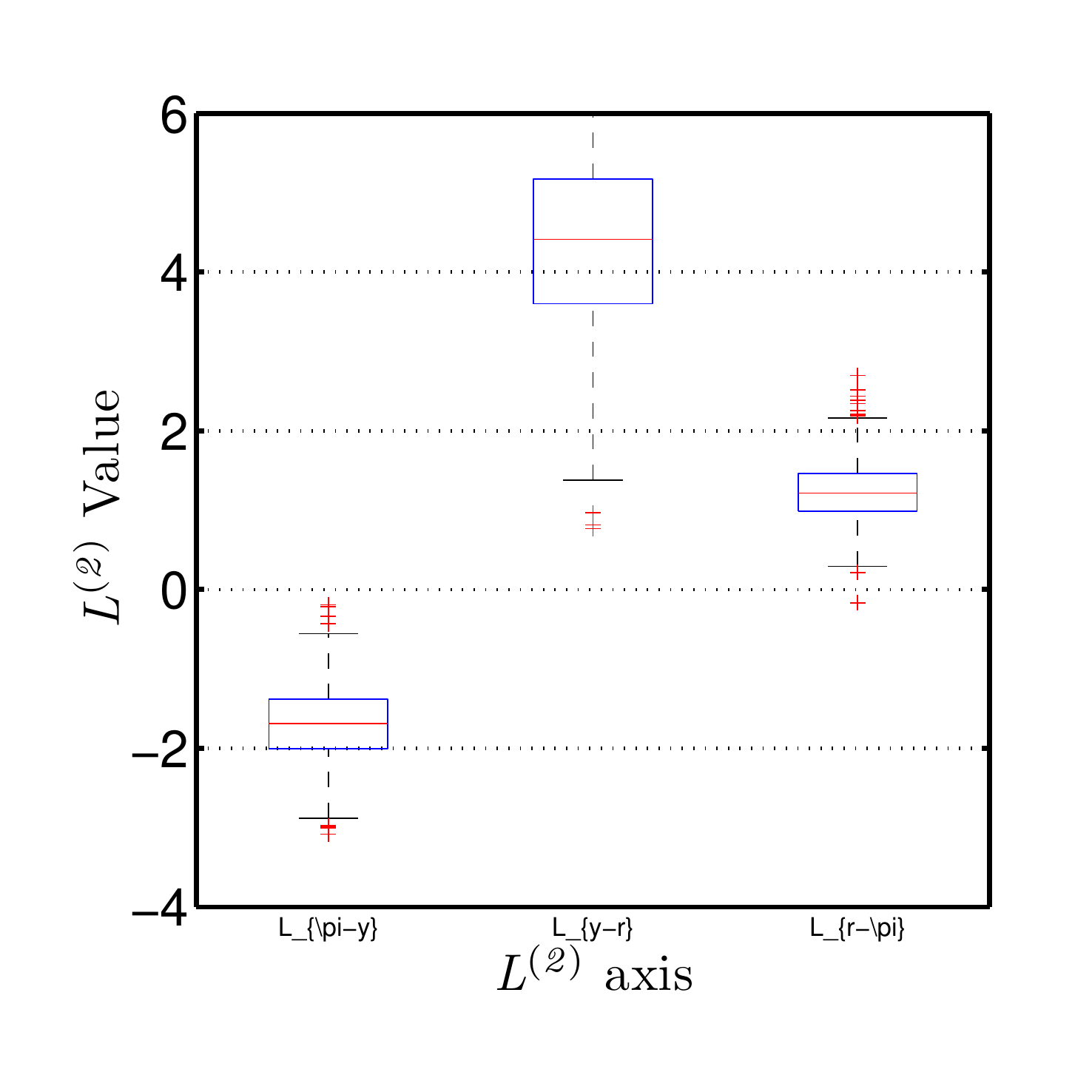}
     \includegraphics[width=.30\linewidth]{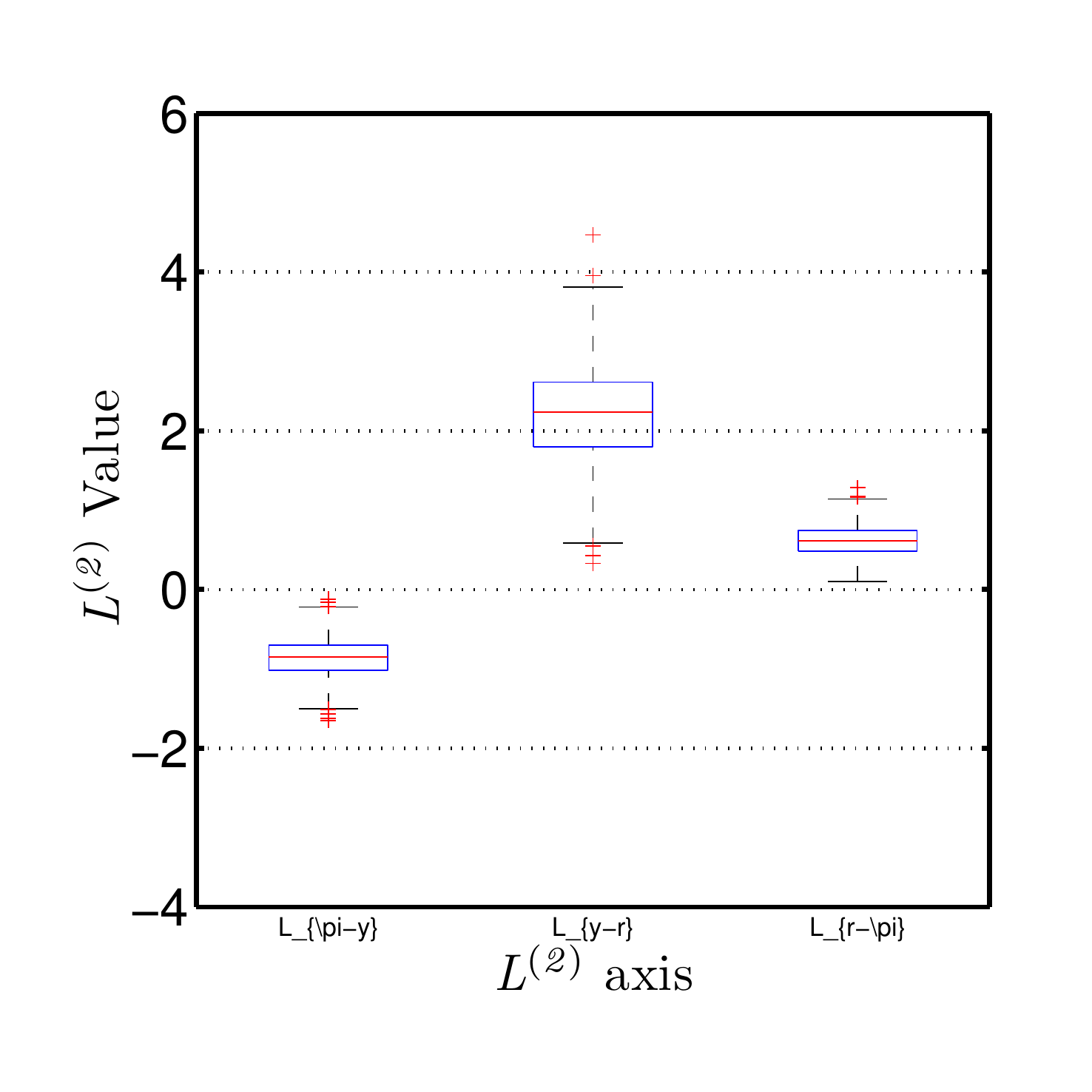}
     \includegraphics[width=.30\linewidth]{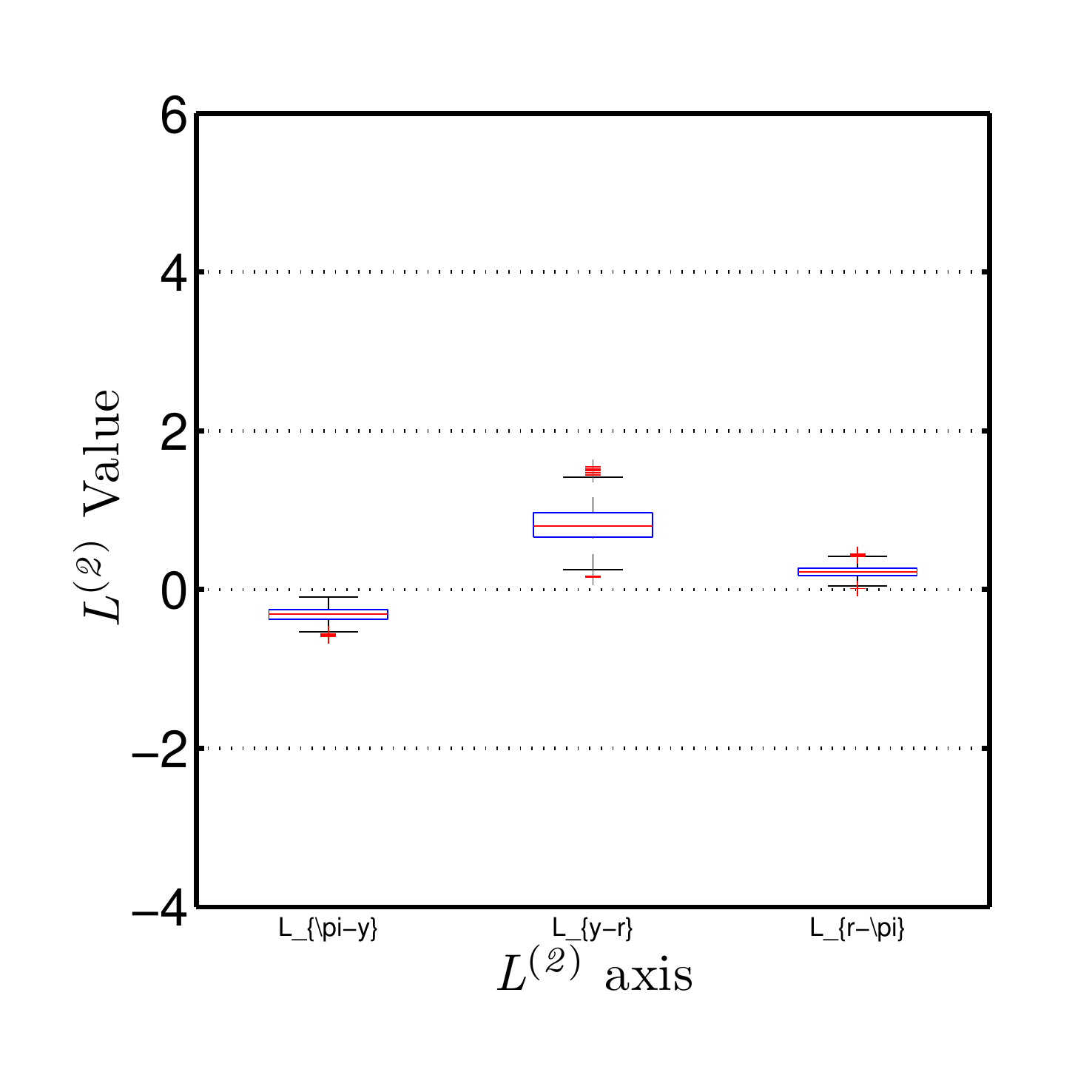}\\
    \end{center}
    \caption{
        \label{fig:noise_strength}  The  quartiles box  of distribution of the observed $L^{(2)}$ having  $\alpha$ = [0.7, 0.5, 0.3] (from left to right) shocks to the general equilibrium. We can see $L^{(2)}$ are more smaller when noise is smaller, because the median (red line) is smaller and closer to zero. Accordingly we have, the strength of cyclic motion, denoted by $L^2$, positively depends on noise strength. 
       }
  \end{figure}

 We control the amplitude of noise in DSGE to test the dependence of cycles on noise. In the original DSGE model, the noise is constantly
appearing in each month period (see the the right most
term in Eq.~\ref{eq:agg3x3Eq2}). We can see that,
  cycles will be smaller when noise amplitude being smaller. Main results are shown in Fig \ref{fig:noise_strength} and explained following.

  According to \cite{de2012lectures}, the standard deviation of normally distributed noise ($\alpha$) shock on output $y$, inflation $\pi$ and nominal interest rate $r$ are assumed to be 0.5. We set this value to be $\alpha$ = 0.7, 0.5 and 0.3. Assuming the system in equilibrium [0, 0, 0] at $t-2$,  adding an $\alpha$ shock at $t-1$, and then at $t$ and then at $t+1$. We use $F$ (see Eq. \ref{f12}) as the action at each transition. In formula, one  monthly  transition can be is presented by
  $$ x_{\tau}  = F x_{\tau-1} + \alpha B^{-1} \epsilon, $$
in which, $\tau$ is time in month,  $\epsilon$ is a 3D vector of the standard deviation of normally distributed noise, and $B$ is the left most $3\times 3$ matrix in Eq. \ref{eq:agg3x3Eq2}. Then, repeat such monthly  transition 12 times, a (year) transition  $x_{t} \rightarrow x_{t+1}$ can be obtained. At each sample ($x_{t-1} \rightarrow  x_{t} \rightarrow x_{t+1}$), we can calculate its $L^{(2)}$. To mimic the empirical data which having 53 years, we use 53 years repeated as a sample group in which we can obtained its average $L^{(2)}$. We repeat such procedure for 1000 (group sample) times and plot the distribution of the  average $L^{(2)}$. 

In Fig \ref{fig:noise_strength},
having shock $\alpha$ = $[0.7, 0.5, 0.3]$ respectively,
we can see all of the median ($L^{(2)}$, in red line) become smaller when the amplitude of noise ($\alpha$) become smaller.
That is to say, the strength of cycles, when denoted by $L^{(2)}$, positively depends on noise amplitude.

\section{reference}


\end{CJK}
\end{document}